\documentclass{jfm}

\usepackage{graphicx}

\graphicspath{{./figures/pdf/}}

\usepackage{mathtools}
\usepackage{natbib}
\usepackage[section]{placeins}
\usepackage{units}
\usepackage{enumerate}
\usepackage{multirow}   
\usepackage{stfloats}   
\usepackage{booktabs}   
\usepackage[mathlines]{lineno}  

\usepackage[final]{changes}

\newcommand{\cmt}{cm$^3$}
\newcommand{\degree}{\ensuremath{^\circ}}

\newcommand{\pdiff}[2]{\dfrac{\partial #1}{\partial #2}}

\newcommand{\rd}{\mathrm{d}}

\newcommand{\St}{St}  

\newcommand{\U}{\ensuremath{\ug/\umf}} 
\renewcommand{\u}{\boldsymbol{u}}
\newcommand{\ug}{\ensuremath{w_g}}
\newcommand{\umf}{\ensuremath{u_{m\!f}}}

\renewcommand{\v}{\boldsymbol{v}}

\newcommand{\xf}{\ensuremath{x_{\!f}}}

\providecommand\bnabla{\boldsymbol{\nabla}}

\newcommand{\pd}[2]{\ensuremath{\frac{\partial #1}{\partial #2}}}

\newcommand{\pdtwo}[2]{\ensuremath{\frac{\partial^2 #1}{\partial {#2}^2}}}
\newcommand{\tsf}[2]{\ensuremath{\textstyle\frac{#1}{ #2}}}

\newcommand{\phibar}{\overline{\phi}}

\newcommand{\hvis}{h_{\mathrm{vis}}}
\newcommand{\hmax}{h_{\mathrm{max}}}
\newcommand{\vmax}{v_{\mathrm{max}}}
\newcommand{\phimeas}{\bar{\phi}_{\mathrm{meas}}}
\newcommand{\phiest}{\bar{\phi}_{\mathrm{est}}}
\newcommand{\Qnom}{Q_{\mathrm{nom}}}

\title{Steady \& unsteady fluidised granular flows down slopes}

\author{D.~E.~Jessop\aff{1, 2}\corresp{\email{d.jessop@opgc.fr}},
  \and A.~J.~Hogg\aff{3}, 
  \and M.~A.~Gilbertson\aff{4},
  \and C.~Schoof\aff{2}
}
  
\affiliation{\aff{1}Laboratoire Magmas et Volcans, Universit\'e Clermont-Auvergne-CNRS-IRD, OPGC, Clermont-Ferrand, France
  \aff{2}Department of Earth, Ocean and Atmospheric Sciences, University of British Columbia, Vancouver, Canada
  \aff{3}School of Mathematics, University of Bristol, University Walk, Bristol, BS8 1TW, United Kingdom
  \aff{4}Department of Mechanical Engineering, University of Bristol, University Walk, Bristol, BS8 1TR, United Kingdom,
}

\newcommand{\thedate}{\today}
\date{\thedate}

\begin{document}
\maketitle

\begin{abstract}
  Fluidisation is the process by which the weight of a bed of particles is supported by a gas flow passing through it from below.  \added{When fluidised materials flow down an incline, the dynamics of the motion differ from their non-fluidised counterparts because the granular agitation is no longer required to support the weight of the flowing layer.  Instead, the weight is borne by the imposed gas flow and this leads to a greatly increased flow mobility. In this paper, a framework is developed to model this two phase motion by incorporating a kinetic theory description for the particulate stresses generated by the flow.  In addition to calculating numerical solutions for fully developed flows, it is shown that for sufficiently thick flows there is often a local balance between the production and dissipation of the granular temperature.  This phenomenon permits an asymptotic reduction of the full governing equations and the identification of a simple state in which the volume fraction of the flow is uniform.}  The results of the model are compared with new experimental measurements of the internal velocity profiles of steady granular flows down slopes.  The distance covered with time by unsteady granular flows down slopes and along horizontal surfaces and their shapes are also measured \added{and compared with theoretical predictions developed for flows that are thin relative to their streamwise extent}. For the horizontal flows, it was found that resistance from the sidewalls was required in addition to basal resistance to capture accurately the unsteady evolution of the front position and the depth of the current and for situations in which side-wall drag dominates, similarity solutions are found for the experimentally-measured motion. 

\end{abstract}

\section{Introduction}
\label{sec:introduction}

Particles are often transported in the form of dense currents under the influence of gravity.  Their bulk flow rate is greatly enhanced if part or all of their weight is supported by a gas flow through them.  When particles are poured onto a slope that is less than their angle of repose, they are held stationary by the action of contact friction and merely flow down the surface of the pile in a thin layer as more grains are successively added.  When they are poured onto a slope that is steeper than the angle of repose, a thin, dense current forms in which the particles move in bulk down the slope \citep[see, for example,][]{ishida_flow_1980, gdr_midi_dense_2004}.  If a gas is passed vertically through the particles then the drag it exerts on the particles bears some of their net weight and hence the frictional forces decrease.  Consequently, the effective angle of repose of the particles decreases as does the minimum slope angle at which bulk flow flow takes place \citep{nott_frictional-collisional_1992}.  When the gas flow is sufficiently large for the entire weight of the particles to be supported (i.e.\ the particles are fluidised), then bulk frictional forces are insignificant and very mobile currents form, even on horizontal surfaces.  The influence of a fluidising gas flow through particles on their mobility is exploited widely in industrial settings  where it is necessary to transport bulk materials either to move them from one place to another using air slides (which can be several kilometres long), or to keep  horizontal surfaces clear of particles in pieces of processing equipment such as circulating fluidised beds \citep{Savage201335}.  There are also features in many particulate  environmental flows in which there is significant upward gas flow  and this  enhances  their speed and range \citep[e.g.][]{druitt_pyroclastic_1998,roche_experimental_2004}.  

There have been extensive studies of the flow of particles down a slope and of some of the effects of fluidisation.  A common approach to mathematically modelling these motions is based on a continuum description that couples expressions of mass conservation with expressions of the balance of momentum within each phase \citep[e.g.][]{nott_frictional-collisional_1992}.  Under this approach, the  fluidised material is treated as two inter-penetrating phases that interact with each other.  The models do not resolve the motion of individual particles, but rather the evolution of averaged, bulk properties, which depend upon the net effect of direct interactions between particles within the current and between the particles and their surroundings.  The duration of  contacts between the constituent particles has important consequences:  if the contacts are sustained then they are likely to be frictional in nature; if they are instantaneous then they are collisional in nature \citep[e.g.][]{campbell_granular_2006}.   The stresses induced by instantaneous collisions between pairs of particles (i.e.\ in dilute and rapid granular flows) can be evaluated through the use of granular kinetic theories \citep{jenkins_theory_1983}, in which a key dependent variable is the granular temperature, $T$, a measure of the variance of the instantaneous velocity field.  Hydrodynamic equations of motion have then been derived for granular materials that are much like those for dense gases except there is substantial energy dissipation through inelastic collisions \citep[see][for example]{jenkins_theory_1983,lun_kinetic_1984,haff_grain_2006}.  It is possible, of course, for there to be collisions over a range of durations and these ideas do not translate to dense and slowly-shearing flows where contacts are prolonged and thus, in part, frictional.  The action of the interstitial fluid is a further factor that needs to be considered when modelling granular flows.  For example, in contrast to the original studies of granular kinetic theories, \cite{koch_particle_1999} proposed that interaction with the fluid could generate agitation within the flows and that fluctuating viscous forces could be the generators of particle temperature.  

There have been relatively fewer studies that report granular flows that are aerated or fluidised.  An early approach was to treat the fluidised particles as a non-Newtonian fluid of power-law rheology, sometimes with a yield stress \citep{botterill_flow_1973,botterill_open-channel_1979,ishida_flow_1980,Savage201335}.  This approach can be made to work well in specific practical situations \citep{singh_flow_1978}, but is entirely reliant on empirical methods to determine the effective rheology in each circumstance since such approaches do not capture the fundamental dynamics of the particle motion.  

A more fundamental approach is to model the evolution of averaged properties of the inter-penetrating phases. \citet{Ogawa1980} modelled steady, one-dimensional fully-fluidised currents down slopes.  They derived constitutive relations based on the collisions between a particle and its neighbours, which were represented by an imaginary spherical shell surrounding it.  This resulted in a balance between collisional stresses and gravitational forces.  \citet{nott_frictional-collisional_1992} coupled a kinetic theory for collisional grain flows with a Coulomb-like model for frictional effects to predict the bulk mass flow rate of aerated grains down an inclined channel.  The experiments and model featured gas flow rates up to the minimum required to fully fluidise the particles.  Their mathematical model of friction in the flows followed \citet{johnson_frictional-collisional_1987} and \citet{johnson_frictional-collisional_1990} and assumed that the frictional component (dominant at high particle volume fractions, $\phi$) was simply added to the collisional component (dominant at low $\phi$).  They pursued a similar approach to the interaction term between the gas and the particles adding together a contribution based on the Ergun equation (dominant at high $\phi$) and one from the Richardson and Zaki equation (dominant at low $\phi$).  No contribution was included from slip between the two phases in the direction of the slope.  \citet{Oger201322} took a similar approach (although with some different closures of the models), again retaining a frictional term, and solved the resulting equations using the MFIX numerical code to study the dynamics of granular motion within air slides, computing the steady, fully-developed velocity and granular temperature fields for flows within a channel of rectangular cross-section.   Finally, \citet{eames_aerated_2000} reported the unsteady flow of fluidised materials along horizontal surfaces.  For their system, they showed that collisional stresses would be small compared with those associated with fluid drag and so when fully fluidised, the force balance set hydrostatic pressure gradient against fluid drag terms.   We will show below how our work differs from their modelling framework and yet is able to reproduce features of their experimental results.  

Key to furthering our understanding of the dynamics of fluidised flows is direct and detailed experimental evidence against which theoretical models can be validated.  However, there are few measurements of fluidised granular currents, especially down slopes.  Previous experimental studies have presented bulk properties such as total flow depth and mass flow rate \citep[e.g.][]{nott_frictional-collisional_1992,eames_aerated_2000}.  Some measurements of local properties such as velocity have been made though this has often been with instruments such as optical probes or turbine elements \citep[e.g.][]{botterill_flow_1973,botterill_flow_1976,ishida_flow_1980,nott_frictional-collisional_1992}.  Whilst providing important information, the disadvantages to these techniques are that they lack spatial resolution, are intrusive (especially in fluidised particles \citealp{Rowe1981} and offer only point measurements i.e.\ traverses are necessary to build velocity profiles and they are therefore only suited to steady flows.  More recently, Particle Image Velocimetry (PIV) has been applied to fluidised systems such as static beds \citep{bokkers_mixing_2004} and dam-break experiments over horizontal surfaces of initially fluidised, fine natural volcanic ash \citep{GIR:BV2010}.  PIV has the advantage of offering high spatial resolution, and allows instantaneous velocity fields to be calculated.  The experiments of \cite{GIR:BV2010} had a short-lived phase of quasi-constant flow following the initial release of material; however \deleted{they also experienced compaction of the solid phase during the flow because}the grains were not continuously fluidised along the apparatus.  \added{This meant that }even though the materials were highly expanded initially because of the very small particle size, they decelerated rapidly due to the loss of mobility associated with compaction in the terminal flow phase.  As such, they are not representative of fully-fluidised flows.

The aim of the present work is to understand better the dynamics of fluidised granular flows by providing further experimental evidence and proposing a new unsteady model of these flows that fully takes into account the interaction between the particles and the fluid and incorporates collisional stresses.  Both of these processes play a crucial role in the dynamics of fluidised granular flows in which the gas flow bears most of the weight of the particulate layer and the particle interaction contribute significantly to the shear stresses developed by the flow.  This implies that the dynamics are different from `dry' granular flows in which the role of the interstitial fluid is negligible \citep[e.g.][]{lun_kinetic_1984,forterre_flows_2008,woodhouse_rapid_2010}.  

In this work, experimental measurements were made of granular currents over a range of slope inclinations and conditions and the experimental arrangement is described in \S~\ref{sec:experiments}. The measurements were made in an apparatus that confined the flow between two walls, which enabled the overall size and shape of currents to be measured over time. In addition, PIV was used to measure the velocity profiles of the particles within the currents, enabling their overall behaviour to be linked to their rheology. \S~\ref{sec:model} develops the general continuum model and the equations of motion for the flowing state.  This builds upon the `two-fluid' approach in which the gas and grains are treated as two inter-penetrating phases \citep{jackson_dynamics_2000}.  Fully developed flows are tackled in \S\ref{wide} and compared with experimental observations.  The continuum model in this section is analysed in the regime for which the properties of the flowing layer vary only with distance from the underlying boundary and the solutions are computed numerically and asymptotically in a regime where the flow thickness far exceeds the diameter of an individual grain.  Unsteady and transient effects found in flows along inclined channels are investigated in \S\ref{unsteady} and a new model developed in the `lubrication' regime where the downslope length-scale is much large than that perpendicular to the slope.  Flows along horizontal surfaces differ their counterparts along inclines (\S\ref{horizontal}) and measurements of their inherently unsteady motion are reproduced well by a new self-similar solution to the flow model in the lubrication regime.  Finally our findings are summarised and discussed in \S\ref{discussion}.  \added{We also include two appendices.  In the first we analyse the consequences of an extended kinetic theory, following the constitutive laws of} \citet{jenk07}.  \added{In the second} the effects of the side walls are analysed in the regime that the flow depth is much less than the channel width.

\section{Experimental approach}
\label{sec:experiments}

\subsection{Experimental setup}
\label{sec:setup}
\begin{figure}
  \centering
  \includegraphics[width=.8\textwidth]{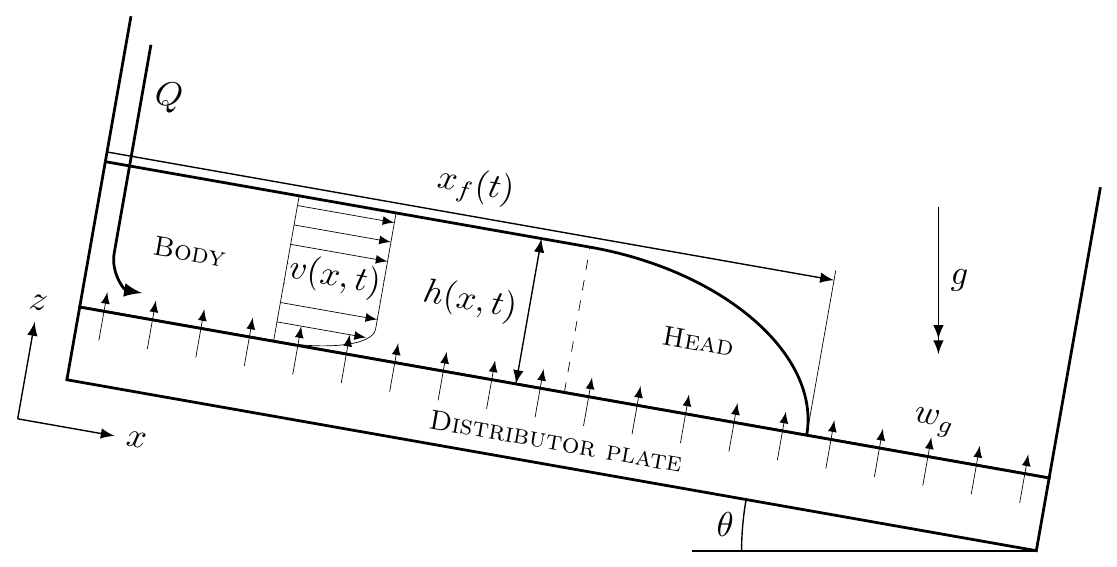}
  \caption{Schematic of flows and experimental setup.  Material is introduced from the raised end of the apparatus at a constant flux, $Q$.  A flow of fluidising gas enters the apparatus at speed $w_g$ through a porous distributor plate and is constant along the entire length of the apparatus and duration of an experiment.  The apparatus can be included to some angle, $\theta$.  The resulting flow has a height profile, $h(x,t)$, length (front position), $\xf(t)$, and longitudinal velocity profile, $v(x,t)$.  For flows down inclined channels, the height of the current increases near to the front (head) to a constant value which is obtained towards the rear of the flow (body).  }
  \label{fig:FlowSchematic}
\end{figure}
The experimental arrangement conformed to that shown in Figure~\ref{fig:FlowSchematic}.  The apparatus was a long, narrow channel (\unit[1]{cm} $\times$ \unit[100]{cm}, \unit[50]{cm} in height) which could be inclined to some angle, $\theta$, to the horizontal. The bottom of the channel was a porous plastic distributor material (Vyon `D') through which dry air was passed from a windbox below at a speed $w_g$, but for which the pressure drop over it was much larger than that through the granular flow.  This ensured that the gas flow was evenly distributed  i.e.\ the gas flow entering the apparatus was uniform and perpendicular to the distributor plate so that at the base of the granular layer, the gas velocity is $\u|_{z=0}=(0,w_g/(1-\phi(0)))$ where $\phi(0)$ is the particle volume fraction evaluated at the base of the flow. The particles were constrained between vertical parallel walls, so that the motion is effectively two-dimensional and the motion of particles within the current could be seen.  The front wall was was made from a glass sheet allowing the flows to be viewed and the other sides were made of aluminium plate.  The rear plate was painted black to increase the contrast between it and the white particles.   Particles entered the apparatus at one end (the uppermost when inclined) through a funnel giving a constant volumetric flux, $Q$,  which could be changed between experiments by changing the aperture of the funnel \citep{Nedd1992}.  \added{The flow rate $Q$ was the flow rate of the current based on the bulk volume when the particles were at rest; so, the volume flow rate per unit width of particles} $q_0=\phi_m Q/B$ where $\phi_m$ is the particle volume fraction of a static bed of particles, and the distance between the front and back of the flows is $B=1$ \protect\unit{cm}.  It could be controlled by using funnels of different sizes, each of which could then be associated with a bulk flow rate, $Q_{nom}$; however, this is a nominal flow rate as the actual flow rate could vary from occasion to occasion.   The apparatus had closed ends; so, to avoid `backing-up' when running experiments with a non-horizontal slope, particles were removed from the downslope end using a vacuum cleaner.  This had no measurable effect on the height profiles obtained but allowed experiments to be run for longer.  No removal of particles was necessary for the slower-moving horizontal flows.  The value of $q_0$ was accurate to within $\pm 3\%$. 

The material used for all the experiments was approximately spherical, glass beads (Potters Ballotini) with particle diameters in the range \unit[250--425]{$\mu$m} and a mean diameter $d\approx375~\mu$m.  We measured the particle volume fraction of densely packed, static material (i.e.\ a maximum) as $\phi_m = 0.610 \pm 0.005$, which is close to the maximum value of  $0.64$ for random, close-packed, mono-sized spheres \citep{JAE:SCI1992}, and the (unfluidised) bulk density was \unit[$1.520 \pm 0.008$]{g/\cmt}.  The powder corresponds to a class~{B} powder according to \cite{geldart_types_1973}, so no bubble-free expansion when fluidised is expected within a `static' bed.  The minimum fluidisation speed, $\umf$, was found by independent experiment where the gas flow rate through a  static bed of material was gradually increased and the resulting pressure drop through the bed measured \citep{davidson_fluidised_1963}.  We found that the entire weight of the bed was supported when  $w_g(=\umf)=10.77$\unit{cm/s}.  

\subsection{Shape of currents and front position extraction}
\label{sec:data}

The flows, viewed from the side, were recorded using a digital video camera.  Calibration was performed using an image of a block of known dimensions placed in the apparatus once the camera was set up in position for a given experiment.  Still images from the recorded experiments were analysed by transforming the RGB images to grey scale.  These were then turned into binary images through thresholding.  Though the threshold value was calculated automatically, the contrast between black back wall and white particles meant that the resulting binary images were robust and consistent.  The upper and lower surfaces of the outline of a current were defined as the first and last white pixels when descending a column of the binary image. \added{95\% confidence intervals for height measurements are $\pm 0.1$ cm.}  The front position was taken as the point where the top surface met the bottom one.  

\subsection{Velocity measurements}
\label{sec:PIV}

PIV was used to make measurements of the velocity fields of the flows using a high-speed video camera capturing at 500 frames per second, close up to a particular region of the flow.   The PIV measurements required the flow to be seeded with marker particles for which we used the same-sized particles as for our other experiments but approximately one third of which were dyed black.  The properties of the dyed particles ($\umf$, angle of repose etc) were identical to the non-dyed particles.  

Two-dimensional velocity fields were calculated by processing image pairs (two consecutive frames) from the video taken by the high-speed camera using the open-source Matlab-based DPIVSoft2010 code.  The software makes an initial estimation of the velocity field on a coarse grid and then uses this to translate and deform the interrogation window in the second image in keeping with the deformation of the flow field.  Errors associated with image pattern distortion, as is the case when velocity gradients are large, are greatly decreased using this method \citep{MEU:EIF2003}.  Several initial iterations were run to get a good approximation for the flow field.  A final run was performed with an interrogation window of 32$\times$32 pixels ($\approx 5d$) and velocity vectors were calculated using a 50\% overlap between adjacent windows.  A median filter was then applied with a limit of 0.5 to remove spurious vectors \citep[e.g.][pp. 406]{ADR:PIV2010}.  

Instantaneous velocity profiles may not be representative of the flow as a whole.  In particular, the bubbles of gas that could form spontaneously in the flows often disrupted the instantaneous velocity profiles.  However, flows down the steeper slopes in our experiments, 10$\degree$ and 15$\degree$, reached a steady state very quickly, and for these flows an ensemble average of the flow velocity could be found by averaging over both many points in time and at several positions along the flow.  The quality, and hence the accuracy, of time-averaged velocity fields has been shown to be greatly improved when the average instantaneous correlation function is used to calculate the velocity field \citep{MEI:JFE2000}.  We therefore modified the PIV routines accordingly to produce a single time-averaged velocity field per experiment using an interval of twenty frames (\unit[0.04]{s}) between image pairs, and fifteen image pairs per experiment.  This interval is larger than characteristic time for shear ($(\rd v/\rd z)^{-1} \approx 0.01$ s), so the velocity fields at successive intervals are uncorrelated.  Velocity profiles were then formed from the stream-wise vectors of the time-averaged velocity field lying on a depth-wise transect at points separated at intervals of 1 cm and averaged to form the ensemble average velocity profile.  The resolution of PIV measurements can be expressed as \citep{ADR:ARFM1991},

\begin{equation}
  \sigma_u \approx \frac{c_1 M d}{\Delta t},
  \label{eq:pivUncertainty}
\end{equation}
where $c_1$ is the uncertainty of locating the centroid of the correlation peaks, $M$ is the magnification factor of the lens, and $\Delta t = 1/500$~s is the time step between images.  For our setup, $M = 1/2$~, and $c_1 \approx 10\%$ so that $\sigma_u = O(1)$ cm/s.  

For the steady flows it is more useful to define error based on the sum of variances of the all the $m$ profiles used to calculate the ensemble averaged \added{standard deviation} over the $n$ images given by
\begin{equation}
  \sigma_{\mathrm{ens~ave}}^2(z) = \dfrac{1}{m \times n}\sum_{i=1}^{m \times n} \sigma_i^2(z).
  \label{eq:ensAveUncertainty}
\end{equation}
\added{This average standard deviation was then used to calculate 95\% confidence intervals for the velocities.  }

\section{Equations of motion}
\label{sec:model}

We investigate the motion of granular currents down an inclined surface when the particles are fluidised, as shown in figure~\ref{fig:FlowSchematic}.  These flows are gravitationally-driven, but do not accelerate unboundedly; instead the principle action of particle interactions is to contribute to the shear stresses that balance the down slope acceleration and potentially lead to steady motion.   We formulate a mathematical model of the two-phase motion that couples mass conservation for each phase with expressions for the balances of momentum and we show how this formulation may be applied to steady fully-developed flows that vary only with distance from the underlying boundary (\S\ref{wide}), and to unsteady, relatively thin flows for which the acceleration perpendicular to the underlying boundary is negligible (\S\S\ref{unsteady},\ref{horizontal}).  

The mathematical model is built upon a continuum description of two inter-penetrating phases which interact with each other.  These models do not resolve the motion of individual particles; rather, they allow the computation of the evolution of averaged properties.  Such approaches have been employed often for confined, horizontal fluidised beds \citep[e.g.][]{bokkers_mixing_2004,Goldschmidt2004}, but these studies differ from the dynamics of the flows analysed in this contribution where there is persistent shear flow down the inclined surface.  The flows analysed here also differ in an essential way from non-fluidised granular motion down inclines since the support of the weight of the grains by the imposed gas flow significantly reduces resistive forces and increases mobility.  Nevertheless, we find that steady flows are admissible and thus the motion must develop sufficient shear stresses to balance gravitational acceleration.  Our model assumes that these stresses arise from particle interactions and are collisional and the particle fluctuations may be characterised by a granular temperature since friction as a bulk property is virtually eliminated by fluidisation and the viscous forces associated with interstitial gas flow are negligible.  The granular temperature will be shown to be relatively small and thus the interactions generate only relatively weak shear stresses, but these are sufficient to balance the gravitational acceleration.

The collisional nature of the motion is justifiable in all but some small regions of the currents, for example close to the surface of the slope.  The model captures only the relatively slow evolution of averaged quantities.  In particular bubbles (i.e.\ volumes largely evacuated of particles that travel through fluidised particles) are not explicitly resolved.  Bubbles are an important feature of deep, static fluidised beds as apart from strongly affecting the local instantaneous volume fraction of particles and they are the primary source of granular temperature in such a bed \citep{Menon1997}.  \added{There are several processes that might lead to the generation or suppression of bubbling, most notably including the dissipation of granular temperature through collisions, which is prone to clustering instabilities} \citep{gold93,full17}.  \added{Some studies have sought to predict the onset of bubbling in static beds through linear stability analysis} (e.g.\ see the review by \citealp{jackson_dynamics_2000}).  
\added{The flows of fluidised materials analysed here are somewhat different from these stability analyses, however, due to the persistent production of granular temperature by work done by the velocity field shear against the shear stresses, a process absent in static beds; hence, by means of a} scaling analysis \citet{eames_aerated_2000} showed that the contribution of these bubbles to the overall balance for granular temperature is likely to be negligible for this downslope motion. Furthermore, shallowness in the bed is thought to suppress bubbling \citep{Botterill1972, Tsimring1999}, as is shear \citep{botterill_open-channel_1979,ishida_flow_1980}. \added{We therefore assume that bubbling is likely to have a limited influence on the fluidised currents.  Extensive bubbling was not observed in the currents.  The photograph shown later in figure~\ref{profile} is typical with no apparent bubbles.  While agitation was visible at the top of the currents, bubbles sufficiently large to fill the width of the bed were hardly ever seen.}

Most of the theoretical developments in this study will be for two-dimensional flows and the effects of the front and back walls of the apparatus are neglected.  The use of this planar set-up allows the structure of the system to be seen and measured (as described in \S\ref{sec:experiments}), but at the expense of it being bounded by walls not present in realistic, three-dimensional systems.  Arguably, because fluidisation eliminates internal friction, a large part of the effect that the presence of these walls might also be eliminated.  Here, in most of what follows, we analyse the motion in the regimes that the side-walls play a negligible role; however in \S\ref{narrow}, we also analyse the case when the side-walls have a dominant effect on horizontal currents and in Appendix~\ref{sidewall}, we derive the extra, weak retardation on flows down slopes that arises from side wall drag when the depth of the flow is much smaller than the width.

The general equations of motion for a continuum model, known as a `two-fluid model', of a gas-particle system have been developed by \cite{jackson_dynamics_2000}.  The conservation of mass in each phase is given by

\begin{gather}
  \pd{}{t}\left(1-\phi\right)+\nabla\cdot\left((1-\phi)\u\right)=0\qquad
  \\
  \pd{\phi}{t}+\nabla\cdot\left(\phi\v\right)=0,
\end{gather}
where $\phi$ denotes the volume fraction of solids and $\u$ and $\v$ the velocity field of the gas and solid phase, respectively. 

Following \cite{jackson_dynamics_2000}, the balance of momentum for the gas is given by
\begin{equation}
  \label{genf}
  \left(1-\phi\right)\rho_g\frac{D_g\u}{Dt}=\left(1-\phi\right)\bnabla\cdot\boldsymbol{S}^{g}-\boldsymbol{F}_D+\left(1-\phi\right)\rho_g\boldsymbol{g}
\end{equation}
and for the particles
\begin{equation}
  \label{gens}
  \phi\rho_s\frac{D_s\v}{Dt}=\bnabla\cdot\boldsymbol{S}^{s}+\phi\bnabla\cdot\boldsymbol{S}^{g}+\boldsymbol{F}_D+\phi\rho_s\boldsymbol{g},
\end{equation}
where $\rho_g$ and $\rho_s$ are the densities of the gas and solid phase respectively, $\boldsymbol{S}$ is the spatially-averaged stress tensor of each phase with the superscript ($g$, $s$) denoting the gas or solid phase, respectively, $\boldsymbol{F}_D$ is the drag force exerted by the particles on the fluid due to the difference in their velocities and $\boldsymbol{g}$ denotes gravitational acceleration.  The material derivatives, $D_g/Dt$ and $D_s/Dt$ denote the rate of change moving with the gas and the solid phase respectively. A number of researchers, including \cite{ergun_fluid_1952} and \cite{jackson_dynamics_2000}, have suggested that $\boldsymbol{F}_D=\beta\left(\u-\v\right)$, where $\beta$ is a drag coefficient.  Virtual mass and particle shear forces are neglected.  

These equations will be solved for the situation shown schematically in figure~\ref{fig:FlowSchematic}. The slope is inclined at angle, $\theta$, to the horizontal with the underlying boundary at $z=0$ and the upper surface of the current at $z=h$, while the $x$-axis is aligned with the basal boundary.  A mixture of solid particles and gas runs down the slope under the influence of gravity. 

\section{Fully developed flows}
\label{wide}

\subsection{Model for fully-developed flows}

First, fully developed flows are investigated, in which the dependent variables are functions only of the distance from the boundary, $z$, and the velocity fields of the gas and solids are given by $\boldsymbol{u}=(u(z),0,w(z))$ and $\boldsymbol{v}=(v(z),0,0)$, respectively.  Conservation of mass for the solid phase is automatically satisfied by this form, but for the fluid phase we deduce that
\begin{equation}
  (1-\phi)w=w_g,
  \label{fluidmass}
\end{equation}
where $w_g$ is the fluidising gas flux per unit area normal to the boundary.

The expressions for the balance of momentum follow those proposed by \cite{johnson_frictional-collisional_1987} and \cite{agrawal_role_2001} where for the gas phase down the slope
\begin{equation}
  \rho_gw_g\pd{u}{z}=(1-\phi)\rho_g g\sin\theta+(1-\phi)\pd{}{z}\left(\mu_g\pd{u}{z}\right)+\beta(v-u),
  \label{fhmom}
\end{equation}
where $g=|\boldsymbol{g}|$ denotes gravitational acceleration and $\mu_g$ is the gas viscosity.  
Perpendicular to the slope, we find
\begin{equation}
  \rho_gw_g\pd{w}{z}=-(1-\phi)\pd{p}{z}-(1-\phi)\rho_gg\cos\theta-\beta w+ (1-\phi)\frac{4}{3}\mu_g\pdtwo{w}{z},
  \label{fvmom}
\end{equation}
where $p$ is the pressure within the fluid phase. In \eqref{fhmom} and \eqref{fvmom} we have assumed that the gas phase is incompressible and can be treated as Newtonian with constant viscosity, $\mu_g$.  

For the solid phase, the balance of down slope momentum is given by
\begin{equation}
  0=\phi\rho_sg\sin\theta+\pd{\sigma_{xz}}{z}-\beta(v-u)+\phi\pd{}{z}\left(\mu_g\pd{u}{z}\right),\label{shmom}
\end{equation}
while normal to the slope,
\begin{equation}
  0=-\phi\rho_s g \cos\theta +\pd{\sigma_{zz}}{z}-\phi \pd{p}{z}+\beta w + \phi\frac{4}{3}\mu_g\pdtwo{w}{z}.
  \label{svmom}
\end{equation}
In \eqref{shmom} and \eqref{svmom}, $\sigma_{xz}$ and $\sigma_{zz}$ are components of the solid phase stress tensor, $\boldsymbol{S}^{s}$, and at this stage we have not yet invoked any constitutive model for these stresses.  Further, from \eqref{shmom} the driving force for the current is gravity and within this framework, currents over horizontal surfaces are inherently unsteady as they decelerate. 

The downslope balance of momentum \eqref{shmom} differs from previous contributions. \cite{nott_frictional-collisional_1992} implicitly assumed that there was no relative component of velocity downslope between each of the phases and thus there was no drag force (i.e.~$\beta(u-v)=0$).   \cite{eames_aerated_2000} did not consider momentum balance for the fluid phase and imposed $u=0$; thus within their model, the drag force $\beta v$ is dominant and by assumption the shear stress associated with the solid phase is negligible. We do not invoke either of these assumptions at this stage, instead maintaining the various dynamical processes until their relative magnitudes have been fully assessed below. 

Adding the normal momentum equations \eqref{fvmom} and \eqref{svmom}, we find that
\begin{equation}
  \pd{}{z}\left(p-\sigma_{zz}\right)=-\left(\rho_g(1-\phi)+\rho_s\phi\right)g \cos \theta+\frac{4}{3}\mu_g\pdtwo{w}{z}-\rho_gw_g\pd{w}{z}.
  \label{hydrostatic}
\end{equation}
When the current is homogeneous so that particle volume fraction $\phi$ is constant, from \eqref{fluidmass} the vertical component of the gas velocity is also constant; so, \eqref{hydrostatic} expresses the hydrostatic balance between the vertical gradient of the normal stress from both solid and fluid phases and the weight of the fluidised grains.  

It is also insightful to eliminate the fluid pressure field between \eqref{fvmom} and \eqref{svmom} to find that
\begin{equation}
  \pd{\sigma_{zz}}{z}+\frac{\phi}{(1-\phi)}\rho_g w_g\pd{w}{z}=\phi(\rho_s-\rho_g)g\cos\theta-\frac{\beta w}{1-\phi}.
  \label{fluidised}
\end{equation}
This expression reveals the fundamental dynamical role played by fluidisation.  The slope-normal component of the inter-phase drag, incorporated into the model by $\beta w$, can balance the weight of the grains and thus it is possible for the normal stress tensor of the solid phase $\sigma_{zz}$ to be much reduced from its non-fluidised magnitude.  This in turn reduces the magnitude of the solids shear stress $\sigma_{xz}$ and thus the mobility of the fluidised flows is greatly enhanced. Equation~\eqref{fluidised} is different from the classical model of a static fluidised bed because velocity gradients lead to normal stresses in the solid phases and these may contribute in a non-negligible way to the balance between weight and drag as shown below.

\subsubsection{Inter-phase drag and constitutive equations}

The drag on the solid phase due to the fluidising gas flow is given by $\beta({\bf u}-{\bf v})$, where the drag coefficient, $\beta$, may be written
\begin{equation}
  \beta=\frac{\mu_g}{d^2}f_0(\phi)+\frac{\rho_f}{d}|{\bf u}-{\bf v}|f_0^\ast(\phi),\label{drag_exp}
\end{equation}
where $f_0$ and $f_0^\ast$ are given in Table~\ref{grantab} \citep{ergun_fluid_1952}. The first term on the right-hand side of the equation represents the drag associated with viscous processes, and the second term with inertial processes.  For the regime of interest in this study, the inertial effects are negligible since the Reynolds number, based on gas velocity and particle size, is sufficiently small $(\Rey\equiv\rho_gw_gd/\mu_g< 10)$; however for completeness at this stage we maintain it in the model formulation.  Other expressions for the drag coefficient have been used \citep[e.g][]{agrawal_role_2001,Oger201322} and these could replace \eqref{drag_exp} within this modelling framework.

\begin{table}
  \begin{align*}
    f_0=&\frac{150\phi^2}{(1-\phi)}\\
    f_0^\ast=&\frac{9\phi}{4}\\
    f_1=&\frac{5\sqrt{\pi}}{4g_0} \frac{\left( 1-\frac25\, \left( 1+e \right)  \left( 1-3\,e
 \right) \phi g_0  \right)  \left( 1+\frac45 \left( 1+e \right) \phi g_0 \right) }
 {\left( 13+12e-{e
}^{2}-c^* \left( \frac23+\frac{3}{16}e-\frac {9}{16}\,{e}^{2} \right)  \right)}+\frac{4}{5\sqrt{\pi}} \left( 1+e \right) {\phi}^{2}g_0 \left( 1-\frac{c^*}{32} \right)\\
       f_2=&\phi\, \left( 1+2\left( 1+e \right) \phi g_0  \right) \\
    f_3=&\frac{12}{\sqrt{\pi}}\left(1 -{e}^{2}\right) {\phi}^{2}g_0 \left( 1+{\frac {3c^*}{32}} \right)\\
   f_4=&{\frac {25\sqrt {\pi }}{4(1+e)g_0}}\frac{\left( 1+\frac{6}{5}\left( 1+e \right) \phi g_0
  \right)}{\left( 9+7\,e+{\frac {1}{64}}\left( 12\,e-221 \right) c^* \right)}  
  \left\lbrack 1+\frac{3}{5}\left( 1+e
 \right) ^{2} \left( 2\,e-1 \right) \phi g_0 +\right. \\
 &\qquad\qquad\qquad\left.\left( 1+\frac{3}{10} \left( 1
+e \right) ^{2}\,\phi g_0  \right) c \right\rbrack+\frac{2}{\sqrt{\pi}} \left( 1+e \right) {
\phi}^{2}g_0\left( 1+{\frac {7c^*}{32}} \right)\\
 f_4^*=&{\frac {75\sqrt {\pi }}{8(1+e)\phi g_0}}
 \frac{\left( 1+\frac{6}{5}\left( 1+e \right) \phi g_0\right) }
{\left( 19-3\,e+ \left( \frac{177}{64}\,e-\frac{161}{64} \right) c^* \right)}
 \left\lbrack{\frac {40(1-e)}{3}}
\frac{ \left( 1+\phi\,{\frac {\rd}{\rd\phi}}\log g_0  \right)   \left( 1+{\frac {3c^*}{32}}\right)}
{\left( 9+7e+{\frac {1}{64}}\, \left( \frac{237}{64}e-\frac{221}{64}
 \right) c^* \right) }
 \left(
 \vphantom{\frac{1}{e}}
  1+ \right.\right.\\
 &\left.\left.\left( 1+
 \left( 1+e \right) \phi g_0 \right) c^*+\frac{3}{5}\phi\left( 1+e \right) ^{2}g_0 \left( 2e-1+ \left( \frac{1}{2}+\frac{e}{2}-\frac{5}{3(1+e)}\right) c^*
 \right)  \right)\right. \\
 &\left.+\frac{1}{3} \left( 1+ 4
\left( 1+e \right) \phi g_0 +2\left( 1+e
 \right) \phi^2 \frac {\rd g_0}{\rd\phi} \right) c^*\right.\\
 &\left.
 -\frac{4\phi g_0}{5}  \left( 
1+\frac{1}{2}\phi\frac {\rd}{\rd\phi}\log g_0 
 \right)  \left( 1+e \right)  \left( e
 \left( 1-e \right) +c^* \left( \frac{1}{3}+\frac{e \left( 1-e \right)}{4}  \right) 
 \right)  
 \vphantom{\frac{ \left( 1+\phi\,{\frac {\rd}{\rd\phi}}\log g_0  \right)   \left( 1+{\frac {3c^*}{32}}\right)}{\left( 9+7e+{\frac {1}{64}}\, \left( \frac{237}{64}e-\frac{221}{64}
 \right) c^* \right) }}\right\rbrack\\
    f_5=&f_1\left(\frac{\pi}{2\sqrt{3}}\frac{\phi}{\phi_m}{g_0}\psi\right)^{-1}\\
    f_6=&\frac{\pi\sqrt{3}}{6}\frac{\phi}{\phi_m }g_0\psi\\
    f_7=&\frac{\pi\sqrt{3}}{4}\frac{\phi}{\phi_m}(1-e_w^2)g_0
  \end{align*}
  \caption{The constitutive laws for granular kinetic theory applied to fluidised systems. Here  $e$ denotes the coefficient of restitution characterising collisions between particles; $e_w$ is the coefficient of restitution between the walls and the particles; $\phi_m$ is the volume fraction at maximum packing; $\psi$ is the specularity coefficient \citep[after][]{johnson_frictional-collisional_1987}; and $g_0$ is the radial basis function which accounts for particle packing \eqref{g0}.  The coefficient $c^*=32(1-e)(1-2e^2)/(81-17e+30e^2(1-e))$ \citep{garz99}.  $f_0$  contributes to the inter-phase drag in the viscous regime and $f_0^*$ in the inertial regime \citep[see, for example,][]{Hoef2005}. $f_1$, $f_2$ model the volume fraction dependence in the collisional contributions to stresses \citep{garz99}; $f_3$, $f_4$ and $f_4^*$ model contributions to the granular energy balance \citep{garz99}.  $f_5$ to $f_7$  determine boundary conditions at the base of the flow:  $f_5$ contributes to the boundary condition for momentum balances and $f_6$ and $f_7$ to that for fluctuation energy \citep[see][]{johnson_frictional-collisional_1987,johnson_frictional-collisional_1990}.  }
  \label{grantab}
\end{table}

It was noted above that particle interactions are dynamically important because of the momentum transfer arising from particle collisions \citep{lun_kinetic_1984,garz99}.  Here we examine  the collisional stresses and follow \cite{nott_frictional-collisional_1992} and \cite{agrawal_role_2001} amongst others who incorporate these effects into models of fluidised and aerated flows, to write the shear and normal components of stress in terms of a granular temperature $T$, which measures the fluctuations of velocity about the mean, the volume fraction of solids $\phi$, and the coefficient of restitution $e$, which characterises dissipation in the instantaneous collisions.  While the constitutive laws invoked here have been validated in some scenarios by simulation and experimentation, there remains some uncertainty about their generality.  Hence we pose the model quite generally so that the constitutive relations could be updated as required.

For fully developed flows, we write
\begin{linenomath}
\begin{align}
  \label{stresstemp}\sigma_{xz}&=f_1(\phi,e)\rho_sd T^{1/2}\pd{v}{z},\\
  \sigma_{zz}&=-f_2(\phi,e) \rho_sT,
\end{align}
\end{linenomath}
where $f_1$ and $f_2$ are dimensionless functions given in Table~\ref{grantab}.  In this study, we employ the constitutive formulae derived by \citet{garz99} and recently used for modelling dense avalanches by \citet{jenk10}.

Granular temperature may be generated and `conducted' via the flow processes and dissipated in the collisions.  Following \citet{lun_kinetic_1984} and \citet{garz99} amongst others, these effects are encompassed in the following expression of energy balance within the flow
\begin{equation}
0=-\pd{\Psi}{z}+\sigma_{xz}\pd{v}{z}-f_3(\phi,e)\rho_s\frac{T^{3/2}}{d},\label{thermal}
\end{equation}
where  the flux of granular temperature is given by
\begin{equation}
\Psi=-f_4\rho_sd T^{1/2}\frac{\partial T}{\partial z}-f_4^*\rho_s T^{3/2}d\pd{\phi}{z},
\end{equation}
and $f_3$, $f_4$ and $f_4^*$ are dimensionless functions given also in Table~\ref{grantab}.  In posing this balance of granular temperature we have neglected generation and dissipation of the granular temperature mediated by viscous interactions with gas \citep{koch_particle_1999}.  

Following the formulation of \citet{agrawal_role_2001}, dissipation by viscous processes is much smaller than dissipation through inelastic collisions when
\begin{equation}
  \frac{\mu_g T}{d^2}\ll\frac{\rho_s(1-e^2)T^{3/2}}{d}.
  \label{dissip}
\end{equation}
Furthermore, the generation of granular temperature by viscous processes is much smaller than that by granular interactions when
\begin{equation}
  \frac{\mu_g^2|u-v|^2}{\rho_s d^3T^{1/2}}\ll \rho_s d T^{1/2}\left(\frac{\p v}{\p z}\right)^2.
  \label{generate}
\end{equation}

The constitutive laws, $f_1-f_4^*$, as well as those involved the boundary conditions ($f_5-f_7$, see \S\ref{boundary}) feature the radial basis function, $g_0(\phi)$.  Various authors have suggested forms for $g_0$ and we employ an expression that is close to the suggestion of \cite{vesc14}, who empirically fitted a function to match data from discrete element simulations.  Importantly, the radial basis function diverges as the volume fraction approaches maximum packing (as established by \cite{torq95}) and following \cite{vesc14} we write
\begin{equation}
g_0=\hat{g}\frac{2-\phi}{2(1-\phi)^3}+(1-\hat{g})\frac{2}{\phi_m-\phi},
\label{g0}
\end{equation}
where the weighting function is given by
\begin{equation}
\hat{g}=\left\{
\begin{tabular}{ll}
$1$&$\phi<\phi_*$,\\
$1-\left(\frac{\phi-\phi_*}{\phi_m-\phi_*}\right)^n$&$\phi_*<\phi<\phi_m$.
\end{tabular}\right.
\label{weighting}
\end{equation}
Thus, when $\phi<\phi_*$ the radial basis function is given by the formula proposed by \cite{carn69}, but it exceeds this value when $\phi_*<\phi$ and diverges as maximum packing is approached.  \cite{vesc14} suggest that $n=2$ and that $\phi_*=0.4$.  While this choice ensures that $g_0$ and its derivative are continuous at $\phi=\phi_*$, the second derivative is discontinuous.  This is problematic for the system of differential equations that we will integrate numerically; therefore, we employ the values $n=3$ and $\phi_*=0.4$, which ensure that $g_0$ is sufficiently smooth.  Moreover, our expression \eqref{weighting} with these values is close to those proposed by \cite{vesc14} and \cite{torq95} and appears to match the simulation data adequately.

\added{The energetic balance encompassed in \eqref{thermal} assumes that the particles are sufficiently agitated so that the particle diameter is the appropriate correlation length scale over which dissipated occurs (and the rate of dissipation is then given by $\rho_sf_3T^{3/2}/d$).  Recently, however,} \cite{jenk07} \added{has suggested that at relatively high concentrations, clusters of particles begin to form and thus the correlation length increases to $L_c$ ($>d$) and then the rate of dissipation is given by $\rho_sf_3T^{3/2}/L_c$. This extended kinetic theory has been applied to unfluidised flows of grains down inclined planes by} \cite{jenk10,jenk12}\added{, where an empirical formula for $L_c$, informed by comparison with simulations and experimental measurements, is proposed in terms of the dependent flow variables.  In our study there is potentially the need to include this phenomenon into the modelling framework to obtain good comparison between the predicted and measured results. However, as shown in appendix~\ref{extendke}, we find that extended kinetic theory makes negligible difference to the model predictions for the fluidised flows in our regime of interest and so we do not include it in the calculations that follow. } 

\subsubsection{Coefficient of restitution}

The dynamical effects of collisions between particles and between particles and the underlying boundary are characterised in the model by three parameters:  the coefficients of restitution between the particles, $e$, and between the particles and the boundary, $e_w$, and the specularity coefficient $\psi$,  which governs the dynamic interaction between the particles and the bottom surface \eqref{bc_base}.  These parameters are relatively difficult to measure directly.

The coefficient of restitution, $e$, plays an important role in continuum models and in  Discrete Particle (or Element) Models (DPM), which endeavour to calculate the motion of large ensembles of particles and to resolve individual particle collisions.  In continuum models, the difference of $e$ from unity is proportional to the rate at which the collisions dissipate energy (see the definition of $f_3$ in table~\ref{grantab}), whereas in DPMs it controls the ratio of normal velocities before and after binary collisions and in these models, there are potentially additional means of energy dissipation. Often values for the coefficient of restitution are adopted without independent experimental confirmation and for DPM studies,  typical values are relatively high (for example, $e=0.90$ and $0.97$ respectively in the studies of \citet{Goldschmidt2004} and \citet{van_der_hoef_numerical_2008}).  These values are close to measured values of discrete collisions \citep[see, for example,][]{Kharaz2001}.  When used in kinetic theory models, commonly adopted values of $e$ are rather lower and \citet{Jenkins2002} suggest a means by which the the appropriate value for kinetic theories can be derived from  directly measured normal and tangential coefficients of restitution and the tangential coefficient of friction.  For glass spheres of 3mm diameter, the measured data of \citet{Foerster1994} corresponds to an effective coefficient of $e=0.85$ if the method of \citet{Jenkins2002} is employed and this is the value we employ in this study.   We have no direct measurements of the appropriate coefficient of restitution for the collisions between the particles and underlying boundary; we choose $e_w = 0.75$, but note that its magnitude has very little influence upon the computed flow profiles apart from within thin basal boundary layers.

\subsubsection{Boundary conditions}\label{boundary}

The boundary conditions for this problem follow the formulation of \citet{johnson_frictional-collisional_1987} and \citet{johnson_frictional-collisional_1990}.    At the base, there is no-slip for the fluid phase, a slip condition for the particle phase, and a condition specifying the flux of granular temperature. These are respectively given by
\begin{equation}
  u=0,\quad f_5d\pd{v}{z}=v\quad\hbox{and}\quad \Psi=\rho_sT^{1/2}(f_6v^2-f_7T)\quad\hbox{at}\quad z=0.\label{bottombcs}
  \end{equation}
What this means physically is that solid-phase slip is allowed and stress is transmitted in the down-slope direction by specularity i.e.~the degree to which  the angle of exit of a particle after collision with the base is different from the entry angle.  This is mathematically represented by the specularity coefficient $\psi$ $(0<\psi<1)$.  Furthermore, fluctuation energy (granular temperature) is generated at the bottom surface and potentially dissipated by inelastic collisions, encompassed through a coefficient of restitution, $e_w$.
  
At the top of the current $z=h$, there are the free-surface boundary conditions that fluid shear and normal stresses vanish, given by  
\begin{equation}
  \pd{u}{z}=0,\quad\hbox{and}\quad p-\frac{4\mu}{3}\pd{w}{z}=0\quad\hbox{at}\quad z=h. \label{topbcs}
\end{equation}
However in addition, the flux of granular temperature vanishes and the solid phase normal and shear stresses adopt small values, representing the surface as being the location where collisional behaviour ends and instead the particles follow ballistic trajectories \citep[see][]{johnson_frictional-collisional_1990}.  Thus we enforce
\begin{equation}
  \Psi=0\quad\hbox{and}\quad (\sigma_{xz},\sigma_{zz})=\frac{\pi}{6}\rho_s\left(\frac{\phi}{\phi_m}\right)^{2/3}gd(\sin\theta,-\cos\theta)\quad\hbox{at}\quad z=h.\label{topbc2}
\end{equation}
The boundary conditions \eqref{bottombcs}-\eqref{topbc2} are of the same character as those employed by researchers in other flow regimes (see, for example, \citealp{jenk10}) and as for the constitutive laws, the framework for analysing these flows is robust to variations in the closures used for these conditions.

\subsubsection{Non-dimensionalisation of equations}
\label{nondim}

We now identify typical dimensional scales for the dependent variables and  assess the magnitude of the various terms in the governing equations.  It is convenient to sum the down-slope momentum equations of each phase \eqref{fhmom} and \eqref{shmom} to eliminate the inter-phase drag so that
\begin{equation}
  \rho_gw_g\pd{u}{z}=\left(\rho_s\phi+\rho_g(1-\phi)\right)g\sin\theta
  + \pd{}{z}\left(\rho_s f_1dT^{1/2}\pd{v}{z}\right)+\mu_g\pdtwo{u}{z}.
  \label{hcombined}
\end{equation}
In this expression, the key driving force is the down-slope gravitational acceleration and it is this term that the other terms must balance.  Since the density of the gas is much smaller than that of the solid phase and the effects of gas viscosity are negligible away from boundaries in this streamwise balance, we deduce that the dominant resistance is provided by the shear stress associated with the solid phase.  Coarsely scaling the variables and assuming the volume fraction and the constitutive functions of it are of order unity, 
$$\rho_sg\sin\theta\sim \rho_sd T^{1/2}v/h^2.$$
Furthermore, if the granular temperature is in local equilibrium between production and dissipation (an assumption that will be tested in the numerical solutions that follow), then from \eqref{thermal}, 
$$ d T^{1/2}(v/h)^2\sim T^{3/2}/d;$$
whence, the scaling for the velocity field is given by
\begin{equation}
  v\sim\left(\frac{g h^3 \sin\theta}{d^2}\right)^{1/2}.
\end{equation}
It is now convenient to introduce dimensionless variables, given by
\begin{equation}
\begin{aligned}
  &\hat{z}=z/h,\quad
  \hat{u}=u\left(\frac{g h^3 \sin\theta}{d^2}\right)^{-1/2},\quad
  \hat{v}=v\left(\frac{g h^3 \sin\theta}{d^2}\right)^{-1/2}\label{Usize}\\
  &\hat{w}=\frac{w}{w_g},\quad\hat{p}=\frac{p}{\rho_s g h \cos\theta}\quad\hbox{and}\quad
  \hat{T}=\frac{T}{g h \sin\theta}.
  \end{aligned}
\end{equation}
This set of scalings for the dependent variables of fluidised flows differs from those for flows of unfluidised, collisional granular media \citep{woodhouse_rapid_2010}.  For non-fluidised flows, the granular agitation must provide sufficient normal stress to support the weight of the overlying layer.  This would require the granular temperature to be of magnitude $gh\cos\theta$, which is considerably larger than the estimate deduced here \eqref{Usize} unless the motion is along relatively steep inclines $(\hbox{i.e.~when} \tan\theta\sim 1)$. For fluidised flows, however, granular temperature is generated by collisions but the imposed gas flow through the underlying particles provides most of the normal stress to balance the weight of the flowing layer.  The granular temperature, therefore, is lower and consequently the shear stresses are lower, which in turn significantly increases the mobility of these flows.  Hence these fluidised flows are characterised by relatively high flow speeds and relatively weak resistance.

The model is characterised by five dimensionless groups:

\begin{equation}
	\begin{aligned}
 	 & S=\tan\theta,\quad
	  R=\frac{\rho_g}{\rho_s},\quad
	  \delta=\frac{d}{h},
		  \\
	  & W_g=\frac{\mu_g w_g}{\rho_sd^2g\cos\theta},\quad
	  \hbox{and}\quad
 	 \St=\frac{\rho_sd(g\sin\theta h)^{1/2}}{\mu_g} \delta^2,
	  \label{parameters}
	\end{aligned}
\end{equation}
which represent respectively the inclination of the underlying boundary, the relative density of the gas to the solid phases, the size of the particles relative to the flow depth, the magnitude of the drag exerted by the fluidising gas flow relative to the weight of the granular layer and the reduced Stokes number, which compares particle inertia to fluid viscous effects.  

It is also possible to define a particle Reynolds number,
\begin{equation}
  Re=\frac{\rho_g w_g d}{\mu_g}=\frac{R W_gSt^2}{S\delta^3}.
  \label{eq:Re}
\end{equation}
Notably, the Reynolds number defined in this way is independent of the inclination of the slope.  The magnitude of the various model parameters for the experiments are set out in Table~\ref{exptpara}.    
\begin{table}
  \centering{ 
    \begin{tabular}{cccc} 
      \toprule
    $\mu_g$ & $1.7\times 10^{-5}$kg/ms & $S$ & $5.2 \times 10^{-2} - 2.7 \times 10^{-1}$ \\
    $\rho_g$ & $1.2$kg/m$^3$ & $R$ & $4.8 \times 10^{-4}$ \\
    $\rho_s$ & $2500$kg/m$^3$ & $\delta$ & $O(0.01)$ \\
    $d$ & $3.75 \times 10^{-4}$m & $W_g$ & $2.7\times 10^{-4}- 1.65 \times 10^{-3}$\\
     & & $St$ & $5.6-12.4$ \\
    $\theta$ & $3\degree-15\degree$ & $Re$ & $1.43-8.55$\\
    $w_g$ & \begin{tabular}{c} $0.09-0.34$m/s\\$(0.5-3.0 u_{mf})$\end{tabular} &  &  \\
    $\Qnom$ & $15-80$cm$^3$/s & & \\
      \bottomrule 
    \end{tabular} 
    \caption{Range of values of the physical parameters in the experiments and the dimensionless groups derived from them as defined by \eqref{parameters}.  }
    \label{exptpara}
  }
\end{table}
These scales may be used to show that the granular temperature dissipation and generation by viscous processes are negligible compared with direct particle interactions:  \eqref{dissip} is satisfied when $\delta^2/\St\ll1$ and \eqref{generate} when $W_g\ll 1$. 

We have five governing equations:  mass conservation \eqref{fluidmass}, down-slope fluid momentum conservation \eqref{fhmom}, the combined normal momentum equation from which the fluid pressure has been eliminated \eqref{fluidised}, the down-slope solids momentum equation \eqref{shmom}, and the equation for the conservation of granular temperature \eqref{thermal}.  Non-dimensionalised, these become
\begin{linenomath}
\begin{gather}
 	 (1-\phi){\hat w}=1,
  \label{ndmass}\\
 	 \delta Re\pd{{\hat u}}{{\hat z}}= R St(1-\phi)+(1-\phi)\delta^2\pdtwo{{\hat u}}{{\hat z}}+\left({\hat v}-\hat{u}\right)\left(f_0+f_0^{\ast}Re{\cal U}\right),
  \label{ndfluidmom}\\
  	-S \pd{}{{\hat z}}\left(f_2{\hat T}\right)+\frac{\delta W_g Re\phi}{(1-\phi)}\pd{{\hat w}}{{\hat z}}= \phi(1-R)-W_g\frac{{\hat w}}{1-\phi}\left(f_0+f_0^{\ast}Re{\cal U}\right),
  \label{ndnormalstress}\\
  	0=\phi+\pd{}{{\hat z}}\left(f_1{\hat T}^{1/2}\pd{{\hat v}}{{\hat z}}\right) -\frac{\left(f_0+f_0^{\ast}Re{\cal U}\right)}{St}\left({\hat 	v}-{\hat u}\right) +\frac{\phi\delta^2}{St}\pdtwo{{\hat u}}{{\hat z}},
  \label{ndsolidmom}\\
 	0=\delta^2\pd{}{{\hat z}}\left(f_4{\hat T}^{1/2}\pd{{\hat T}}{{\hat z}}\right)
 +\delta^2\pd{}{{\hat z}}\left(f_4^*{\hat T}^{3/2}\pd{\phi}{{\hat z}}\right)
+f_1{\hat T}^{1/2}\left(\pd{{\hat v}}{{\hat z}}\right)^2- f_3{\hat T}^{3/2},
  \label{ndthermal}
\end{gather}
\end{linenomath}
where ${\cal U}$ measures the magnitude of the dimensionless relative velocity between the phases and is given by ${\cal U}^2=S^2({\hat u}-\hat{v})^2/(W_gSt)^2+{\hat w}^2$.  From \eqref{bottombcs}, the dimensionless boundary conditions at the base $({\hat z}=0)$ are given by
\begin{equation}
  {\hat u}=0,\quad
  f_5\delta\pd{{\hat v}}{{\hat z}}={\hat v}\quad\hbox{and}\quad
  -\delta f_4\pd{{\hat T}}{{\hat z}}=\frac{f_6}{\delta^2}{\hat v}^2-f_7{\hat T},
  \label{bc_base}
\end{equation}
while from \eqref{topbcs}, we enforce at the top surface $({\hat z}=1)$
\begin{equation}
  \pd{{\hat u}}{{\hat z}}=0,\quad
  \pd{{\hat T}}{{\hat z}}=0\quad\hbox{and}\quad
  \left(f_1 {\hat T}^{1/2}\pd{{\hat v}}{{\hat z}},f_2{\hat T}\right)=\frac{\pi\delta}{6}\left(\frac{\phi}{\phi_m}\right)^{2/3}\left(1,S^{-1}\right)
  \label{bc_top}
\end{equation}

\begin{figure}%
  \centering
  \includegraphics[trim=4cm 9cm 4cm 9cm,width=0.8\columnwidth]{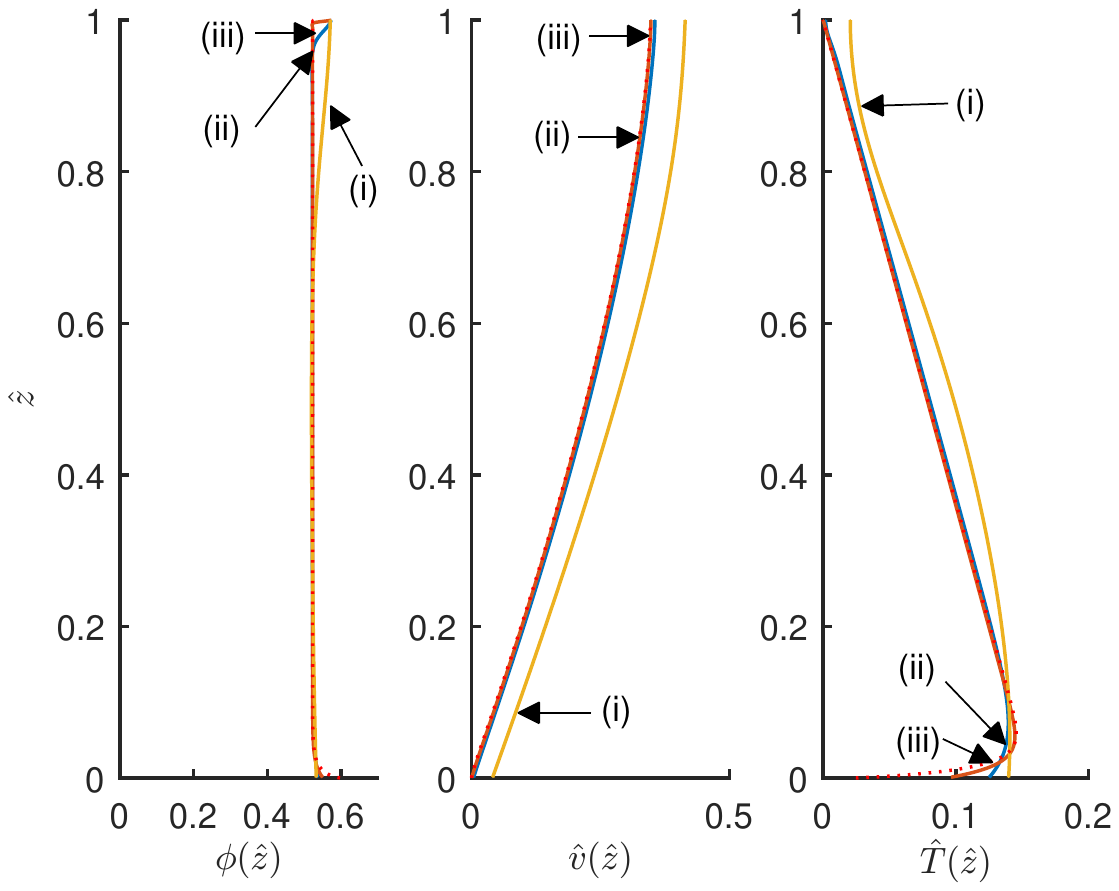}%
  \caption{The volume fraction, $\phi(\hat{z})$, velocity of the solid phase,  $\hat{v}({\hat z})$ and the granular temperature, $\hat{T}({\hat z})$, as functions of the dimensionless depth within the current for parameter values $R=10^{-3}$, $\psi=0.5$, $\phi_m=0.63$, $e=0.85$, $e_w=0.75$, $S=0.1$, $St=10^3\delta^2$, $W_g=10^{-3}$ and (i) $\delta=0.1$, (ii) $\delta=0.01$ and (iii) $\delta=0.001$.  Also plotted  are the asymptotic solutions (dotted lines), although these are often overlain by the full solution.}
  \label{fig_dvary}
\end{figure}

\begin{figure}
  \centering
  \includegraphics[trim=4cm 9cm 4cm 9cm,width=0.8\columnwidth]{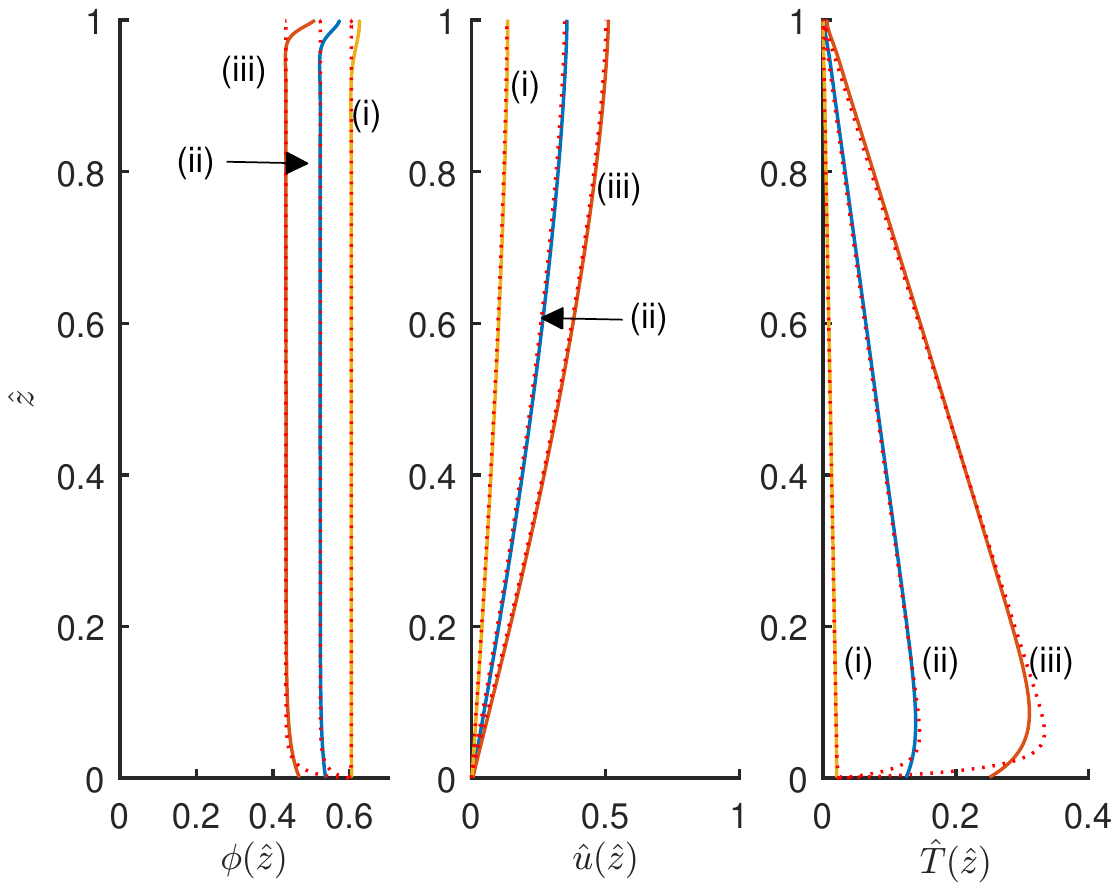}%
  \caption{The volume fraction, $\phi(\hat{z})$, velocity of the solid phase,  $\hat{v}({\hat z})$ and the granular temperature, $\hat{T}({\hat z})$, as functions of the dimensionless depth within the current for parameter values $R=10^{-3}$, $\psi=0.5$, $\phi_m=0.63$, $e=0.85$, $e_w=0.75$, $S=0.1$, $St=0.1$, $\delta=0.01$, (i) $W_g=0.5\times 10^{-3}$, (ii) $W_g=1.0\times 10^{-3}$ and (iii) $W_g=2\times 10^{-3}$.  Also plotted  are the asymptotic solutions (dotted lines).}
  \label{fig_wgvary}%
\end{figure}

\begin{figure}
  \centering
  \includegraphics[trim=4cm 9cm 4cm 9cm,width=0.8\columnwidth]{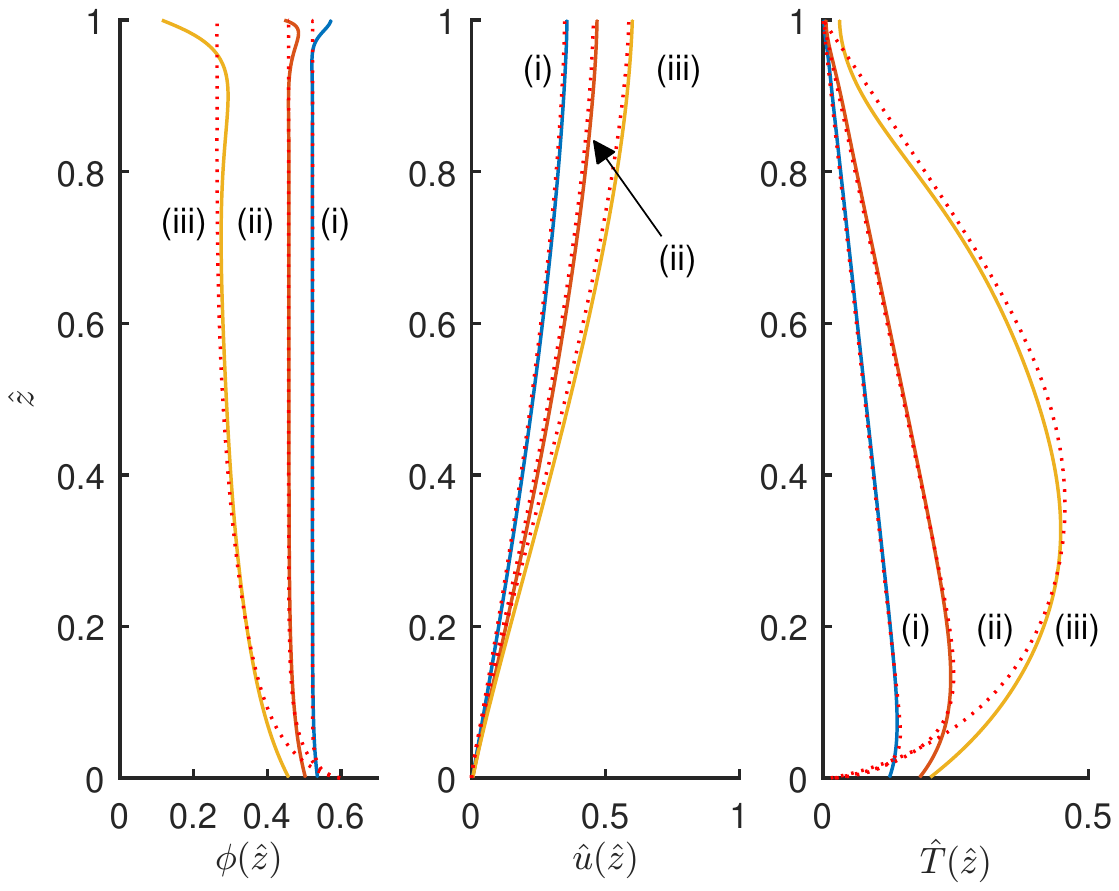}%
  \caption{The volume fraction, $\phi(\hat{z})$, velocity of the solid phase, $\hat{v}({\hat z})$ and the granular temperature, $\hat{T}({\hat z})$, as functions of the dimensionless depth within the current for parameter values $R=10^{-3}$, $\psi=0.3$, $\phi_m=0.63$, $e=0.85$, $e_w=0.75$, $St=0.1$, $\delta=0.01$, $W_g=10^{-3}$, (i) $S=0.1$; (ii) $S=0.2$ and (iii) $S=0.3$.  Also plotted  are the asymptotic solutions (dotted lines).}
  \label{fig_Svary}%
\end{figure}

The system of governing differential equations \eqref{ndmass}-\eqref{ndthermal} and boundary conditions \eqref{bc_base}-\eqref{bc_top} form a seventh order differential boundary value problem.  We use \eqref{ndmass} to eliminate $\hat{w}$ in favour of $1/(1-\phi)$ and we also evaluate $\rd^2\phi/\rd z^2$ in \eqref{ndthermal} by explicitly differentiating \eqref{ndnormalstress}.  The system is then integrated numerically. (For this task we employ the boundary value problem solver bvp4c in MatLab.)  Example solutions for the volume fraction, $\phi(\hat{z})$, the granular temperature, $\hat{T}(\hat{z})$ and the velocity of the solid phase, $\hat{v}(\hat{z})$ are plotted in figures \ref{fig_dvary}--\ref{fig_Svary} for various values of the governing dimensionless parameters.  We do not plot the gas velocity, $\hat{u}(\hat{z})$, because outside of thin basal boundary layers, it is indistinguishable from the solids velocity, $\hat{v}(\hat{z})$.  This basal boundary layer exists because the while the gas phase satisfies a no-slip condition, the solid phase exhibits slip.

The general trends are that the volume fraction is approximately uniform while the granular temperature decreases with distance from the bottom boundary.  Additionally there is a small slip velocity for the solid phase and the velocity shear decreases with distance from the the boundary. There are some systematic deviations from these general trends, most notably in regions close the upper and lower boundaries.  Since the interactions with the basal boundary are more dissipative than interactions with the constituent grains, the granular temperature decreases within a region in the vicinity of the boundary.  This boundary effect is diminished as the flow becomes thicker (i.e.\ as $\delta$ decreases, see figure~\ref{fig_dvary}), but is magnified where either the fluidising gas flow is increased (figure \ref{fig_wgvary}) and the slope is increased (figure~\ref{fig_Svary}).  We also find that throughout the bulk of the domain, away from thin layers adjacent to the upper and lower boundaries,  the production of granular temperature is in close balance with its dissipation, thus confirming the dimensional scales identified above \eqref{Usize}. 

\begin{figure}%
  \centering
  \includegraphics[trim=4.5cm 10cm 4.5cm 10cm,clip,width=0.48\columnwidth]{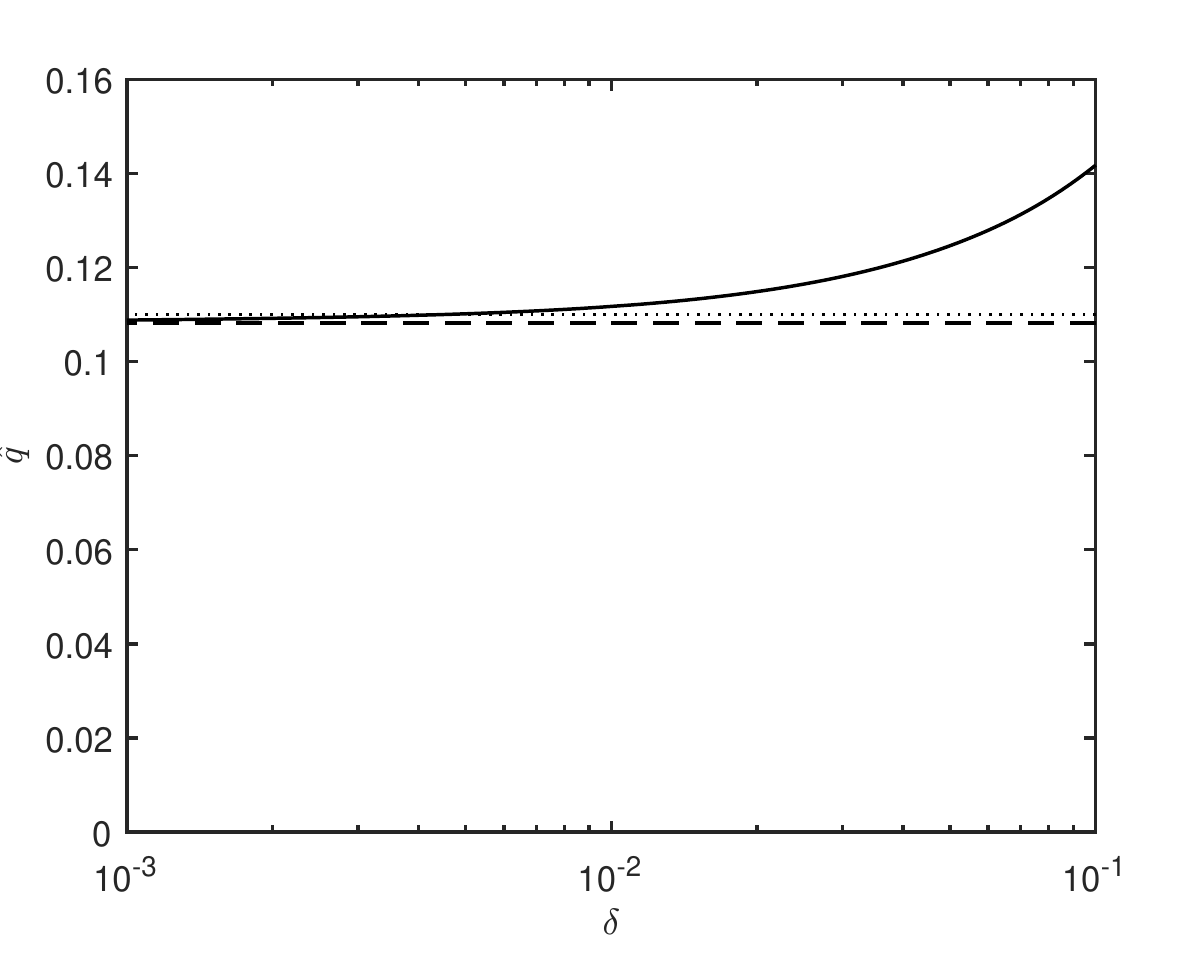}%
  \includegraphics[trim=4.5cm 10cm 4.5cm 10cm,clip,width=0.48\columnwidth]{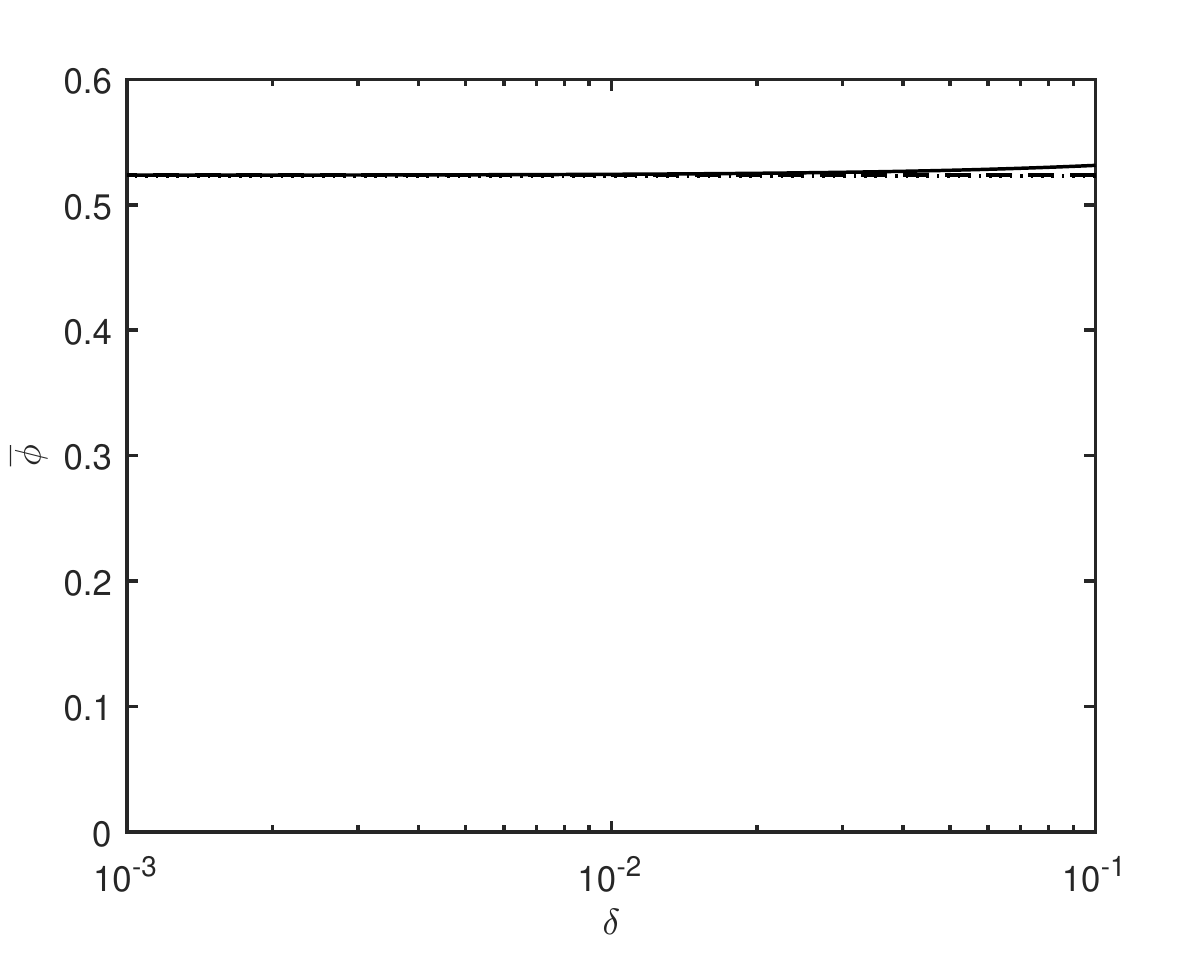}%
  \caption{The dimensionless volume flux per unit width transported by the flowing layer, $\hat{q}$, and the depth-averaged volume fraction, $\phibar$, as functions of the relative particle size for parameter values $R=10^{-3}$, $\psi=0.50$, $\phi_m=0.63$, $e=0.85$, $e_w=0.75$, $S=0.10$, $St=10^3\delta^2$ and $W_g=10^{-3}$.  Also plotted are the asymptotic solutions (dashed) and the simple approximate solutions (dotted).}
  \label{flux_delta}%
\end{figure}

\begin{figure}%
  \centering
  \includegraphics[trim=4.5cm 10cm 4.5cm 10cm,clip,width=0.48\columnwidth]{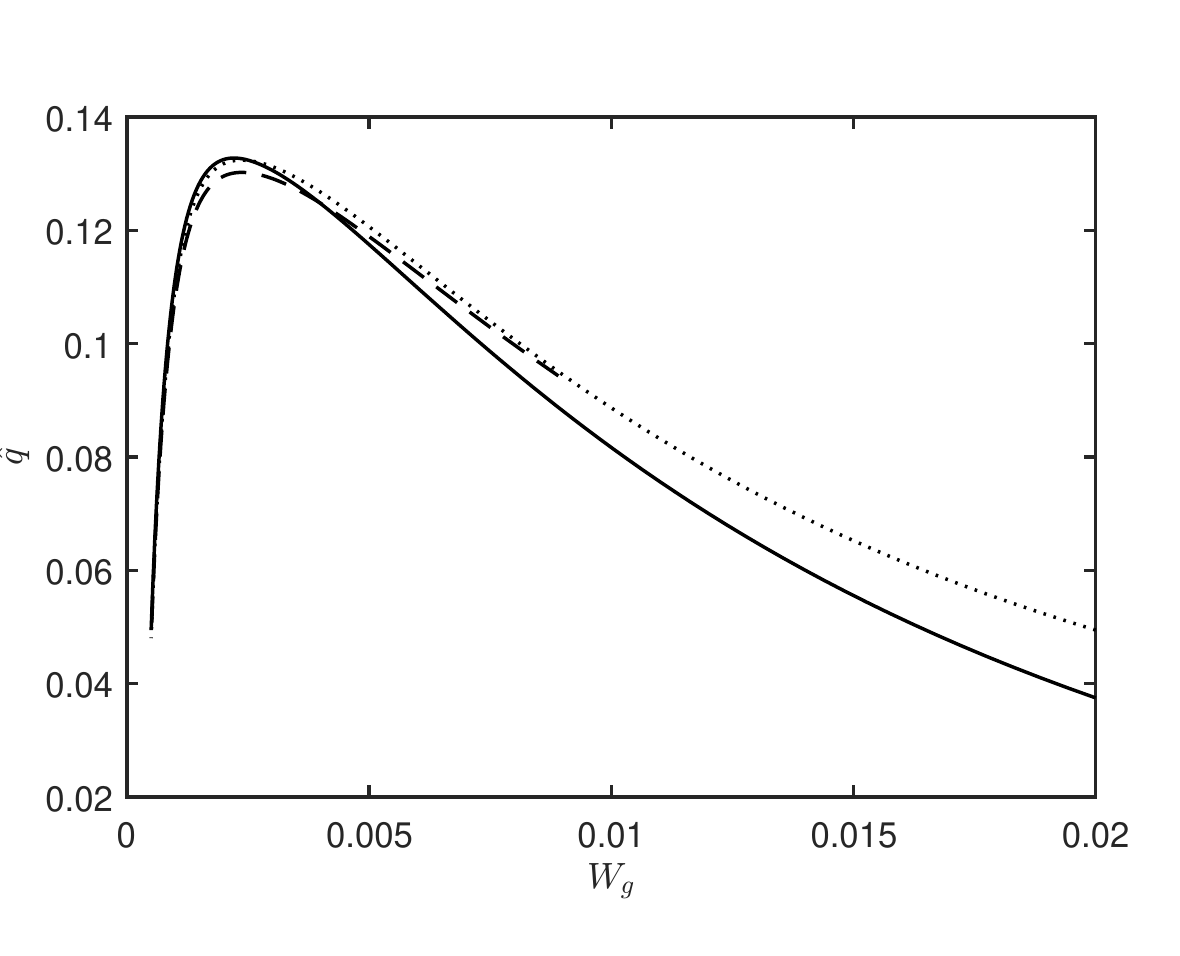}%
  \includegraphics[trim=4.5cm 10cm 4.5cm 10cm,clip,width=0.48\columnwidth]{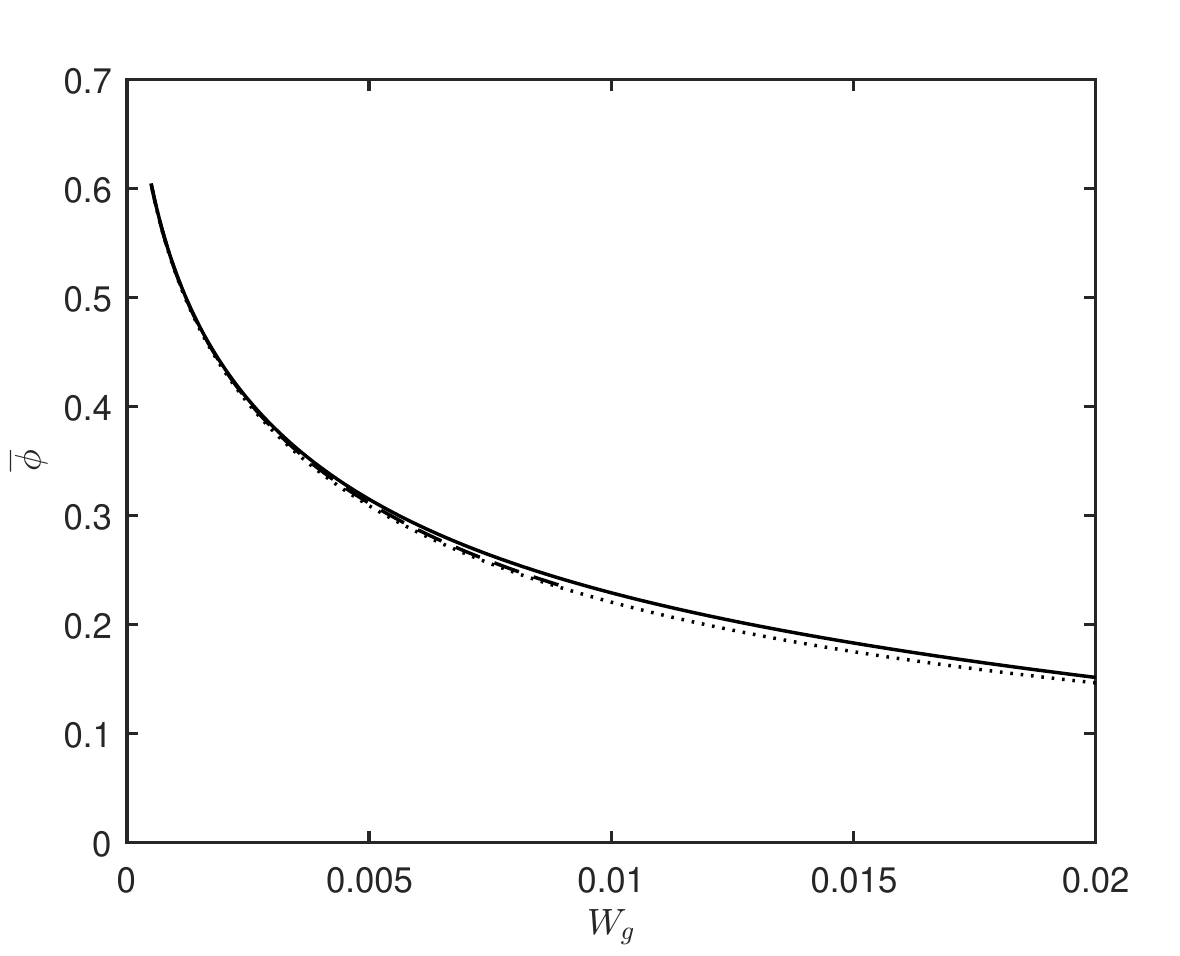}%
  \caption{The dimensionless volume flux per unit width transported by the flowing layer, $\hat{q}$, and the depth-averaged volume fraction, $\phibar$, as functions of the dimensionless strength of the fluidising gas flow, $W_g$, for parameter values $R=10^{-3}$, $\psi=0.50$, $\phi_m=0.63$, $e=0.85$, $e_w=0.75$, $S=0.10$, $St=0.10$ and $\delta=10^{-2}$.  Also plotted are the asymptotic solutions (dashed) and the simple approximate solutions (dotted).}
  \label{flux_wg}%
\end{figure}

\begin{figure}%
  \centering
  \includegraphics[trim=4.5cm 10cm 4.5cm 10cm,clip,width=0.48\columnwidth]{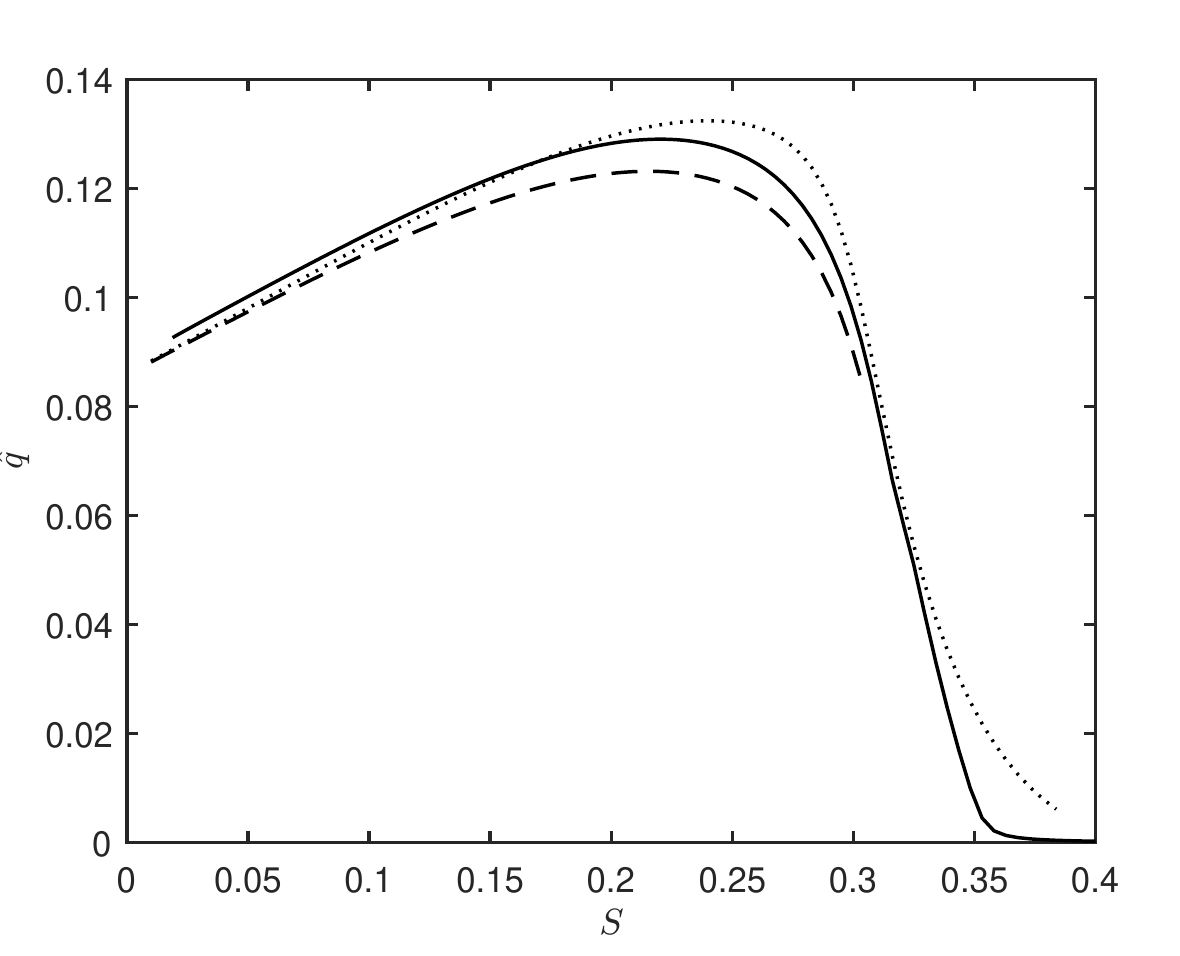}%
  \includegraphics[trim=4.5cm 10cm 4.5cm 10cm,clip,width=0.48\columnwidth]{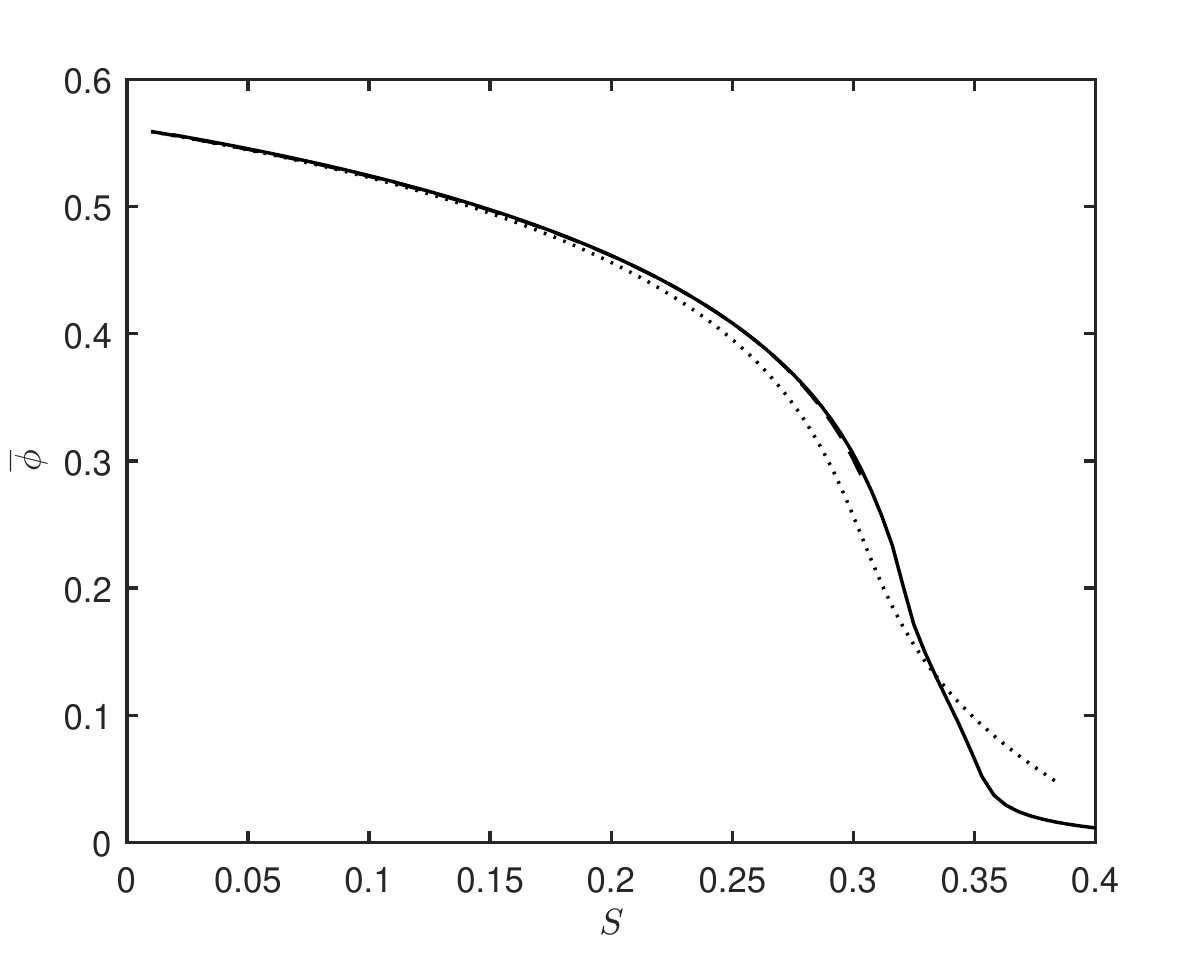}%
  \caption{The dimensionless volume flux per unit width transported by the flowing layer, $\hat{q}$, and the depth-averaged volume fraction, $\phibar$, as functions of the slope of the underlying boundary $S=\tan\theta$ for parameter values $R=10^{-3}$, $\psi=0.50$, $\phi_m=0.63$, $e=0.85$, $e_w=0.75$, $\delta=10^{-2}$, $W_g=9.95\times 10^{-4}(1+S^2)^{1/2}$, $St=0.317S^{1/2}(1+S^2)^{-1/4}$.  Also plotted are the asymptotic solutions (dashed) and the simple approximate solutions (dotted). Note the local maximum at $S=0.21$.}
  \label{flux_S}%
\end{figure}

It is of particular interest to evaluate the dimensionless flux of solids per unit width and the average concentration of particles, respectively given by
\begin{equation}
  {\hat q}=\int_0^1\phi {\hat v}\; {\rm d}{\hat z}\qquad\hbox{and}\qquad
  \phibar=\int_0^1\phi\;{\rm d}\hat{z},
  \label{dimensionless_flux}
\end{equation}
and these are plotted as functions of the dimensionless parameters in figures~\ref{flux_delta}, \ref{flux_wg} and \ref{flux_S}.  

From figure~\ref{flux_delta}, we observe that the dimensionless volume flux per unit width, $\hat{q}$, and the average volume fraction, $\phibar$, do not vary strongly with the relative particle size, $\delta$.  Thus, we deduce that boundary-related effects on the bulk characteristics are negligible for flows that are in excess of twenty particles thick.  This result is of particular significance when unsteady shallow flows are analysed (\S\ref{unsteady}).

The effects of the fluidisation velocity, $W_g$, are rather more subtle (see figure \ref{flux_wg}).  Increasing  $W_g$ increases the normal support of the weight of the granular layer due to the gas flow and this lowers the concentration of the layer.  However, the net volume flux, $\hat{q}$, does not vary monotonically with $W_g$.  Indeed, for the parameters in figure \ref{flux_wg}, it attains a  maximum at a dimensionless gas flow rate $W_g=2.3\times 10^{-3}$ and at that value of $W_g$, $\phi=0.43$. This reflects the trade-off between the increased mobility but lower solids fraction of more dilute currents.  Finally there is also relatively complex behaviour with increasing slope angle (figure~\ref{flux_S}).  For the computations in this figure, as we increase the inclination, we also  adjust $W_g$ and $St$, but maintain $Re$ constant (see \eqref{parameters} and \eqref{eq:Re}). We find that as the slope increases, the normal stress developed by the particle collisions increases with increasing granular temperature and this supplements the fluidising gas flow, leading to a progressively decreasing average volume fraction.  However the dimensionless volume flux exhibits a more complicated dependency because while the velocity fields increase with increasing slope, the volume fraction diminishes and eventually becomes sufficiently dilute for $\hat{q}$ to be maximised at some finite value of $S$.  (For the parameters analysed in figure~\ref{flux_S}, the local maximum in the flux occurs at $S=0.22$.)

\subsubsection{Asymptotic solution}\label{asymptotic}

In the bulk of the flow away from the boundaries, it is possible to deduce an asymptotic solution to the governing equation for the regime $\delta\ll 1$ and $R\ll 1$. This regime will have a widespread validity as $d\ll h$ in order to use a continuum approach, and for gas-solid flows $\rho_g\ll\rho_s$.  From \eqref{ndfluidmom}, we note that to leading order and away from boundaries, the downslope velocities of the two phases must be equal (${\hat u} = {\hat v} + O(1)$). Furthermore, from \eqref{ndthermal} there is a local balance between granular temperature production and dissipation such that
\begin{equation}
f_1\left(\pd{{\hat v}}{{\hat z}}\right)^2=f_3{\hat T},
\label{thermal_local}
\end{equation}
provided the volume fraction of particles is not too small (i.e $\phi\gg\delta$, so that the `conductive' effects of the granular temperature remain negligible).

The governing equations for the normal and perpendicular momentum balances, \eqref{ndnormalstress} and \eqref{ndsolidmom}, are then given by
\begin{eqnarray}
-S\pd{}{{\hat z}}\left(f_2{\hat T}\right)&=&\phi-W_g\frac{f_0}{(1-\phi)^2},\label{red1}\\
\pd{}{{\hat z}}\left((f_1f_3)^{1/2}{\hat T}\right)&=&-\phi\label{red2}
\end{eqnarray}
These reduced governing equations neglect shear stresses in the fluid phase, which become non-negligible as the basal boundary is approached and which allow the velocities of the two phases to differ.  Also the `conduction' of granular temperature is neglected because, to leading order, we find a balance between production and dissipation \eqref{thermal_local}.  When $\delta\ll 1$, the lower boundary layer corresponds to a region within which the velocity of the solid phase is small, while the upper boundary layer to a region within which the granular temperature is small.  The leading order boundary conditions are then given by $\hat{T}(1)=0$ and the volume fraction at the base is given by $\phi(0)=\phi_0$, which is determined by substituting for $\partial \hat{v}/\partial \hat{z}$ from \eqref{thermal_local} into \eqref{bc_base},
\begin{equation}
0=\left(\frac{f_6f_5^2f_3}{f_1}-f_7\right)\hat{T},
\end{equation}
where the constitutive functions $(f_i)$ are evaluated at $\phi=\phi_0$.  The basal volume fraction is thus a function of $e$, $e_w$, $\psi$ and $\phi_m$.

Rearranging \eqref{red1} and \eqref{red2} and denoting $f=(f_1f_3)^{1/2}$, we find that
\begin{eqnarray}
\left(f\dot{f}_2-\dot{f}f_2\right){\hat T}\pd{\phi}{{\hat z}}&=&\left(-\phi+\frac{W_gf_0}{(1-\phi)^2}\right)\frac{f}{S}+f_2\phi,\\
\left(\dot{f}f_2-f\dot{f}_2\right)\pd{{\hat T}}{{\hat z}}&=& \left(-\phi+\frac{W_gf_0}{(1-\phi)^2}\right)\frac{\dot{f}}{S}+\dot{f}_2\phi,
\label{phaseplane}
\end{eqnarray}
where $\dot{}$ denotes differentiation with respect to $\phi$. It is straightforward to integrate numerically these coupled first-order equations subject to the boundary conditions $\hat{T}(1)=0$ and $\phi(0)=\phi_0$. The solutions are plotted in figures \ref{fig_dvary}, \ref{fig_wgvary} and \ref{fig_Svary} and it is evident that these asymptotic solutions accurately reproduce the numerical solution of the complete system (very often in these figures, the asymptotic curves are indistinguishable from the numerical solution of the complete system).  

\begin{figure}
  \centering
  \includegraphics[trim=4.5cm 10cm 4.5cm 10cm,clip,width=0.48\columnwidth]{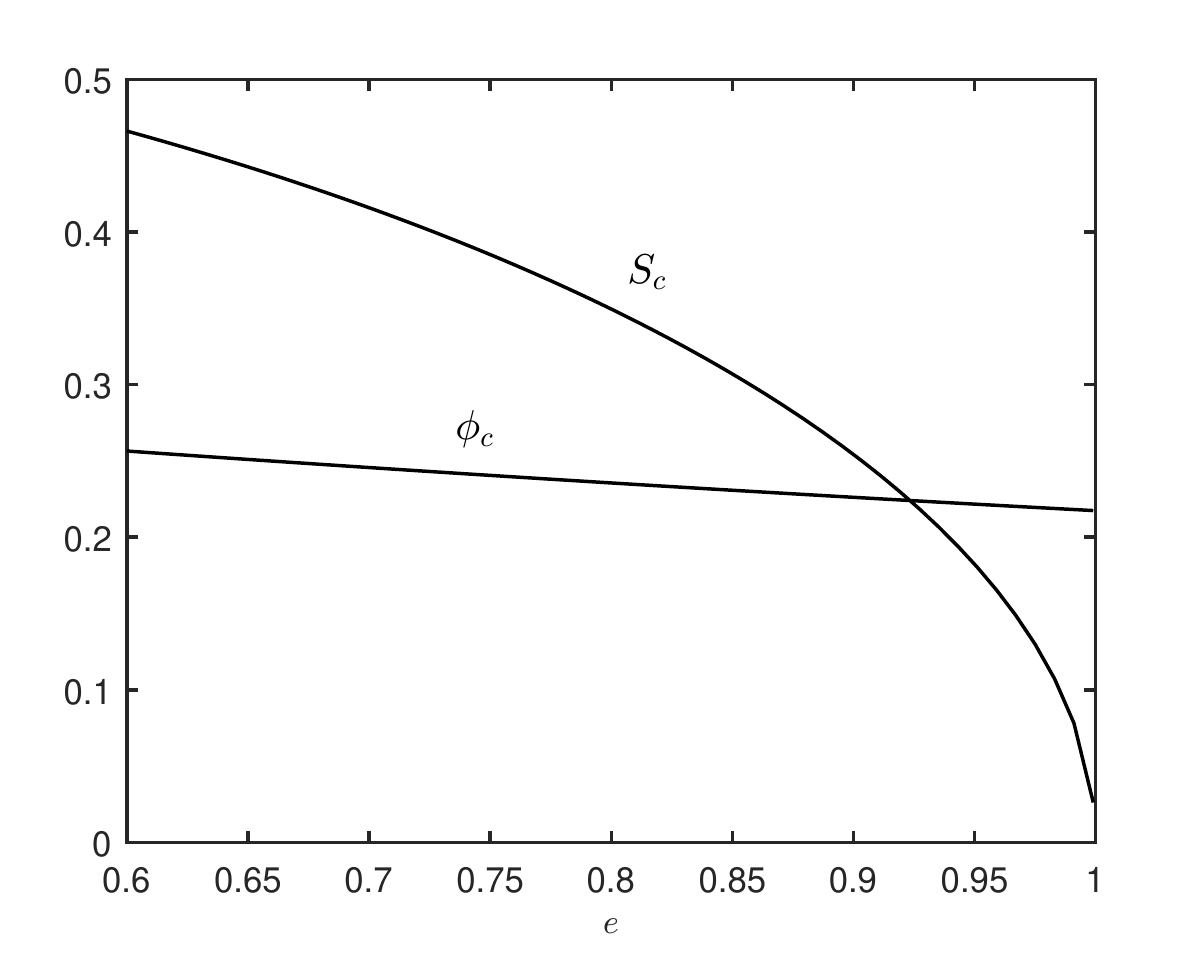}%
    \includegraphics[trim=4.5cm 10cm 4.5cm 10cm,clip,width=0.48\columnwidth]{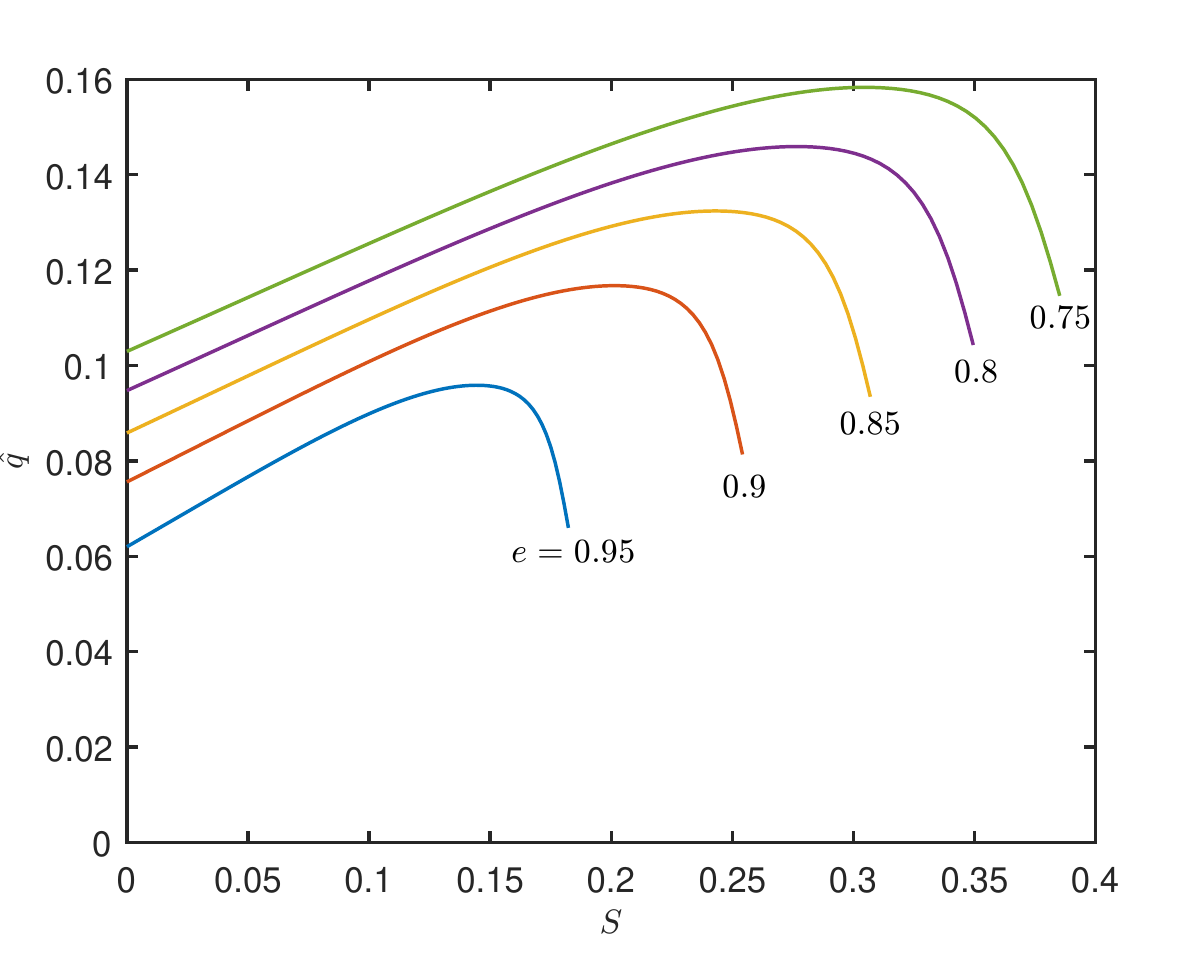}\\
  \includegraphics[trim=4.5cm 10cm 4.5cm 10cm,clip,width=0.48\columnwidth]{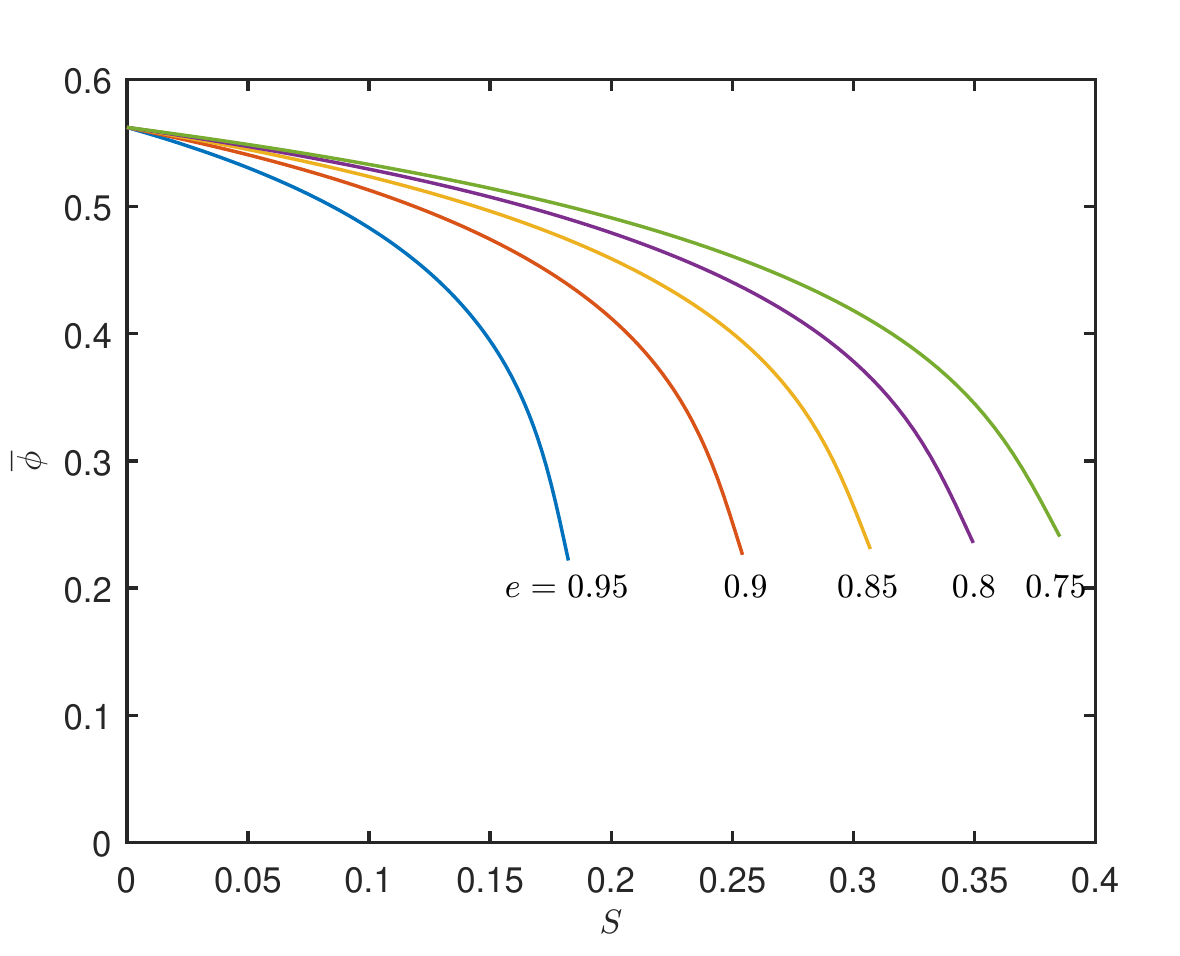}
  \includegraphics[trim=4.5cm 10cm 4.5cm 10cm,clip,width=0.48\columnwidth]{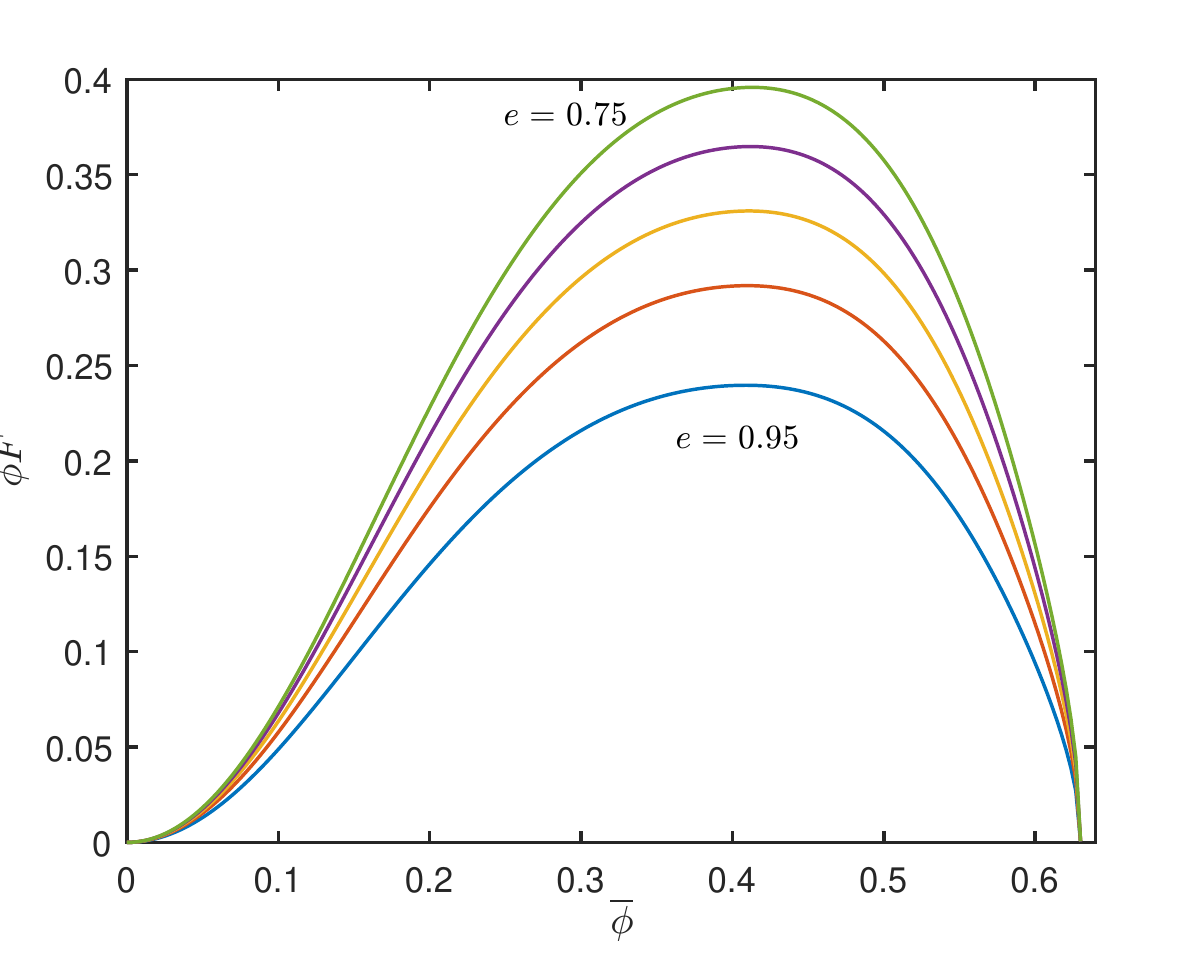}%
  \caption{(a) The limiting volume fraction, $\phibar_c$ and slope $S_c$ for which the uniform volume fraction is the `attracting' solution as functions of the coefficient of restitution;  (b) The volume flux per unit width, $\hat{q}$, (c) the average volume fraction, $\phibar$, as functions of the slope $S$ for $W_g=10^{-3}$ and varying values of the coefficient of restitution and (d) the product of $\phibar$ and the mobility factor, $F$, as a function of $\phibar$ for various values of the coefficient of restitution.}
  \label{flux_phaseplane}
\end{figure}

There is also an even simpler approximate solution.  The coupled system admits a homogeneous solution $\phi(z)=\phibar$ when
\begin{equation}
{\phibar}-\frac{f_0W_g}{(1-\phibar)^2}=\frac{S\phibar f_2}{(f_1f_3)^{1/2}},
\label{phiapprox}
\end{equation}
where the constitutive functions are evaluated at $\phi=\overline{\phi}$. In this case, the temperature gradient is constant, $\partial {\hat T}/{\partial}{\hat z}=-\lambda$, with $\lambda=\phibar/f$.  This solution is `attracting' in the sense that trajectories in the phase plane $(\phi({\hat z}),{\hat T}({\hat z}))$ approach it when
\begin{equation}
  f\dot{f}_2-\dot{f}f_2<0,
  \label{eq:attracting}
\end{equation}
which in turn demands that $\phibar>\phibar_c(e)$ and that $S<S_c(e)$ (see figure~\ref{flux_phaseplane}c). If these inequalities are not held then the reduced system evolves towards a  state different from a uniform volume fraction $(\phi=\phibar)$ and may not admit solutions at all.   Physically, when $S>S_c$, the dissipation of granular temperature, here encapsulated through a coefficient of restitution $e$, is insufficient to allow for a steady balance between the weight of the flowing layer, the fluidising drag and the normal stresses generated through particle interactions (a balance expressed by \eqref{red1}). When $e\lessapprox 0.85$, we find that this limitation does not play for parameter values associated with the flows considered in this study and that a flow with a homogeneous volume fraction of particles provides a good representation of the more complete dynamics (see figures \ref{fig_dvary}-\ref{fig_Svary}).  

When the dimensionless fluidisation velocity $W_g$ and the slope $S$ are set, the average volume fraction within the current, $\phibar$, can be calculated using \eqref{phiapprox}.  Figure~\ref{flux_phaseplane}a shows the effect of $e$ on the curves of $\phibar$ as a function of slope $S$ when the fluidisation flow is constant ($W_g=10^{-3}$).  The curves in this plot are continued up to the maximum value of the slope, $S_c(e)$ for which the reduced model leads to a homogeneous volume fraction and it can be seen that the slope at which this can be achieved is successively reduced as dissipation in the collisions is decreased.  From figure \ref{flux_phaseplane}b, for a given slope, $S$, and fluidisation gas flow rate, $W_g$, flows with lower coefficients of restitution lead to higher dimensionless volume fluxes per unit width.  This is simply rationalised: as a high coefficient of restitution implies reduced dissipation and high granular temperatures.  Consequentially there are higher stresses and greater resistance to the downslope motion.
 
Since the granular temperature must vanish at the surface ${\hat z}=1$ to leading order, we find for the simple approximate solution with uniform volume fraction that the granular temperature is given by
\begin{equation}
  {\hat T}=\frac{\phibar}{(f_1f_3)^{1/2}}(1-{\hat z}),
  \label{Tapprox}
\end{equation}
and the velocity field of the solid phase is given by
\begin{equation}
  {\hat v}=F\frac23\left(1-(1-{\hat z})^{3/2}\right),
  \label{vapprox}
\end{equation}
where $F=(\phibar^2f_3/f_1^3)^{1/4}$.  The scaled slip velocity at the wall $Ff_5\delta$ can be added to \eqref{vapprox}, but when $\delta\ll 1$, it is negligible. The velocity profile \eqref{vapprox} is similar to the dimensionless `Bagnold' velocity profile up to the factor $F$, which controls the mobility of the flowing layer and is influenced by the fluidisation velocity.  In many situations the approximate solution provides a very good representation of the solution to the complete system (see figures \ref{fig_dvary} to \ref{fig_Svary}).

Also in this regime $(\delta\ll 1)$,  the approximate solution yields
\begin{equation}
  {\hat q}= \tsf 25\phibar F,
\end{equation}
and this is plotted in figures \ref{flux_delta}, \ref{flux_wg} and \ref{flux_S}, once again illustrating the utility of this asymptotic solution.  The quantity $\overline{\phi}F$ thus plays a crucial role in determining the dimensionless flux, $\hat{q}$ and in figure \ref{flux_phaseplane}d, we plot its dependence on volume fraction for a range of values of $e$.  We note that $\overline{\phi}F$ is maximised for $\phi\approx 0.41$ (with the precise value weakly dependent on $e$) and vanishes both when $\phi$ vanishes and when it approaches maximum packing.  This variation reflects the balance between fast moving dilute flows and slow moving concentrated flows, leading to a flux maximum at intermediate values $(\phi\approx 0.41)$.  Finally, we note that a dimensional estimate of the depth of a fluidised current may be obtained
\begin{equation}
  \label{hest}
  h=\left(\frac{5}{2}\frac{d}{\phibar F}\frac{q_0}{\sqrt{g\sin\theta}}\right)^{2/5},
\end{equation}
where $q_0$ is the dimensional flux per unit width at the source and the effects of slip at the wall has been neglected.  

\subsection{Experimental measurements of fully developed flows}

\subsubsection{Depth of currents}

\begin{figure}
	\centering
	\includegraphics[width=.8\textwidth]{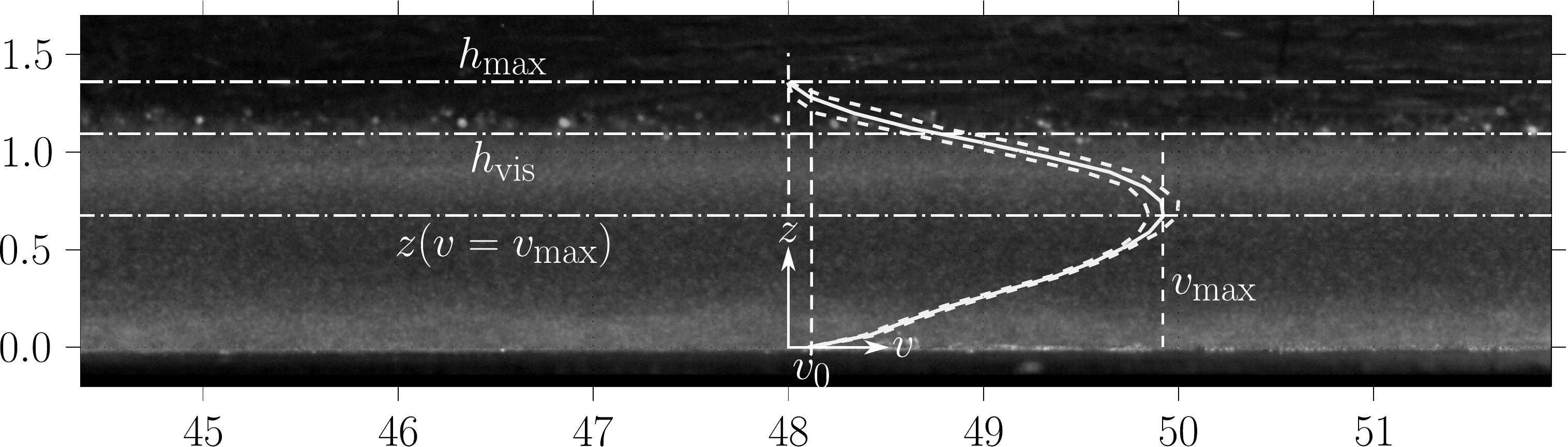}
	\caption{Image of a granular current with the measured velocity superimposed onto it (solid line).  The image is an average-intensity composite of the of the images used in the PIV measurements.  Note the non-zero (slip) velocity at the base ($v_0$).  The dashed lines indicate the 95\% confidence interval using \eqref{eq:ensAveUncertainty}.  Also shown are the height where the velocity profile drops to zero, $\hmax$, the maximum visible height of the current, $\hvis$, and the height of the peak of the velocity profile, $z(v=\vmax)$.  For this figure $\theta = 15^\circ$,  $Q = 42.5$~\cmt/s, $\U = 1.5$.  }
	\label{profile}
\end{figure}

A typical velocity profile is shown in figure~\ref{profile}, superimposed on a captured image from the recording of an experiment. There is a small slip at the lower boundary, an approximately linear increase in velocity with distance from the wall until a maximum velocity is attained and then a progressive drop to zero.   There appears to be a top to the current where the particle volume fraction suddenly drops and there $\hvis(\equiv h)$.  $h_{vis}$ is greater than the height at which the maximum velocity is attained.  Above $h_{vis}$ particles are detected, but their velocity drops with increasing height and it has a large variance.  This is consistent with there being a ballistic region into which individual particles may be projected.  The height at which particle velocity drops to zero is the top of the entire current and is denoted by $\hmax$.     The depth of the current can fluctuate a little with time (see figures~\ref{currentshapes} and \ref{heightTime}, top).  As a result, the averaging process will occasionally include points that are above the average height of the current so that the averaged velocity at these points will be necessarily lower than in the bulk of the current and the variance will be higher.  The measured heights for the different currents are summarised in Table~\ref{tab:heights} compared with the height estimated from (\protect\ref{hest}).  

\begin{table}
	\centering
	\begin{tabular}{cccccc} 
		\toprule
		\multirow{2}{*}{$\theta$/[$^\circ$]} & \multirow{2}{*}{$Q$/[\unit{\cmt/s}]} & \multicolumn{3}{c}{Experimental measurements} & Estimate \\ 
		\multicolumn{2}{c}{}& $z\left( v = v_{\mathrm{max}}\right)$/[\unit{cm}] & $\hvis$/[\unit{cm}] & $\hmax$/[\unit{cm}] & $h$/[\unit{cm}]\\ 
		\midrule
		10 	& 33.1 	& 0.56 & 0.74 & 1.42 & 0.68 \\
			& 33.8	& 0.61 & 0.96 & 1.00 & 0.77 \\
			& 53.5	& 0.76 & 1.35 & 1.35 & 0.93 \\
		15 	& 11.2 	& 0.52 & 0.69 & 0.94 & 0.43 \\
			& 38.2	& 0.73 & 1.07 & 1.19 & 0.71 \\
			& 42.5	& 0.77 & 1.09 & 1.25 & 0.74 \\
		\bottomrule 
	\end{tabular} 
	\caption{Various estimates of height in steady-state currents.  $\hvis$ is measured from photographs of the currents such as that in figure~\ref{profile}. It may be compared with the prediction $h$, calculated from \eqref{hest} with $F$ based on $\phiest$ (see Table~\ref{tab:flows}),  $e=0.85$, and $\psi=0.50$.  $\hmax$ is found directly as the height at which particle velocity drops to zero.  In all cases $w_g=1.5u_{mf}$.  }
	\label{tab:heights}
\end{table}

\begin{figure}
	\includegraphics[width=0.8\textwidth]{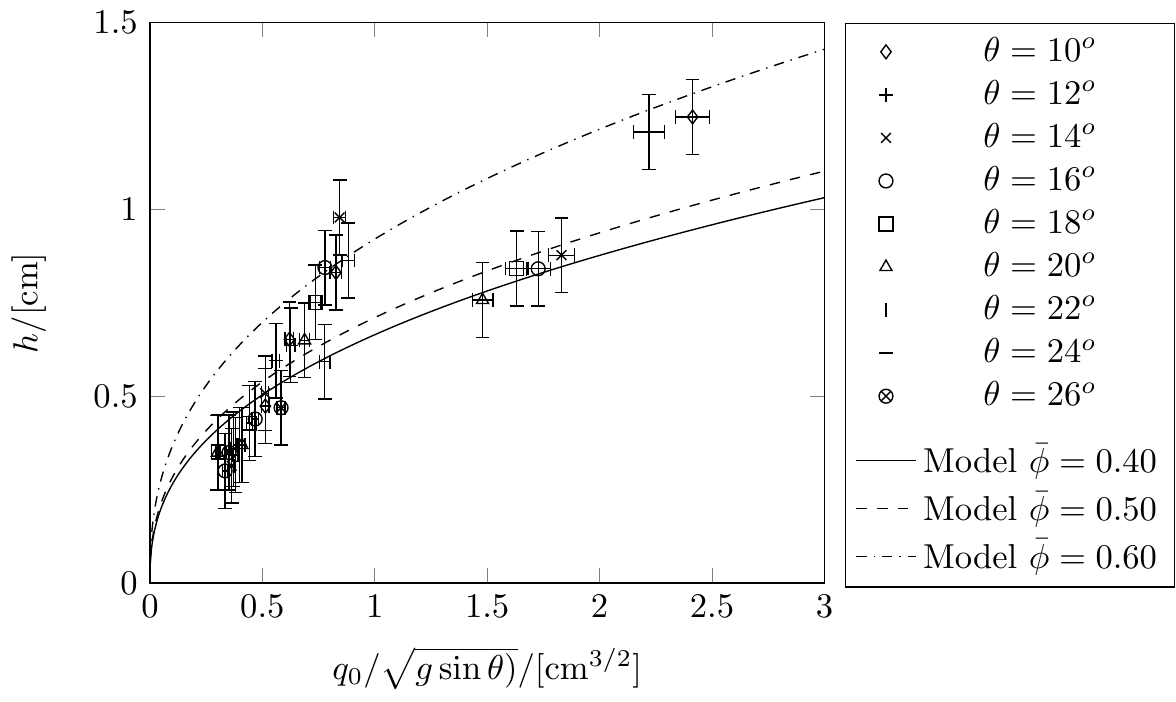}
	\caption{\label{qvsh}  The height of the flowing layer as a function of the scaled volume flux, $q_0/\sqrt{g\sin\theta}$ for a range of channel inclinations.  The model curves are for \eqref{hest} with  $\phi_m=0.63$ and $e=0.85$. }   \label{fig_qvsh}
\end{figure}

The asymptotic solution yielded an approximate formula linking height and source volume flux \eqref{hest}, and this is shown in figure~\ref{fig_qvsh}.  In this figure,  fixed values of  $\phibar$ were used because from \eqref{phiapprox}, $\phibar$ depends on $\theta$ and it was not possible to find a solution for the full experimental range of $\theta$ for a fixed value of $e=0.85$. The decreasing effect of $\phibar$ on mobility as its value approaches $\phibar_c=0.40$ reflects the effect it has on mobility $\phibar F$ shown in figure~\ref{flux_phaseplane}d.  The theoretical formula contains no adjusted parameters and is an approximation to the more complete description, but it yields a reasonable quantitative representation of the relationship between the depth of the flowing layer, the source flux and channel inclination. 

\subsubsection{Velocity profiles}

\begin{figure}
  \centering
  \begin{minipage}[t]{0.8\textwidth}
    \centering
    \includegraphics[width=\linewidth]{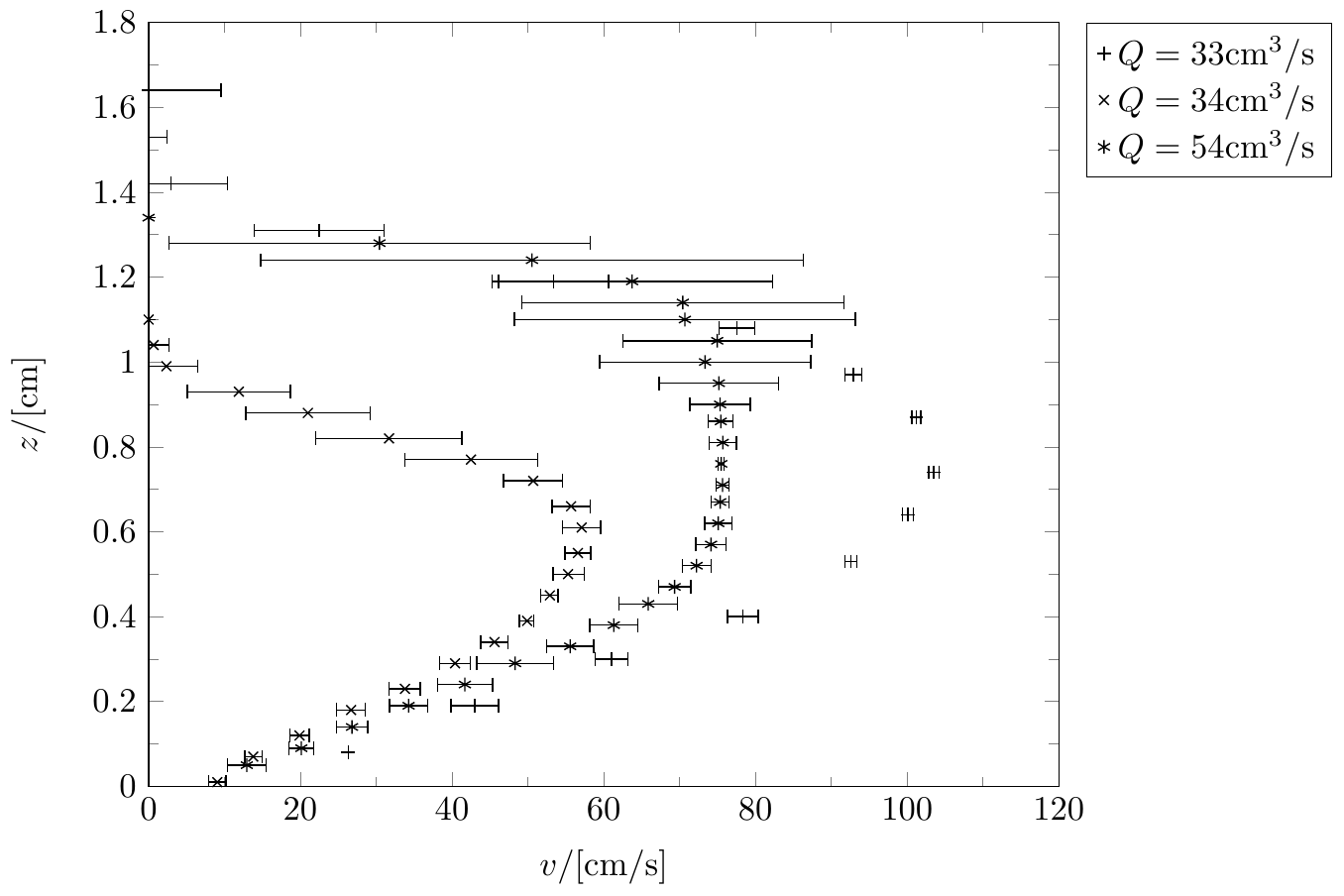}
  \end{minipage}
  \\
  
  \begin{minipage}[t]{0.8\textwidth}
    \centering
    \includegraphics[width=\linewidth]{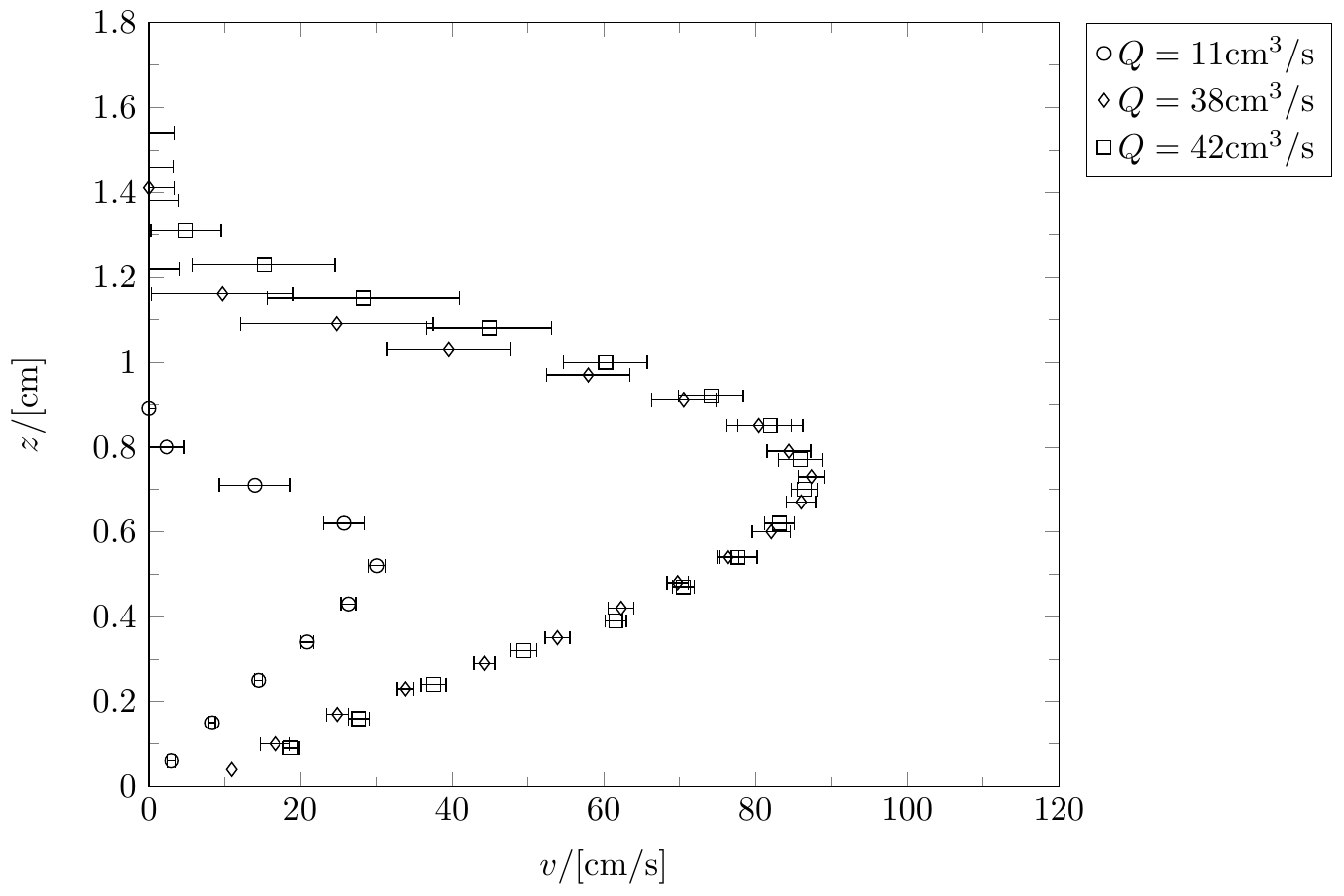}
  \end{minipage}
  \caption{\label{velprofs}Ensemble averaged velocity profiles for steady flows for $\theta=10\degree$ and $\theta=15\degree$.  Error bars are 95\% confidence limits calculated from \eqref{eq:ensAveUncertainty} with $m = 5$ and $n = 16$.  }
\end{figure}

The ensemble averaged velocity profiles for 10\degree~and 15\degree~slopes measured halfway along the tank are shown in figure~\ref{velprofs}.  The features described for figure~\ref{profile} are reflected in each of the profiles.  The structure of the currents posited by the model is similar in structure to the experimental measurements up to the point at which the velocity is a maximum (see figure~\ref{fig_dvary}).  Above this point, the variance of the velocity measurements increases markedly as the velocity drops-off.  As described above this effect could be because the PIV is simply sampling ballistic particle trajectories.

The velocity profiles measured on inclinations of $\theta=10\degree$, with source fluxes $Q=\unit[33]{cm^3/s}$ and $Q=\unit[34]{cm^3/s}$ are distinct from each other with the former forming a current that is deeper and much faster than the latter.  It is not clear why there is such a large difference between the  measured profiles.  One possibility could be that the flowing layer exhibits multiple states for the same imposed flux. This behaviour is known in models of unfluidised granular flows \citep{woodhouse_rapid_2010}; however we failed to find such multiplicity of solutions in the governing equations examined in this study in the parameter regime corresponding to these experiments.  The relatively fast and expanded flow with $Q=\unit[33]{cm^3/s}$ leads to small estimates of particle volume fraction with an excessive portion of the flow where $h>h_{vis}$ (see below, section~\protect\ref{sec:phi}.  Based on $h_{vis}$, $\bar{\phi}=0.33$; based on $h_{max}$, $\bar{\phi}=0.22$), which seem physically unlikely, and  so this experimental run is not reported further.  

\begin{figure}
  \centering
  \begin{minipage}[t]{0.8\textwidth}
    \includegraphics[width=\linewidth]{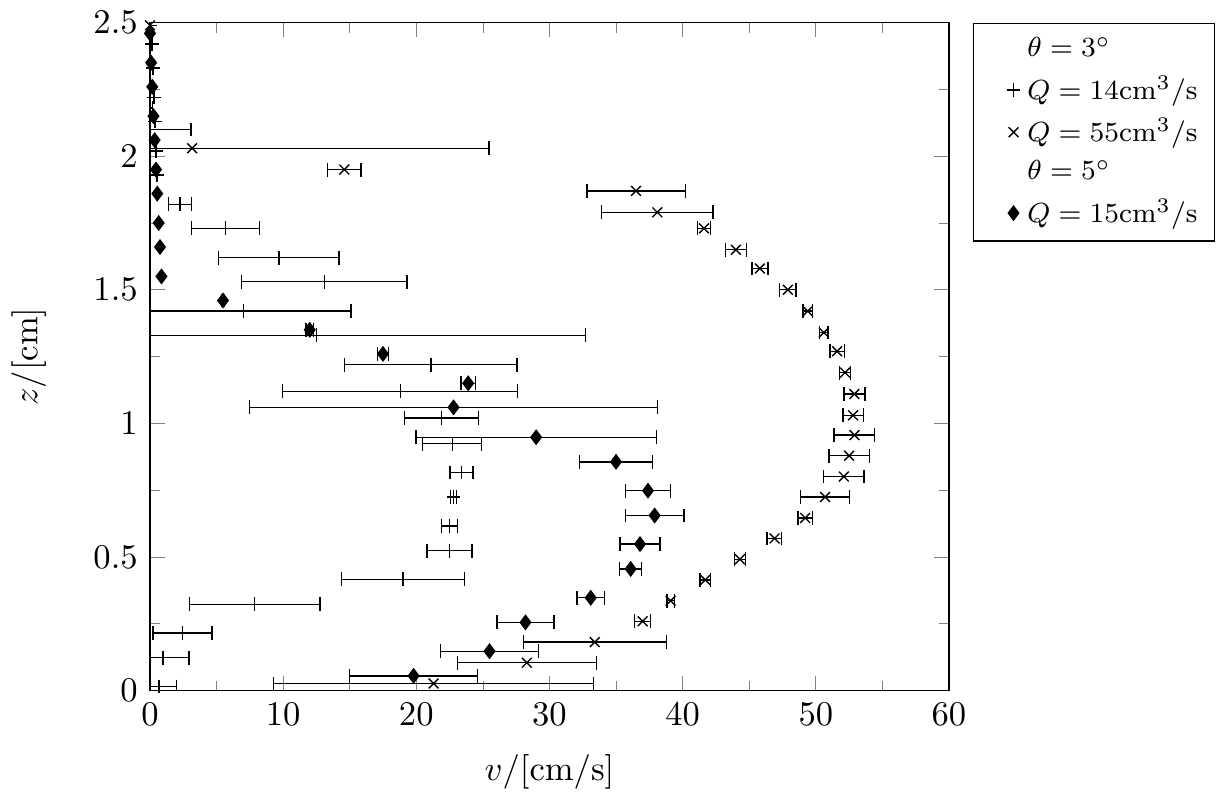}
  \end{minipage}
  \caption{\label{instvelprofs}Measured velocity profiles from experiment for individual fluidised granular flows that are potentially unsteady when they are at shallow angles.  Error bars are calculated from \eqref{eq:pivUncertainty} with values of $c_1$ determined from the correlation functions calculated during the PIV analysis.  }
\end{figure}

Figure~\ref{instvelprofs} shows some instantaneous velocity profiles halfway along the tank for shallower slope angles whose motion may not be steady.  The velocity profiles for the 3\degree~and 5\degree~ slopes are similar in character to the averaged profiles for steeper slopes.

\subsubsection{Particle volume fraction}
\label{sec:phi}

The expectation from the model is that the particle volume fraction $\phi$ would be approximately uniform within the fluidised currents.  It is not possible to analyse the degree to which $\phi$ is a function of position in the currents from our experimental set-up; however, it is evident that towards the top of the current above $h_{vis}$, $\phi$ drops sharply so that the current loses its opacity.  This is consistent with the decreasing particle velocity there.  Some bubbles were seen in the currents, but not many and at a small number of sites, even once the current had traversed the bottom of the container, and those that were seen were small in size.  

From the measured velocity profiles, it is possible to estimate the average volume fraction, $\bar{\phi}$, by integrating the particle velocity profiles and dividing the measured particle flow rate $\phi_0Q$ by the result.  The results $\phi_{meas}$ are shown in Table~\ref{tab:flows}, using $\hvis$ as the overall depth of the current, to give measured volume fraction of the currents ($\phimeas$).  The measured values of particle volume fraction can be compared with estimated values, $\phi_{est}$, which have been calculated using \eqref{phiapprox}.

There can be quite good agreement between $\phiest$ and $\phimeas$ despite several inherent uncertainties in their computation.  Overall, the values of $\phimeas$ were comparable to the values of $\phiest$ and with with the typical values of $\phi=0.50-0.60$ for static fluidised beds \citep{Epstein1962}.  In between $\hvis$ and $\hmax$ particles are present, but in practice the fall-off of velocity above $h_{vis}$ is sufficiently rapid that this makes very little difference to calculations of $\phi_{meas}$:  if $\phi_{meas}$ is calculated on the basis of the top of the current being $h_{max}$ rather than $h_{vis}$, then its value decreases by less than 0.02, except when $\theta=15\degree$ and $Q=\unit[11.2]{cm^3/s}$, when it reduces by 0.09.

\begin{table}
	\centering
	\begin{tabular}{cc cc} 
		\toprule
		$\theta$/[$^\circ$] & $Q$/[cm$^3$/s] 	&   $\phiest$ 	&  $\phimeas$ \\ 
		\midrule
		10					& $33.8$ 				& $0.51$				& $0.59\pm{0.05}$\\
							& $53.5$ 				& $0.51$ 					& $0.45\pm{0.05}$\\
		15					& $11.2$ 				& $0.41$ 					& $0.60\pm{0.02}$\\
							& $38.2$ 				& $0.41$ 					& $0.40\pm{0.02}$\\
							&$ 42.5 $				& $0.41$ 					& $0.40\pm{0.02}$\\
		\bottomrule 
	\end{tabular} 
	\caption{Measured estimates for $Q$ and $\bar{\phi}$ from integration of the velocity profiles.  Values of $\phiest$ are found from solving \eqref{phiapprox}, and those for $\phimeas$ from integration of the measured velocity profiles up to the height $h_{vis}$. The errors are calculated using 95\% confidence limits calculated from \eqref{eq:ensAveUncertainty}.}  
	\label{tab:flows}
\end{table}

\subsubsection{Slip at the wall}

\begin{table}
  \centering{ 
    \begin{tabular}{ccccccc} 
    \toprule
    $\theta$/[$^\circ$] & $Q$/[cm$^3$/s] & $v(0)$/[cm/s]	& $\frac{\partial v}{\partial z}\Big|_{z=0}$/[1/s] & \begin{tabular}{c}Slip\\length\end{tabular} & $\psi$ & $\delta f_5$\\
    \midrule
    10 			   & 33.8 		& 7.1  	& 113 	& 1.68 & 0.22 & 0.022\\
				    & 53.5 		&6.6 	& 147	& 1.20 & 0.30 & 0.014\\
    15 			    & 11.2 		& 0 	& 61		& - & -  & -	\\
				    & 38.2 		& 4.2 	& 134	& 0.83 & 0.37 & 0.020\\
				    & 42.5 		& 6.7 	& 133	& 1.34 & 0.23 & 0.019\\
    \bottomrule 
    \end{tabular} 
    \caption{Measured slip velocities at the wall, $v(0)$, and velocity gradients $\frac{\partial v}{\partial z}\big|_{z=0}$ with the resulting estimate for slip length and for $\psi$ using \eqref{psiest}. The slip length is expressed in terms of particle diameters and is defined as $v(0)/d\frac{\partial v}{\partial z}\big|_{z=0}$. $e=0.85$, $\phi_m=0.63$.  }  
    \label{tab:psiest}
  }
\end{table}

Specularity coefficients have not been measured for fluidised granular currents and some modellers think they should not be used at all for individual collisions \citep{Goldschmidt2004}.  Their value is sufficiently badly defined that in their investigations of bubbling fluidised beds of glass particles \cite{Altantzis2015} used values between $10^{-4}$ and $0.5$ and \cite{Li2010} from $0$ to $0.5$.  Our computations showed that apart from within relatively narrow layers close to the boundary, the magnitude of the specularity coefficient had relatively little effect upon the flow profiles. It is, however, possible  to estimate the value of $\psi$ from the directly measured slip velocities and gradients using the boundary condition \eqref{bottombcs} and the definition of  $f_5$ in Table~\ref{grantab}, so that in terms of dimensional variables
\begin{equation}
\label{psiest}
\psi=\frac{2\sqrt{3}}{\pi}\frac{\phi_m}{\bar{\phi}}\frac{f_1}{g_0}\frac{d}{v(0)}\frac{\partial v}{\partial z}\bigg|_{z=0}.
\end{equation}
The results are shown in Table~\ref{tab:psiest}, and it can be seen that a measured average value of $\psi$ is 0.28. The values for $\psi$ shown in Table~\ref{tab:psiest} should be treated as only indicative as the velocity gradients close to the wall are shallow and the slip velocities small, so small variations in the velocity profiles can result in significant changes in the value of $\psi$; however, despite these uncertainties, the value of $\psi$ is reasonably consistent.  The effect of $\psi$ on the velocity profiles is to introduce a slip velocity proportional to $\delta f_5$.  Even for the relatively large values of $\psi$ estimated here, the magnitude of this dimensionless term is relatively small.  

\subsubsection{Scaling of the velocity profiles}

\begin{figure}
   \centering
   \includegraphics[width=0.8\textwidth]{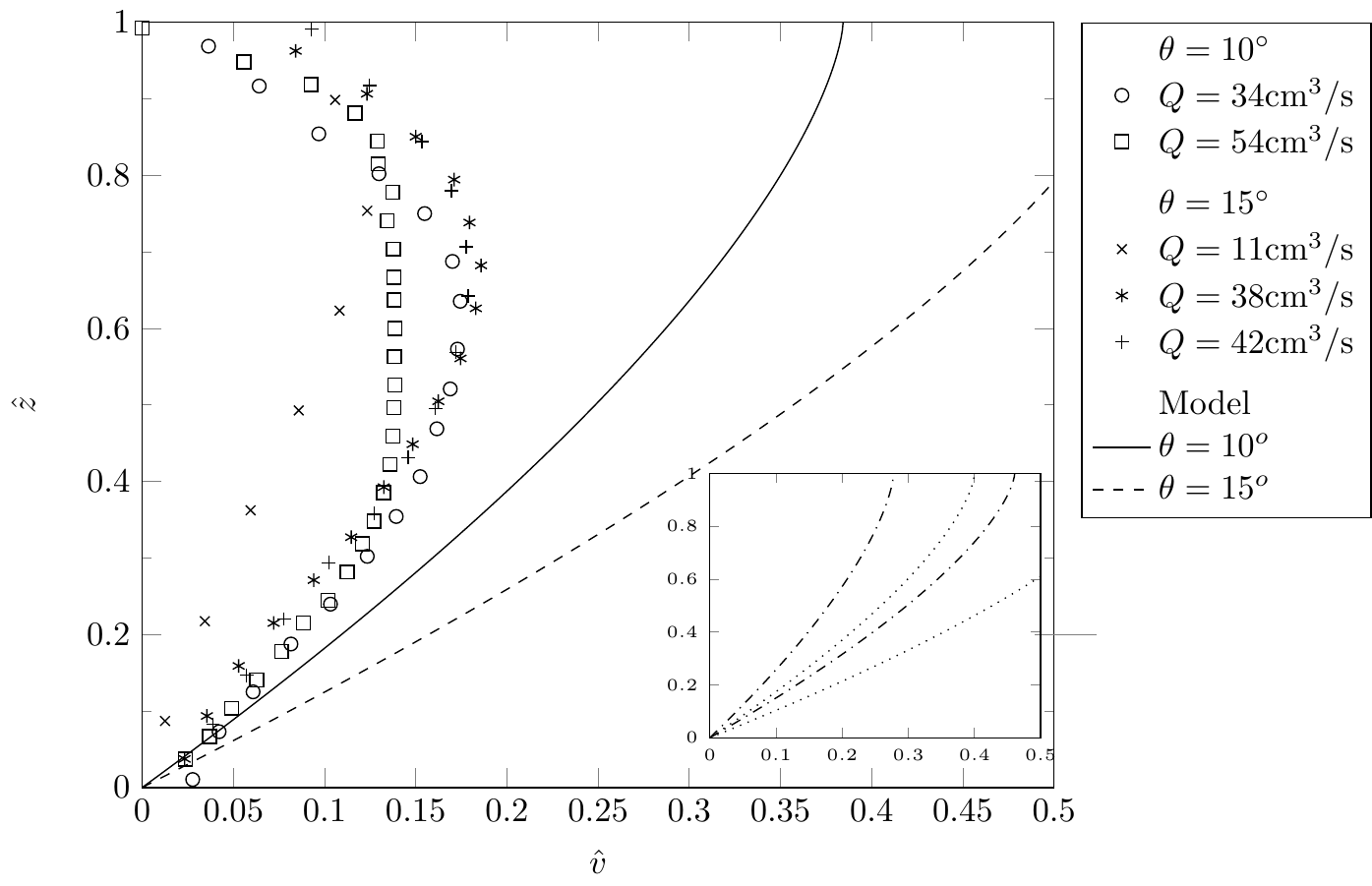}
   \caption{Velocity profiles with measurements scaled as \eqref{Usize} compared with the model scaled velocity profile under steady state \eqref{vapprox} for each angle where $e=0.85$.  The curves correspond to $\bar{\phi}=0.51$ when $\theta=10\degree$ and $\bar{\phi}=0.41$ when $\theta=15\degree$.  $W_g=8.09 \times 10^{-4}$ when  $\theta=10\degree$ and $W_g=8.24 \times 10^{-4}$ when  $\theta=15\degree$.   The inset graph shows the effect of the value of $e$ on the model solutions with the chain-dot curves corresponding to $e=0.95$ and the dotted curves to $e=0.75$.}
   \label{velscale}
\end{figure}

The measured velocity profiles scaled as $\hat{v}$ and $\hat{z}$ are plotted in figure~\ref{velscale}.  With the exception of the lowest flow rate when $\theta=15\degree$, the data collapses well in the region close to the wall.  The model predicts dependence of the theoretical curve on the slope angle for the flow through its influence on $\bar{\phi}$ and hence $F$, but this is not reflected in the experimental curves for which the scaling appears to eliminate the effect of $S$. $F$ is also affected by the value of $e$.  Increasing $e$ causes a decrease in the predicted dimensionless velocity:  the theoretical curves shift towards the left and the difference between the curves for $\theta =10\degree$ and $15\degree$ becomes less (see inset, figure~\ref{velscale}).  

\section{Unsteady, developing flows on slopes}
\label{unsteady}

The mathematical model  may be extended to unsteady, developing flows of fluidised currents down slopes, but it now takes a somewhat different form because stream-wise gradients can no longer be neglected.  In this situation, we analyse the motion in the `lubrication' regime, for which a representative streamwise length scale, $L$, far exceeds a representative length scale perpendicular to the boundary, $H$ ($H/L\ll 1$).  This means that accelerations perpendicular to the boundary are negligible and that to leading order, the normal stresses adopt the `hydrostatic' balance given by \eqref{hydrostatic}.  We again assume that the flows are many particles thick, $\delta\ll 1$, the density of the gas is negligible relative to that of the solids, $R\ll 1$, and the effects of gas viscosity are negligible (see \S\ref{asymptotic}).  The leading order, dimensional momentum equations of each phase parallel with the incline then take the form,
\begin{eqnarray}
  0&=&-(1-\phibar)\pd{p}{x}+\beta(v-u)\label{gasmom}\\
  \rho_s\phibar \frac{Dv}{Dt}&=&-\phibar\pd{p}{x}+\pd{\sigma_{xz}}{z}+\pd{\sigma_{xx}}{x}+\phibar\rho_s g\sin\theta-\beta(v-u),\label{solidmom}
\end{eqnarray}
where the average volume fraction in the flowing layer, determined by the balance between the fluidising gas flow and the particle weight, is given by \eqref{phiapprox}. It is interesting to note from \eqref{gasmom} that now there must be a leading order difference between the downslope velocities of the two phases.  Furthermore, since the flow is spatially and temporally evolving, we must include the inertia of the solid phase, which in \eqref{solidmom} is given by the term $\rho_s\phibar Dv/Dt$ (here $D/Dt$ denotes the material derivative).  Summing these two momentum balances to eliminate the inter-phase drag and assuming further that the stresses in the solid phase are isotropic $(\sigma_{xx}=\sigma_{zz})$ and that the current is in hydrostatic balance \eqref{hydrostatic}, we deduce that
\begin{equation}
  \rho_s\phibar\frac{Dv}{Dt} =\rho_s\phibar g\left(\sin\theta-\cos\theta\pd{h}{x}\right)+\pd{\sigma_{xz}}{z}.
  \label{solidslopemom}
\end{equation}
The granular temperature of the flow in this regime is assumed to be in local balance between its production and dissipation through collisions, and these processes dominate its advective and diffusive transport.  We proceed further by adopting the appropriate dimensionless scales following the distinguished scaling identified in \S\ref{nondim} and embodied in the dimensionless variables of \eqref{Usize} and \eqref{Usize}.  However, here we non-dimensionalise the depth of the flowing the current $h$ by a representative depth-scale $H$, $(\hat{h}=h/H)$, and additionally 
\begin{equation}
  \label{slopescale}
  \hat{v}=\frac{v}{F\left(g\sin\theta H^3/d^2\right)^{1/2}},\quad\hat{x}=\frac{x}{L}\quad\hbox{and}\quad
  {\hat t}=F\left(\frac{g\sin\theta H^3}{d^2}\right)^{1/2}\frac{t}{L}.
\end{equation}
Then, using the approximate form of the solution established in \S\ref{asymptotic}, we find that the depth-integrated, dimensionless momentum equation is given by
\begin{equation}
  {\cal R}\left(\pd{}{\hat t}\int_0^{\hat h}{\hat v}\;\rd {\hat z}+\pd{}{\hat x}\int_0^{\hat h}{\hat v}^2\;\rd {\hat z}\right)+\frac{1}{\Delta}\pd{}{\hat x}\left(\frac{{\hat h}^2}{2}\right)={\hat h}-\left(\pd{\hat v}{\hat z}\right)^2_{z=0},
  \label{solidslopemom_int}
\end{equation}
where $\Delta=L\tan\theta/H$ and 
\begin{equation}
  {\cal R}=\frac{F^2H^3}{d^2 L}.
\end{equation}
In this setting, as for \citet{Kumaran2014}, $\cal{R}$ measures the relative magnitude of the inertial to resistive terms.  Unlike a Reynolds number for viscous fluid flows, it features only the length scales in the problem and $F$, because both the inertial terms and the shear stresses are proportional to the square of velocity.  

To proceed further we assume that the velocity field adopts similar dependence to \eqref{vapprox} on distance from the boundary and this permits the evaluation of the integral and boundary quantities in terms of the average velocity and the depth of the layer:
\begin{equation}
  \int_0^{\hat h}{\hat v}\;\rd{\hat z}={\hat h}{\overline v},\qquad
  \int_0^{\hat h}{\hat v}^2\;\rd{\hat z}=\frac 5 4 {\overline v}^2 {\hat h}\qquad\hbox{and}\qquad
  \pd{\hat v}{\hat z}_{z=0}=\frac{5{\overline v}}{2{\hat h}}.
  \label{depth_av_mom}
\end{equation}
To complete the model, we express conservation of mass, 
\begin{equation}
  \pd{\hat h}{\hat t}+\pd{}{\hat x}\left({\overline v}{\hat h}\right)=0.
  \label{depth_av_mass}
\end{equation}
This system is subject to the boundary condition that we impose a sustained source of particles at the origin
\begin{equation}
  \phibar {\hat h}{\overline v} = {\hat q}_0 \qquad\hbox{at}\qquad \hat{x}=0.
\end{equation}
Additionally, if the flow is supercritical then we must enforce the Froude number at the source.  We impose the initial condition, ${\hat h}({\hat x},0)=0$ and the current forms a front ${\hat x}={\hat x}_f({\hat t})$, such that ${\hat h}({\hat x}_f,t)=0$.  

Before constructing solutions, it is convenient to relate the height and streamwise length scales.  We choose $L = H/\tan\theta$ and thus $\Delta = 1$.  The lubrication regime requires that streamwise lengths far exceed the thickness of the flow; since the current is expanding in the streamwise direction, this regime is inevitably entered after sufficient time.  However the adopted scaling  may imply that in terms of these variables, the initial evolution may not be well captured by the lubrication assumption.  We further choose the dimensional height $H$, using (\ref{hest}) so that 
\begin{equation}
\label{Hdef}
H=\left(\frac{d}{\phibar F}\frac{q_0}{\sqrt{g\sin\theta}}\right)^{2/5}.
\end{equation}
The governing equations now entail the single dimensionless parameter, ${\cal R}=\tan\theta F^2 H^2/d^2.$

We construct travelling wave solutions for the dimensionless height and velocity fields.  We write ${\hat h}({\hat x},{\hat t})\equiv {\hat h}({\hat x}-c{\hat t})$ and ${\hat v}({\hat x},{\hat t})\equiv {\hat v}({\hat x}-c{\hat t})$, where $c$ is the dimensionless wave speed which is to be determined. Conservation of mass then implies that ${\hat v}=c$ and in particular, the front speed is given by 
\begin{equation}
  \label{eq:penslope}
{\hat x}_f=c{\hat t}. 
\end{equation}
Balance of momentum leads to
\begin{equation}
  \frac{{\cal R}c^2}{4}{\hat h}'+{\hat h}{\hat h}'=\hat{h}-\frac{25}{4}\frac{c^2}{{\hat h}^2},
  \label{travel_ode}
\end{equation}
where a prime denotes differentiation with respect to $\eta=\hat{x}-c\hat{t}$. Distant from the front, the flowing layer carries a constant volume flux of material determined by the source conditions ($c{\hat h}\to 1$ as $\eta\to-\infty$).  Here we note that the travelling wave solutions do not satisfy the source condition precisely at $\hat{x}=0$, but instead it is satisfied as $\eta\to-\infty$.  For these flows, we find that the current adjusts over a short distance behind the front to a uniform depth and velocity and thus the travelling wave solution provides an accurate representation of the solution for the flow.  Thus, we deduce that the position of the front and the far-field depth are given by
\begin{equation}
c=\left(\frac 25\right)^{2/5}\qquad\hbox{and}\qquad \hat{h}\to\hat{h}_\infty=\left(\frac 52\right)^{2/5}.
\label{hinfty}
\end{equation}

\begin{figure}
	\centering
  \includegraphics[width=0.9\textwidth]{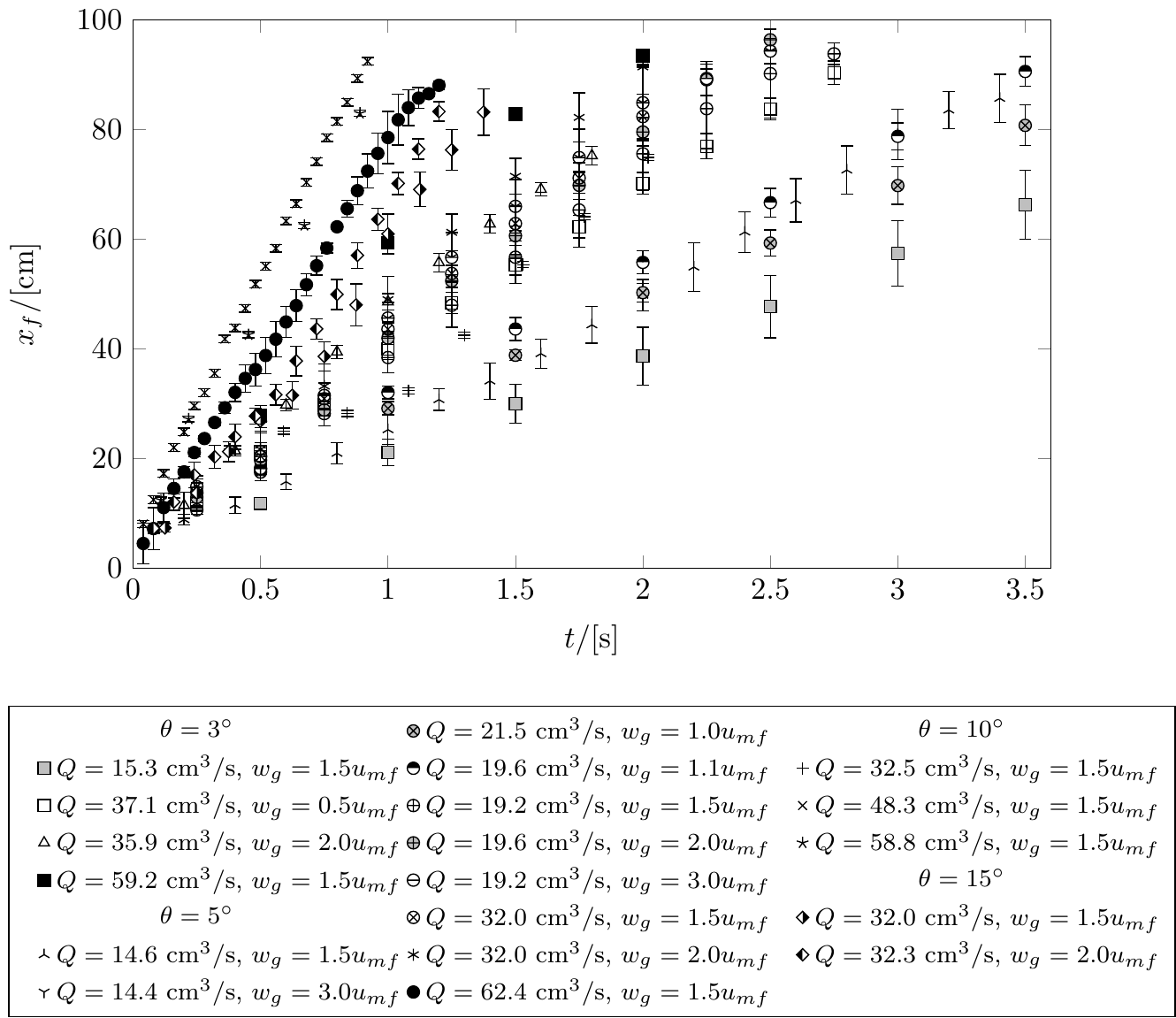}
  \caption{The position of the front of the fluidised current as a function of time for flows along channels of varying inclinations with varying source fluxes and fluidising gas flows.}
  \label{slopepen}
\end{figure}

\begin{figure}
  \centering
  \begin{minipage}[t]{0.8\textwidth}
    	\includegraphics[width=\linewidth]{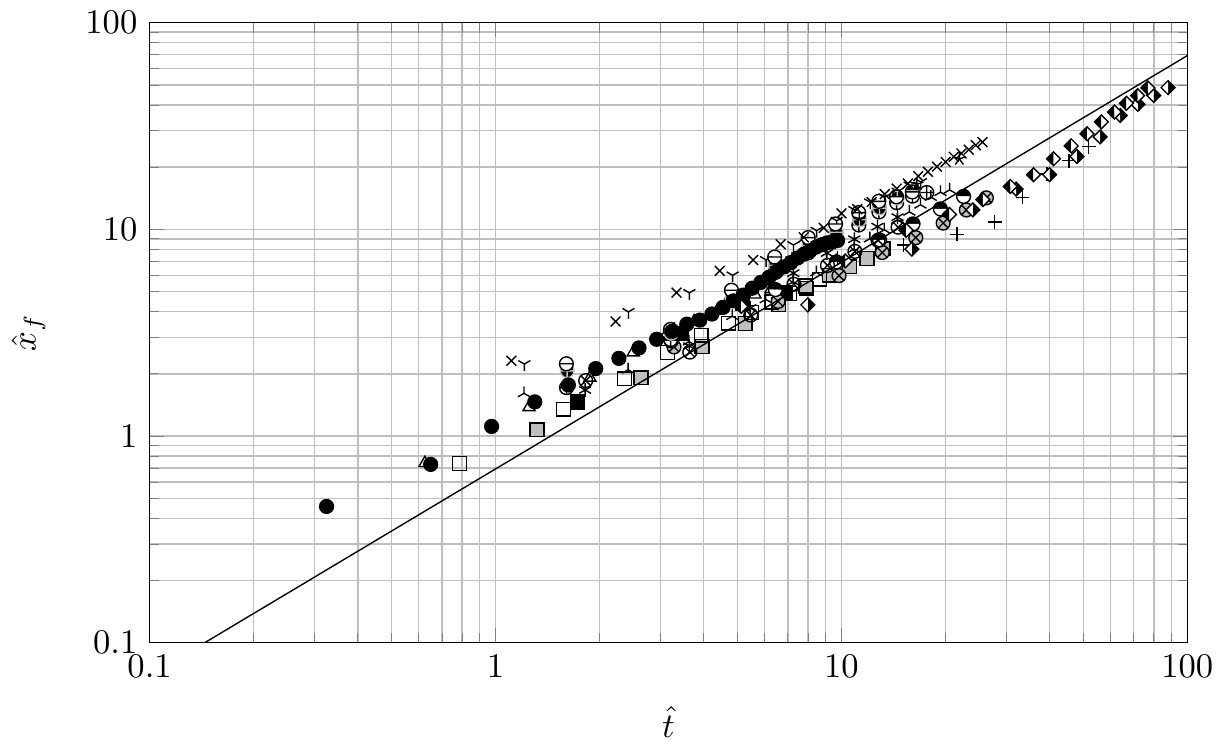}
	\label{logslope}
  \end{minipage}
  \hfill
  \begin{minipage}[t]{0.8\textwidth}
    	\includegraphics[width=\linewidth]{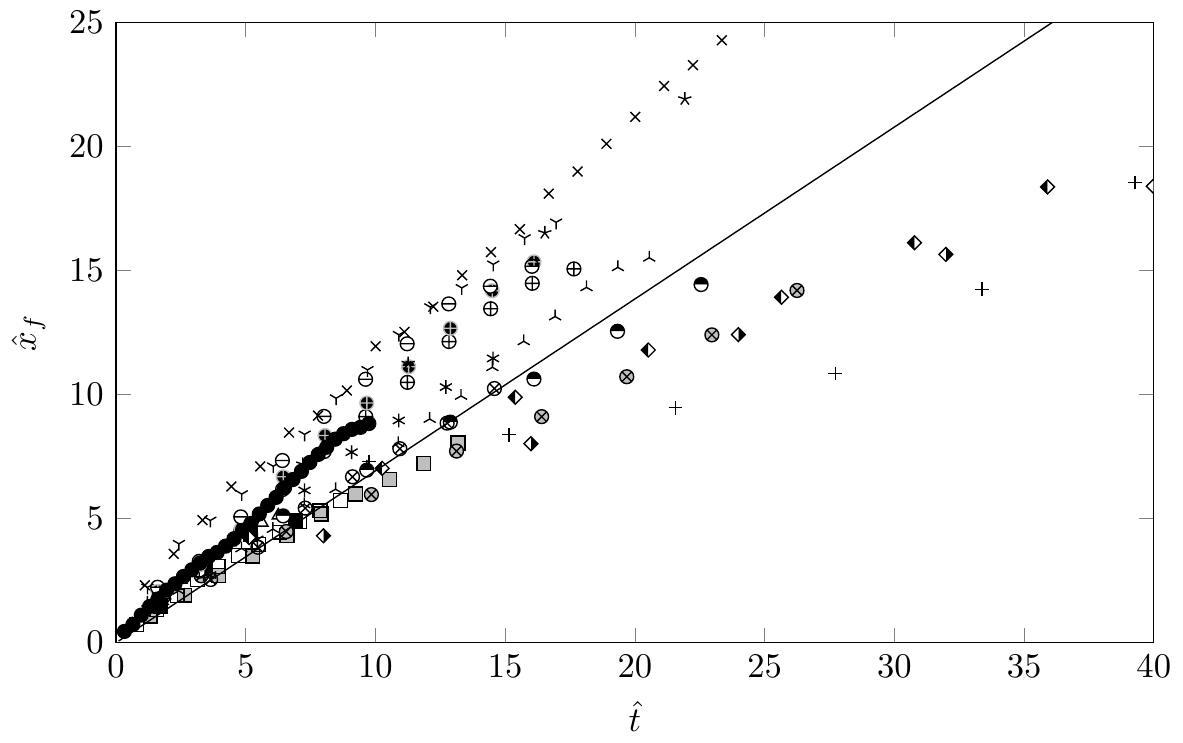}
    	\label{linslope}
  \end{minipage}
  \caption{The rescaled position of the front as a function of rescaled time on logarithmic and linear axes (with scalings given by (\ref{slopescale})). The model curve corresponds to (\ref{eq:penslope}).  Legend as for Figure~\ref{slopepen}.   }
  \label{slopemodel}
\end{figure}

Experimental measurements of the distance travelled by fluidised currents with time are shown in Figure~\ref{slopepen}, in which the inclination of the channel, the source volume flux and the fluidising gas velocity were varied; the measured flow speeds ranged over a factor of five.  The measurements for the fully fluidised currents ($u_g/u_{mf}>1$) after scaling are shown in Figure~\ref{slopemodel}. From \eqref{eq:penslope}, $\hat{x}_f$ should be proportional to $\hat{t}$ and this is true even when the slope angles are small.   Individually, the currents display a constant speed; however, the measured speeds can be significantly different from that expected from the model ($q_0/\phi h$).

\begin{figure}
  \centering
  \begin{minipage}[t]{0.45\textwidth}
    \includegraphics[width=\linewidth]{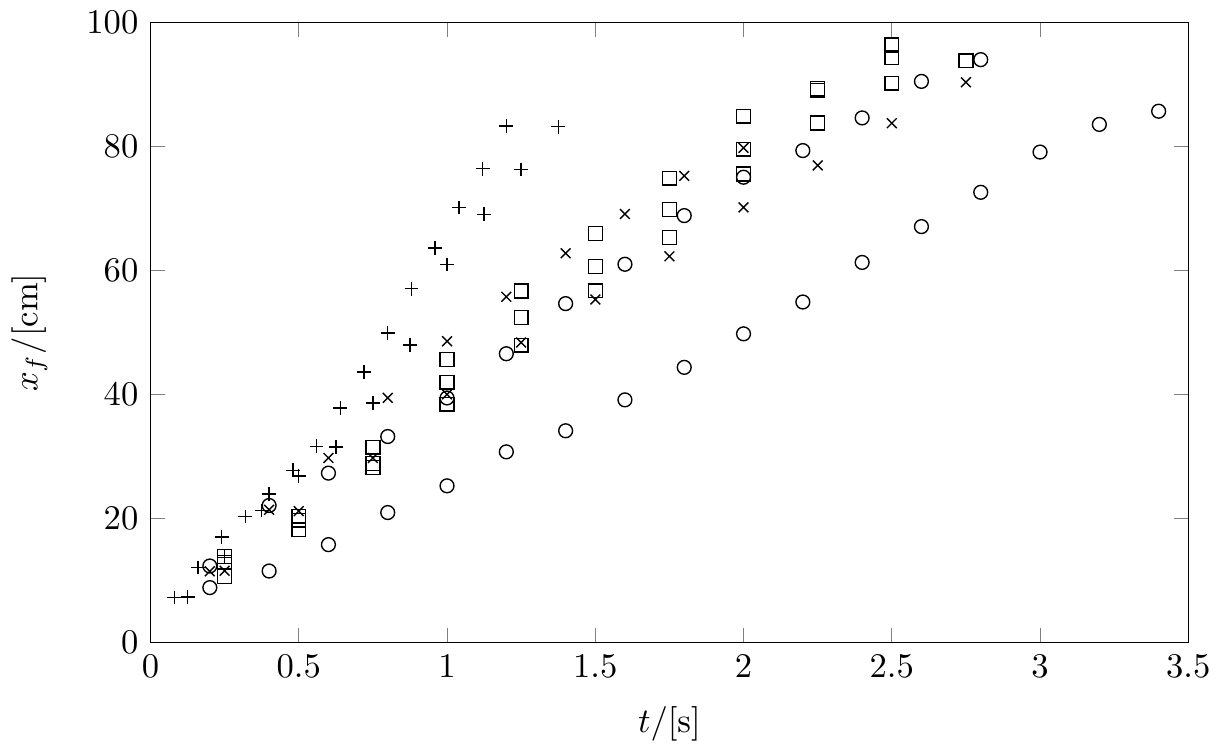}
    \label{umfraw}
  \end{minipage}
  \hfill
  \begin{minipage}[t]{0.45\textwidth}
    \includegraphics[width=\linewidth]{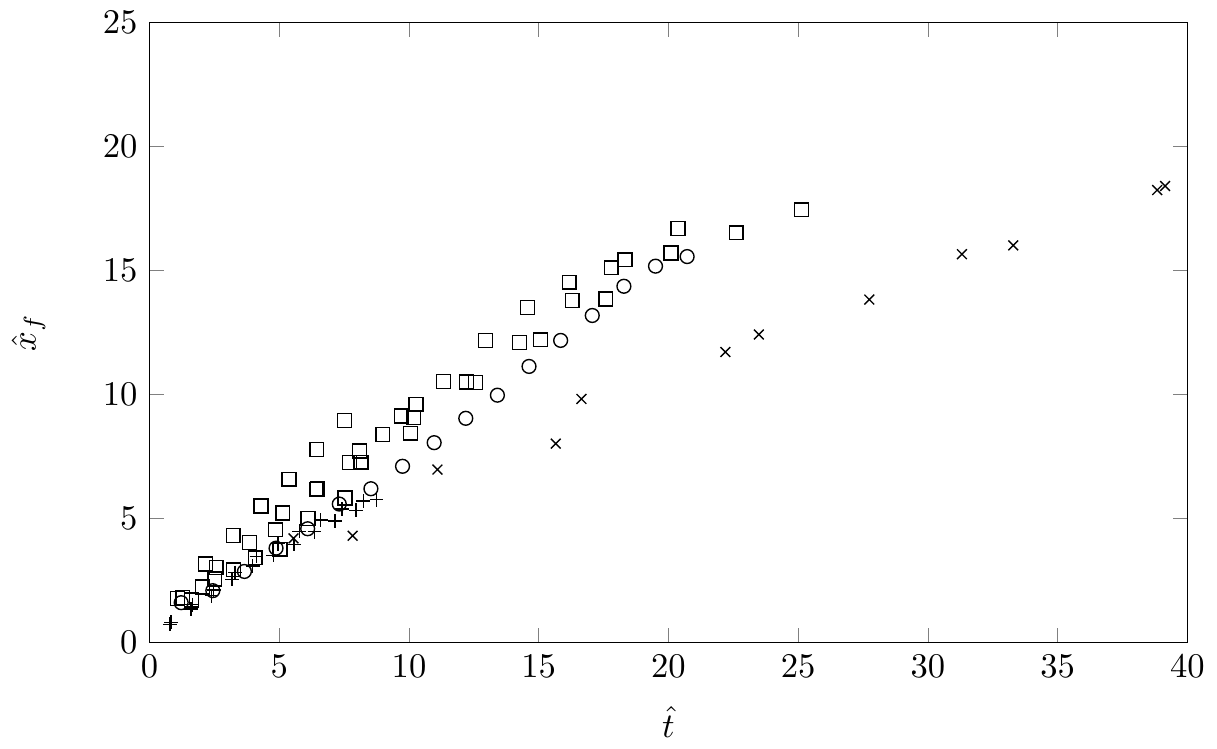}
    \label{umfscaled}
  \end{minipage}
  \\
  \begin{minipage}[t]{0.45\textwidth}
    \includegraphics[width=\linewidth]{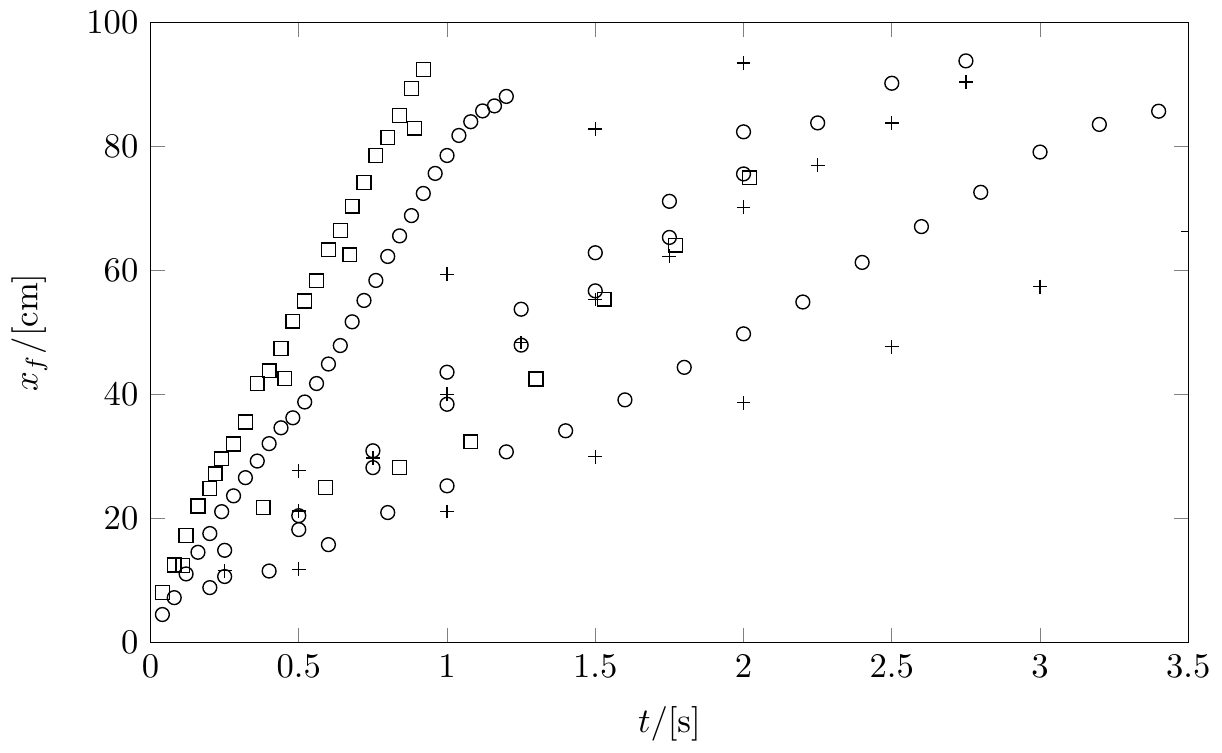}
    \label{Qraw}
  \end{minipage}
  \hfill
  \begin{minipage}[t]{0.45\textwidth}
    \includegraphics[width=\linewidth]{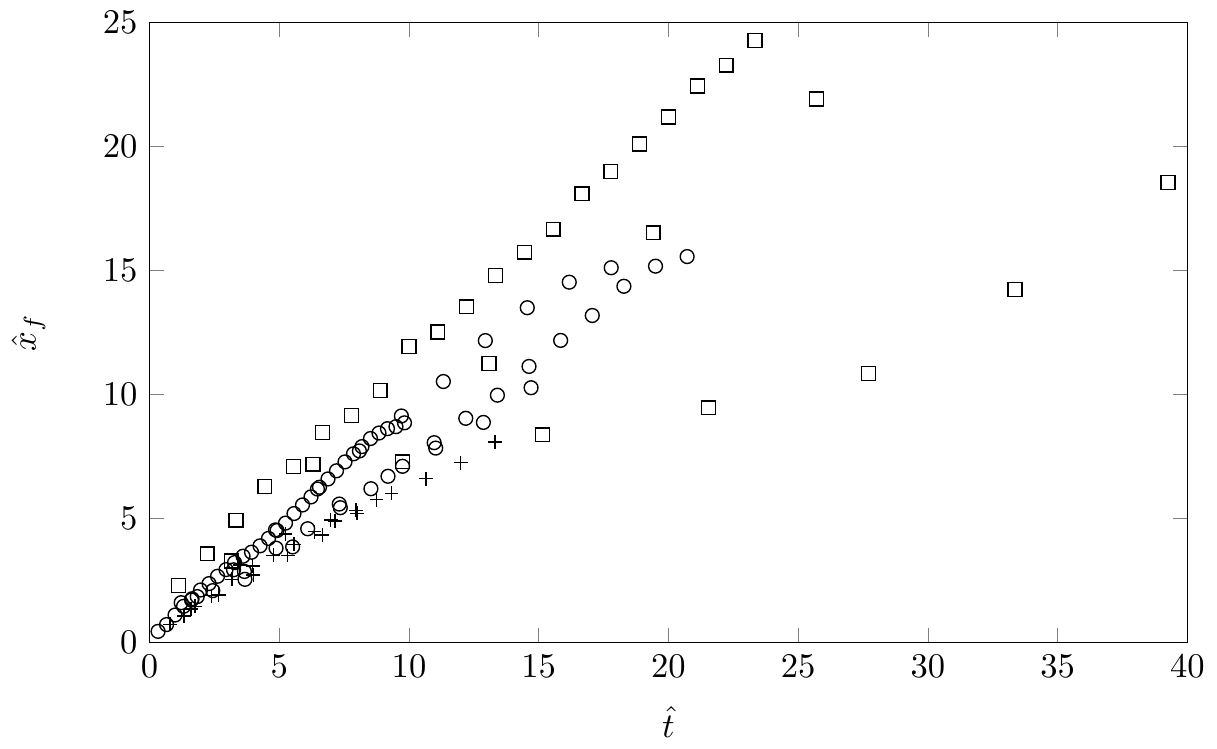}
   \label{Qscaled}
  \end{minipage}
  \\
  \begin{minipage}[c]{0.45\textwidth}
    \includegraphics[width=\linewidth]{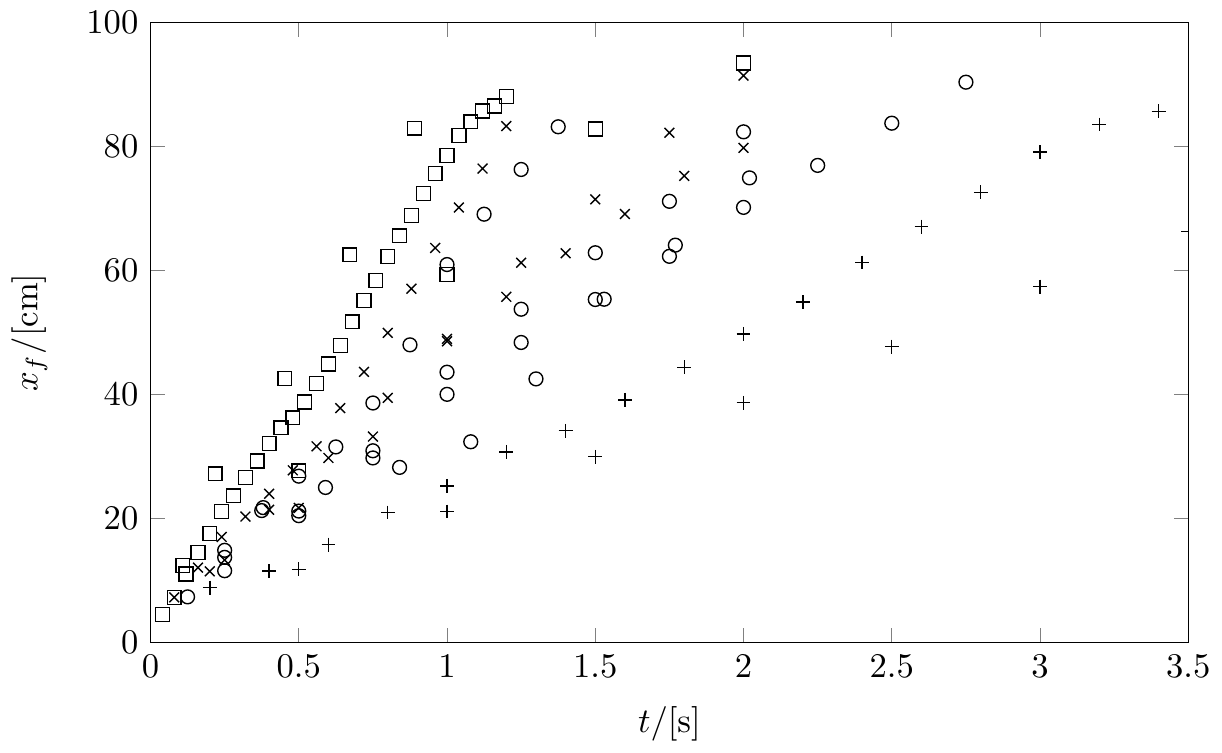}
    \label{thetaraw}
  \end{minipage}
  \hfill
  \begin{minipage}[c]{0.45\textwidth}
    \includegraphics[width=\linewidth]{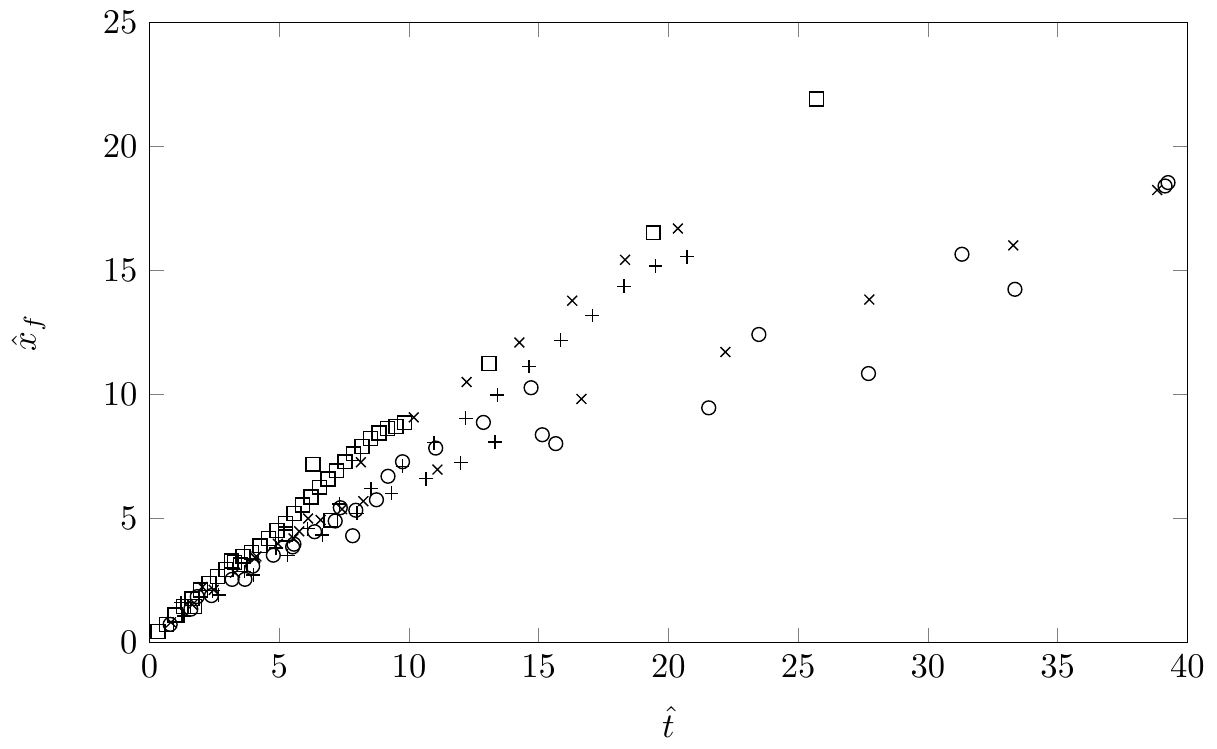}
    \label{thetascaled}
  \end{minipage}
  \caption{The effect of different variables on the distance travelled by flows down slopes before and after scaling.  The graphs from the top downwards show the effects of varying $w_g$, $Q$, and $\theta$.  The points in the graphs correspond to the legend in figure~\ref{slopemodel}.  For $w_g$ there are four families of curves:  $\theta=3\degree$, $\Qnom=35$cm$^3$/s ($+$); $\theta=5\degree$, $\Qnom=15$cm$^3$/s ($\circ$); $\theta=5\degree$, $\Qnom=20$cm$^3$/s ($\square$); $\theta=15\degree$, $\Qnom=35$cm$^3$/s ($\times$).  For $Q$ there are  three families of curves corresponding to $\theta= 3\degree$ ($+$), $5\degree$ ($\circ$), $10\degree$($\square$), with $w_g=1.5u_{mf}$.  For $\theta$ there are four families of curves:  when $w_g=1.5u_{mf}$, $\Qnom=15$cm$^3$/s (+), $35$cm$^3$/s ($\circ$), $60$cm$^3$/s ($\square$), and when $w_g=2.0u_{mf}$, $\Qnom=35$cm$^3$/s ($\times$).
    }
  \label{slopeeffects}
\end{figure}

The degree of data collapse for different parameters - $w_g$, the nominal flow rate $\Qnom$, and $\theta$ - is shown in Figure~\ref{slopeeffects}.  For all three parameters, the scaling eliminates much of the scatter, but it is not fully eliminated.  The collapse of data onto different lines with the same values of $\theta$ and $\Qnom$ for $w_g$ is excellent.  It can be quite good for $\theta$, especially at low $\hat{t}$.  For $\Qnom$ the collapse of data is often incomplete.  There will be some variation reflecting the difference in value of the true value of $Q$ from the nominal value $\Qnom$.  It can be seen in Figure~\ref{umfraw} that the effect of $w_g$ is small, but significant; however, it is eliminated after scaling, as shown in Figure~\ref{slopeeffects}b.  

A systematic omission from our model is the effect of side wall drag and this could provide an additional resistance to motion, thus slowing the speed of propagation.  In appendix~\ref{sidewall}, we analyse the effects of the side walls when the height of the current, $H$, is much less than the breadth of the channel, $B$.  We demonstrate that there is a weak retardation to the dimensionless speed proportional to $(H/B)^2$ when $H/B\ll 1$.  We analysed the speed of the flow from the data plotted in figure~\ref{slopepen} and found no systematic dependence on $H/B$ and thus there is no evidence that these relatively shallow currents were significantly slowed by side wall effects.


\subsection{Time taken to establish steady uniform behaviour}

\begin{figure}
  \centering
  \includegraphics[trim=5cm 10cm 6cm 10cm,clip,width=0.6\columnwidth]{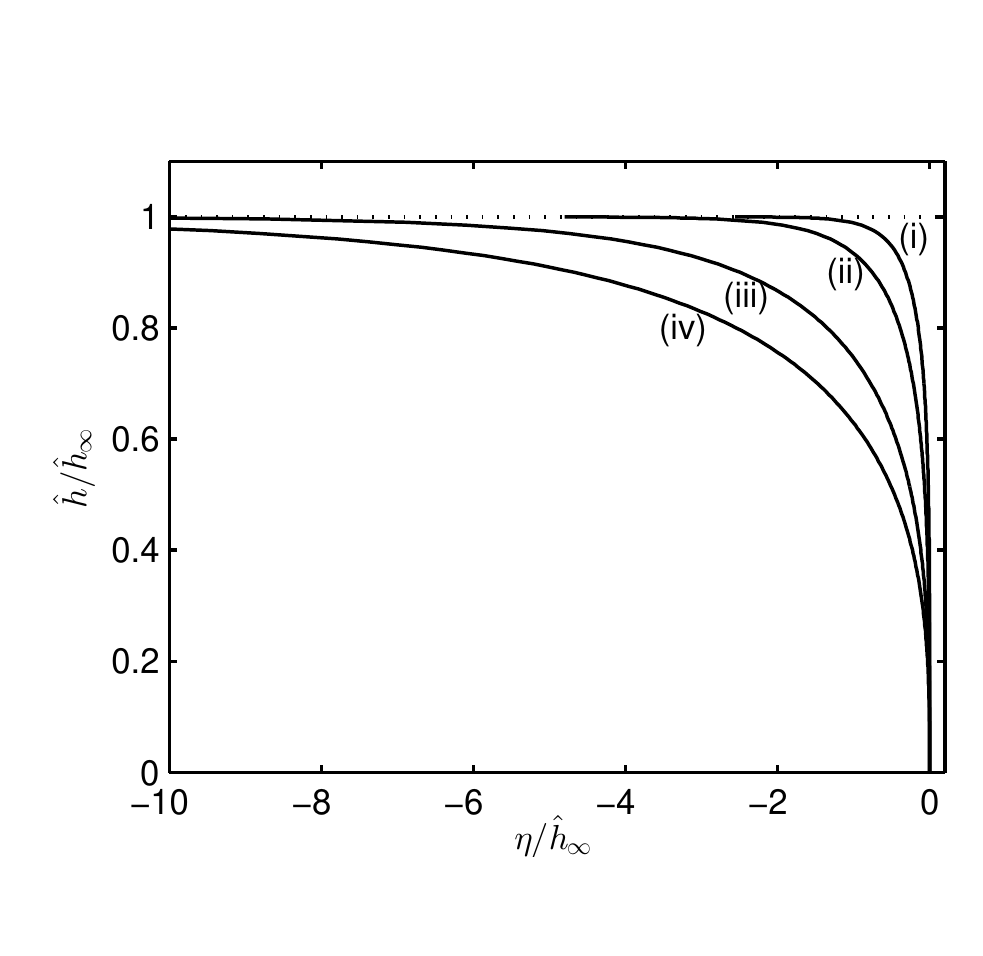}
    \caption{The scaled height  of the current, $\hat{h}/\hat{h}_\infty$, as a function of position $\eta/\hat{h}_\infty=(\hat{x}-c\hat{t})/\hat{h}_\infty$ for parameter values (i) $A = {\cal{R}} / (4 \hat{h}_{\infty}^3) = 0$; (ii) $A = 1$; (iii) $A = 5$; and (iv) $A = 10$. }
  \label{travelling_wave}
\end{figure}

The dimensionless profile of a fluidised current moving down a slope is determined from \eqref{travel_ode} and is implicitly given by
\begin{equation}
\int_0^{\hat{h}/\hat{h}_\infty}\frac{s^2(A+s)}{s^3-1}\;\rd s=\frac{\eta}{\hat{h}_\infty},
\end{equation}
 where $A = {\cal{R}} / (4 \hat{h}_{\infty}^3)$.  We plot in Figure~\ref{travelling_wave} the height  of the travelling wave of material as a function of distance from the front for various values of the inertial parameter, $\cal R$, and note that the length scale over which the flow adjusts to the uniform depth, ${\hat h}_\infty$, increases with increasing ${\cal R}$. One measure of the streamwise length, $\Delta_\epsilon$, over which the flow attains its uniform depth is given by evaluating when $\hat{h}(-\Delta_\epsilon)=\hat{h}_\infty(1-\epsilon)$, which when $\epsilon\ll 1$ is given by
\begin{equation}
-\frac{\Delta_\epsilon}{h_\infty}=
1+\frac13 (A+1)\log\epsilon +\frac16 (2A-1)\log 3-\frac{\sqrt{3}\,\pi}{18}+\ldots
\end{equation}
At a fixed location, it is then possible to evaluate the dimensionless timescale over which the uniform depth is established, $\hat{t}_\epsilon={\Delta_\epsilon}/{c}$. 

\begin{figure}
  \centering
  \includegraphics[width=\textwidth]{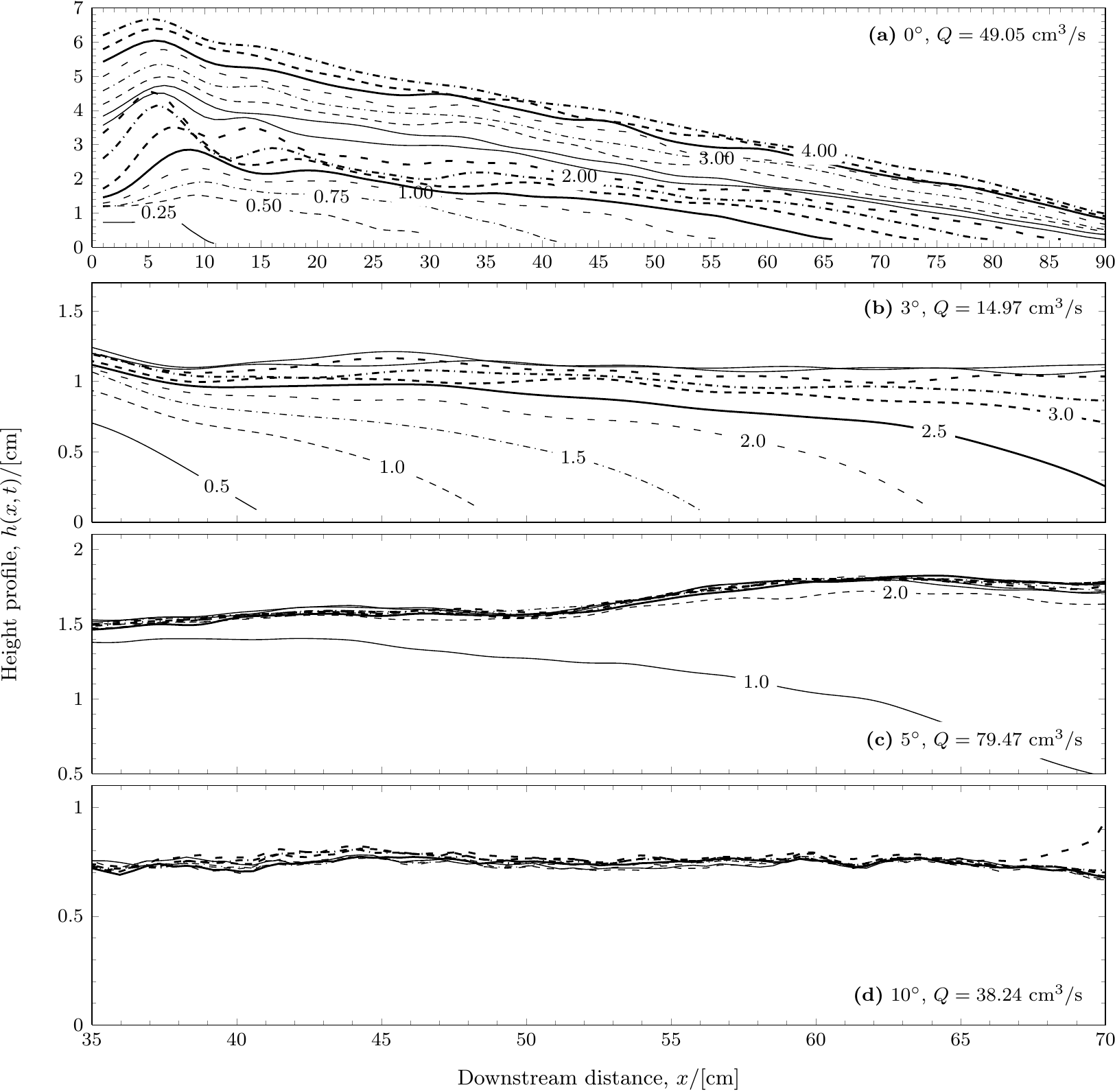}
  \caption{Shape of the currents over time for fluidised flows at different slope angles. The numbers on the contours indicate time in seconds after the release of material.  A spatial Gaussian filter with a kernel size of \unit[0.5]{cm} has been applied to the contours in order to improve clarity.   Note that the vertical scale is different in each plot and the horizontal scale is different in (a) from the other diagrams.  (a) $\theta=0\degree$, $Q=49.1$~\unit{cm$^3$/s}.  Time interval 0.25 s between contours.  Note the growth and decay of the surface waves near the origin.  (b)  $\theta=3\degree$, $Q=15.0$~\unit{cm$^3$/s} ($A=0.09$).  The time interval between contours is \unit[0.5]{s}.  The flow becomes uniform after about \unit[4]{s} in this case.  (c) $\theta=5\degree$, $Q=79.5$~\unit{cm$^3$/s}  ($A=1.57$). The time interval between contours is \unit[1]{s} with a uniform state achieved after about \unit[2]{s}.  (d)  $\theta=10\degree$, $Q=38.2$~\unit{cm$^3$/s}  ($A=0.75$). The time interval between contours is \unit[0.5]{s}  and the current reaches uniform state within \unit[1]{s}.}
  \label{currentshapes}
\end{figure}

Figure~\ref{currentshapes} shows examples of the development of the fluidised currents at different angles of channel inclination.  In figure~\ref{currentshapes}, the flows on the steeper slopes relatively rapidly attain a uniform state in which the current does not vary along the apparatus, whereas those on shallower slopes and with smaller sources fluxes take much longer to approach this state.  Flows along horizontal channels never approach a uniform state; instead currents adopt the shape of a wedge and do not progress at constant speed. These are are analysed in \S\ref{horizontal}.

\begin{figure}
  \centering
  \begin{minipage}[t]{\textwidth}
    \centering
    \includegraphics[width=.8\linewidth]{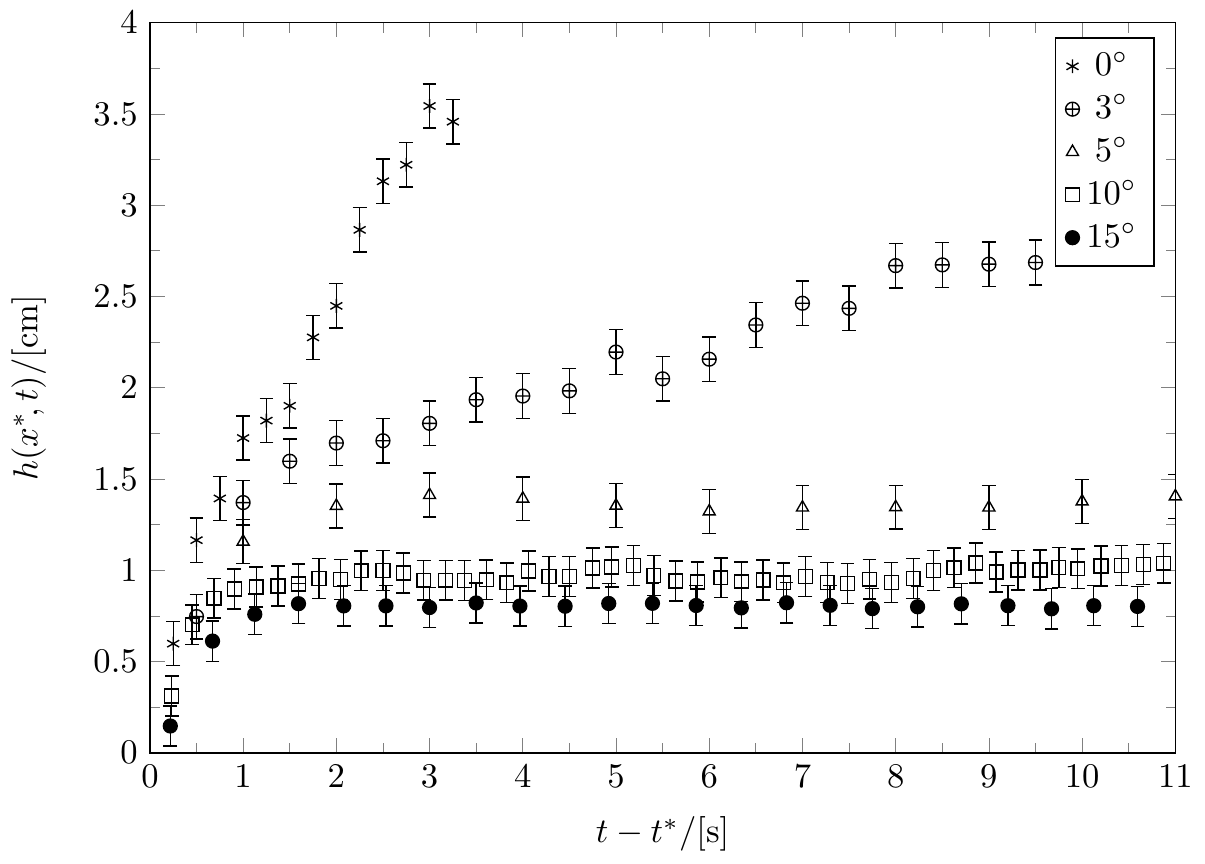}
    \label{fig:currentgrowth}
  \end{minipage}
  \\
  \begin{minipage}[t]{\textwidth}
    \centering
    \includegraphics[width=.8\linewidth]{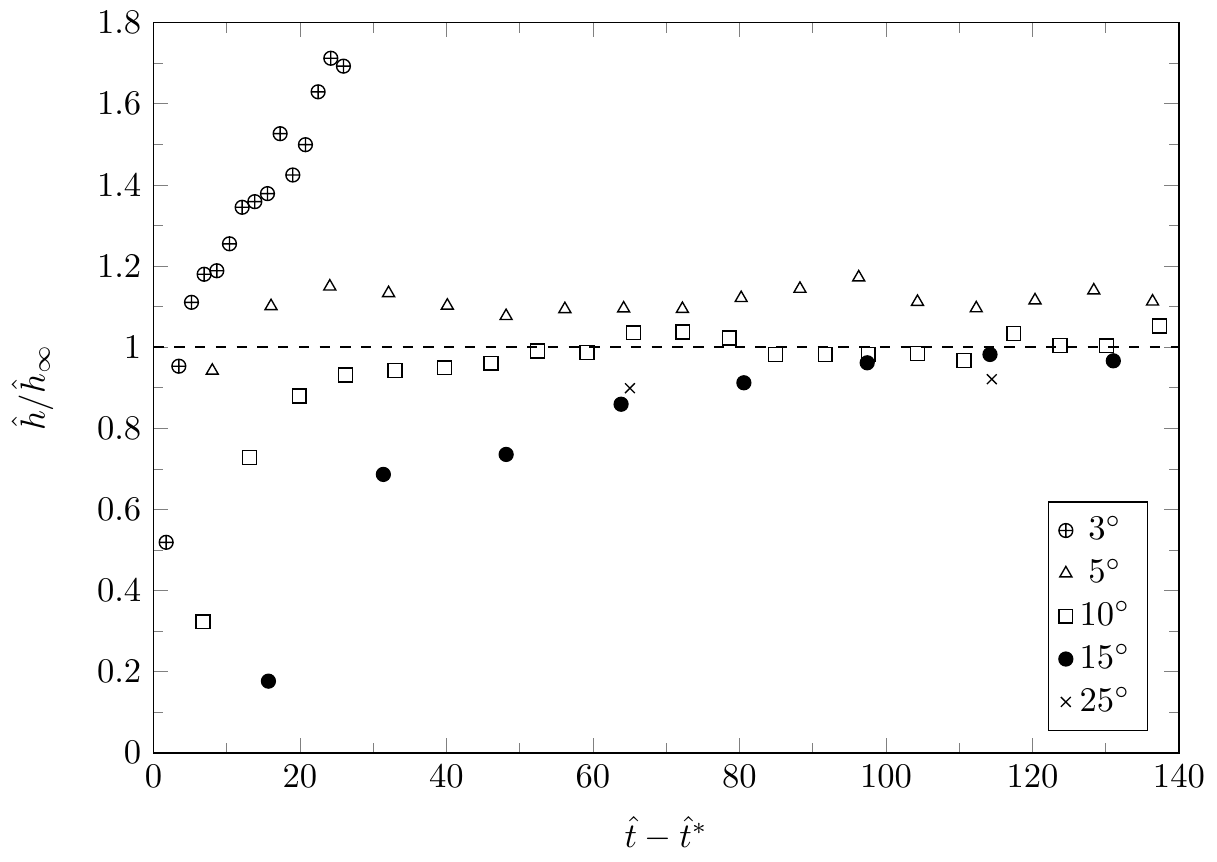}
  \end{minipage}
  \caption{The transition from unsteady to uniform behaviour of fluidised currents down slopes.  In both panels, $t^\ast$ is the time at which the current reaches the measuring position $x^\ast =$ \unit[50]{cm} from the source. The top figure shows the measured depth of the flowing current at a fixed position as a function of time at a fixed point for currents on slopes between 0--15\degree~with the same nominal source flux ($60$~\cmt/s).  Experimental conditions  are: $\theta = 3\degree$, $Q = 59.44$~\unit{\cmt/s}; $\theta = 5\degree$, $Q = 58.75$~\unit{\cmt/s}; $\theta = 10\degree$, $Q = 58.76$~\unit{\cmt/s} and $\theta = 15\degree$, $Q = 56.38$~\unit{\cmt/s}. The bottom figure shows the scaled height of the current, $\hat{h}/\hat{h}_\infty$, as a function of scaled time after the front reaches $x^\ast$ using the scales of (\ref{slopescale}) and using the definition of $H$ in (\ref{Hdef}). $\epsilon=0.05$ and when $\theta=3\degree$, $A=0.28$, $\hat{t}_{\epsilon}=1.38$; $\theta=5\degree$, $A=0.45$, $\hat{t}_{\epsilon}=1.60$; $\theta=10\degree$, $A=1.07$, $\hat{t}_{\epsilon}=2.41$; $\theta=15\degree$, $A=2.56$, $\hat{t}_{\epsilon}=4.38$.  }
  \label{heightTime}
\end{figure}

Figure~\ref{heightTime} shows the change of height with time for five flows with the same nominal flow rate halfway along the apparatus before and after scaling.  It can be seen that the scaled times at which the currents achieve a constant height (and systematically with $A$), as would be expected, but they are an order of magnitude larger than $\hat{t}_{\epsilon}$.  For the expected values of $\hat{t}_{\epsilon}$, the currents would have to achieve their constant height very quickly, almost instantly, and for the values of $A$ corresponding to the experimental flows, from figure~\ref{travelling_wave} the front of the currents would be expected to be `blunt-nosed', with quite steep gradients of height at the front of the current. In fact, the front of the currents (figure~\ref{currentshapes}) had a relatively shallow gradient.  



\section{Horizontal flows}
\label{horizontal}
\begin{figure}
  \centering
  	\includegraphics[width=\linewidth]{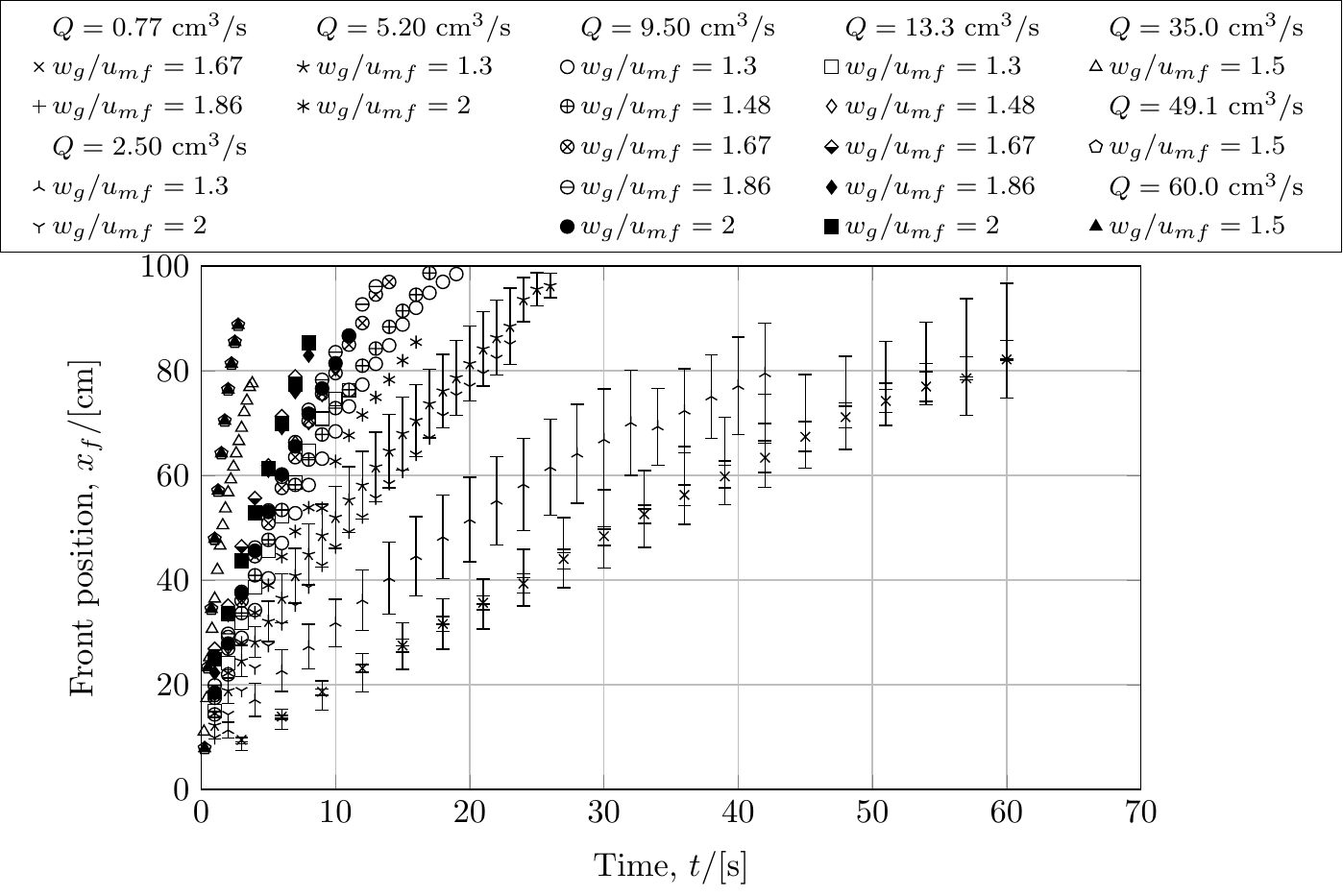}
	\caption{The position of the front of the fluidised current as a function of time for varying source fluxes and fluidising gas flows.}
	\label{horzraw}
  \end{figure}
  \begin{figure}
	\centering
	\includegraphics[width=0.6\linewidth]{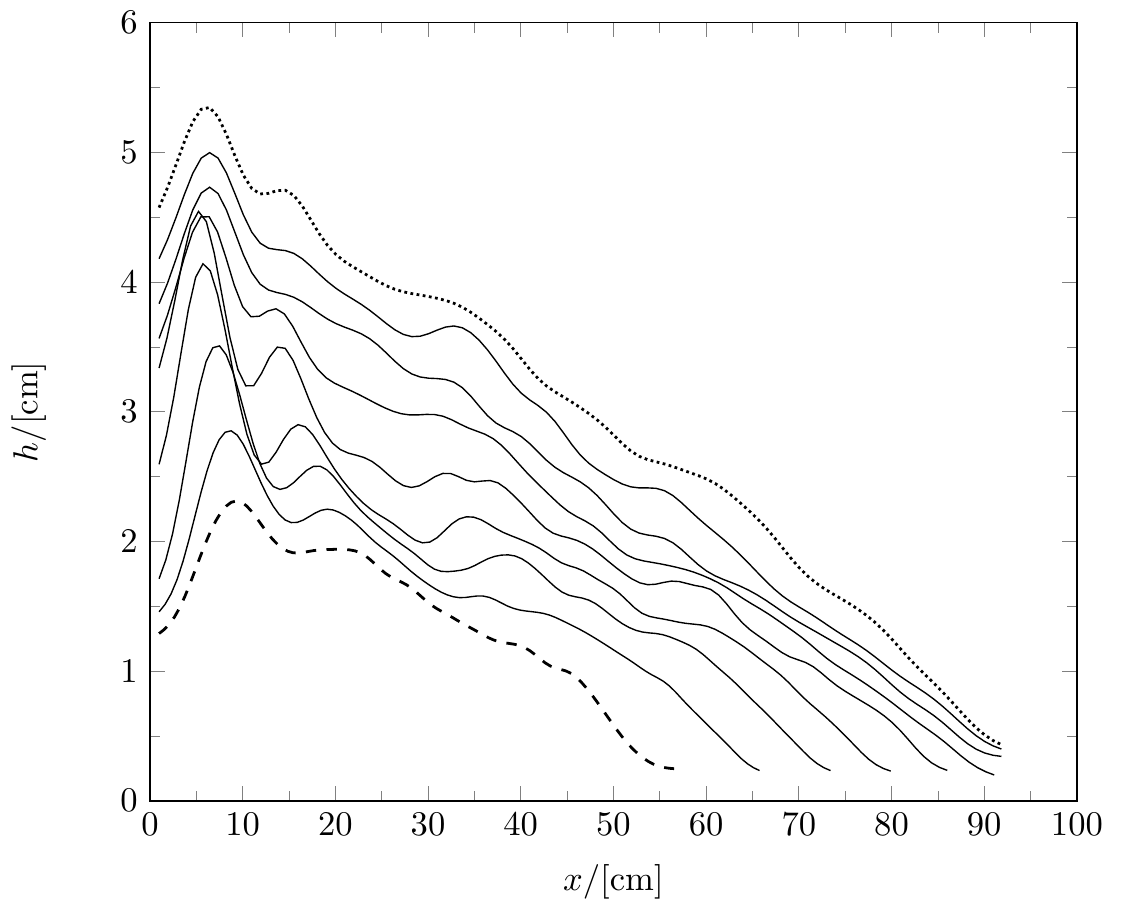}
	\caption{Change in shape of horizontal fluidised currents with no scaling for $Q=49.05$\unit{\cmt/s} and $w_g/u_{mf}=1.5$.  The different profiles are drawn at 0.25s intervals.  The dashed line corresponds to $t=1$~s and the dotted line to $t=3$~s.  }
	\label{horzshape}
\end{figure}

Figure~\ref{horzraw} shows the distance travelled by the front of  a fluidised flow over horizontal surface as a function of time.  It is evident that the currents do not travel at a constant speed.  The shapes of the currents are shown in figure~\ref{horzshape} and, ignoring the disturbance at the start of the currents at the point that they are poured into the system, they have an approximately triangular shape, though one with a low aspect ratio (i.e.\ their extent far exceeds their depth).  Furthermore they flow through `bulk' motion, not through the build-up of lamina arising from the constant avalanching down the current's top surface seen for non-fluidised granular flows.  They must also be scaled differently because the length and time scales introduced in \eqref{slopescale} become singular when $\theta=0$.  To this end, we introduce the characteristic height scale, $\tilde{H}$ and define the following dimensionless variables
\begin{equation}
  \label{horzscale}
  \tilde{h}=\frac{h}{\tilde{H}},\qquad
  \tilde{v}=\frac{v}{F(g\tilde{H}^3)^{1/2}/d},\qquad
  \tilde{x}=\frac{x}{\tilde{H}},\qquad\hbox{and}\qquad
  \tilde{t}=\frac{Ft(g\tilde{H})^{1/2}}{d}.
\end{equation}

\noindent The depth-integrated expression of momentum balance is then given by
\begin{equation}
  \tilde{{\cal R}}\left(\pd{}{\tilde{t}}\left(\tilde{h}\tilde{v}\right)
  +\frac54\pd{}{\tilde{x}}\left(\tilde{h}\tilde{v}^2\right)\right)
  +\frac{\tilde{h}}{2}\pd{\tilde{h}}{\tilde{x}}=-
  \frac{25}{4}\frac{\tilde{v}^2}{\tilde{h}^2},\label{unsteady_hori_mom}
\end{equation}
instead of (\ref{solidslopemom_int}), where the residual dimensionless parameter $\tilde{\cal R}=(FH/d)^2$ is the relative magnitude of inertial to resistive forces.  

Conservation of mass is given by
\begin{equation}
  \pd{\tilde{h}}{\tilde{t}}+\pd{}{\tilde{t}}\left(\tilde{v}\tilde{h}\right)=0.\label{unsteady_mass}
\end{equation}
The appropriate dimensional depth-scale, $\tilde{H}$, is determined from the source flux, 
\begin{equation}
\tilde H= \left(\frac{q_0dg^2}{F\phibar}\right)^{1/5},
\end{equation}
so that the boundary condition is given by 
\begin{equation}
  \tilde{v}\tilde{h}=1\qquad\hbox{at}\qquad x=0.
  \label{eq:H_def}
\end{equation}

Flows over horizontal surfaces decelerate as the basal drag is no longer balanced by a sustained downslope acceleration.  Thus, at sufficiently early times the flow speeds and depths are set by source conditions, and after the flow has propagated for sufficient time, the resistive forces become non-negligible and the motion enters a dynamical regime in which the drag force balance the streamwise gradients of the hydrostatic pressure and the inertial forces are negligible.  Analogously to \cite{hogg01a}, simple scaling shows that this regime is fully attained when $\tilde{t}\gg\tilde{\cal R}$.  In this scenario, we deduce from \eqref{unsteady_hori_mom} that 
\begin{equation}
  \tilde{v}=\frac25\left(-\tilde{h}^3\pd{\tilde{h}}{\tilde{x}}\right)^{1/2}
\end{equation}
and consequentially from \eqref{unsteady_mass}
\begin{equation}
\pd{{\tilde h}}{{\tilde t}} + \pd{}{\tilde{x}}\left(\frac25\tilde{h}^{5/2}\left( -\pdiff{\tilde{h}}{\tilde{x}} \right)^{1/2}\right) = 0,
\label{hori_gov}
\end{equation}
subject to the source condition
\begin{equation}
\frac25\tilde{h}^{5/2}\left(-\pdiff{\tilde{h}}{\tilde{x}} \right)^{1/2}=1.
\label{hori_bc}
\end{equation}
\eqref{hori_gov} may be integrated numerically  to reveal the evolution of the front position as a function of time and the variation of the depth of the current along its length; however,  for these currents flowing over a horizontal surface, we may also construct a quasi-analytical similarity solution for the motion.

\begin{figure}
  \centering
  \includegraphics[trim=5cm 11cm 6cm 11cm, clip, width=0.60\columnwidth]{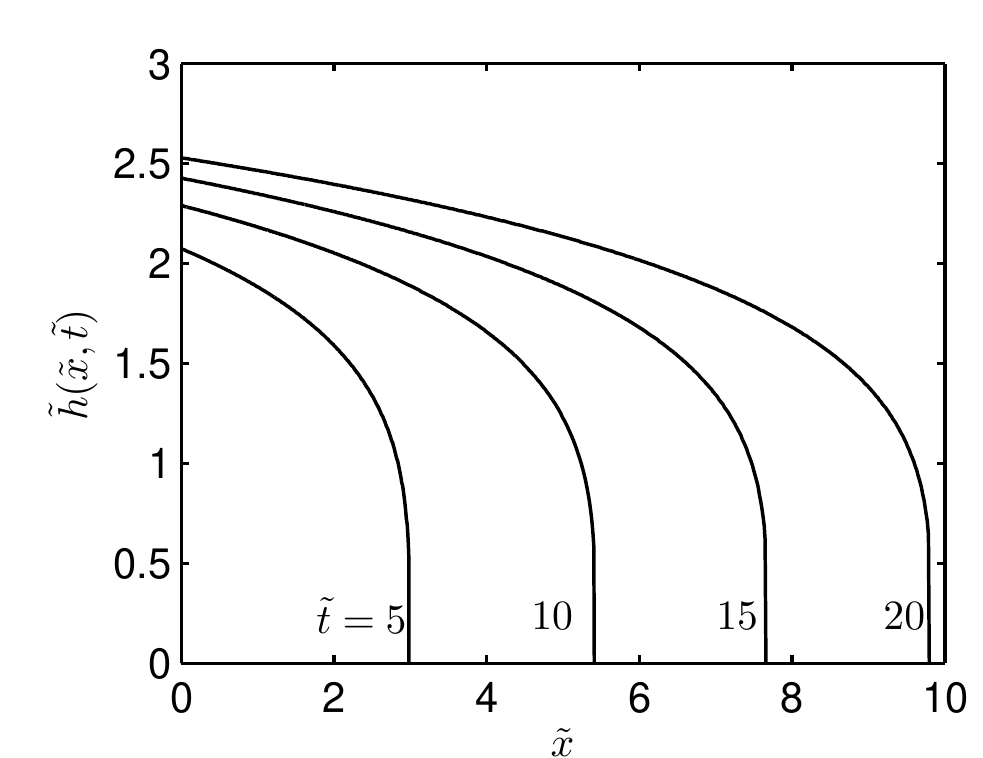}
  \caption{The scaled height of the current as a function of downslope distance at various instances of time from the similarity solution for unsteady propagation along a horizontal channel.}
  \label{fig_unsteady_hori}
\end{figure}

First, we determine the gearing between spatial and temporal scales that underpins the similarity solution for unsteady flow over a horizontal surface.  To do this we scale and balance terms in the governing equation \eqref{hori_gov} and boundary condition \eqref{hori_bc}.  This yields
\begin{equation}
\frac{\tilde h}{\tilde t}\sim\frac{{\tilde h}^3}{{\tilde x}^{3/2}}\qquad\hbox{and}\qquad \frac{{\tilde h}^3}{{\tilde x}^{1/2}}\sim 1.
\end{equation}
Thus we deduce that ${\tilde x}\sim  {\tilde t}^{6/7}$ and ${\tilde h}\sim \tilde{t}^{1/7}$.
We may then seek a similarity solution of the form
\begin{gather}
\label{horzsim}
{\tilde h}=K^{3/4}{\tilde t}^{1/7}{\cal H}(y),\\
{\tilde x}_f({\tilde t})=K{\tilde t}^{6/7},
\end{gather}
where $K$ is a dimensionless constant to be determined as part of the solution and $y={\tilde x}/{\tilde x}_f({\tilde t})$.  On substitution in the governing equation \eqref{hori_gov},  this gives
\begin{equation}
\frac 17 {\cal H}-\frac67 y{\cal H}'+\frac 25\left\lbrack \left(-{\cal H}'\right)^{1/2}{\cal H}^{5/2}\right\rbrack'=0,
\label{hori_sim}
\end{equation}
where a prime denotes differentiation with respect to $y$.  This ordinary differential equation \eqref{hori_sim} is to be integrated subject to the boundary conditions
\begin{equation}
{\cal H}(1)=0\qquad\hbox{and}\qquad
\frac 25 K^{2/7}\left(-{\cal H}'\right)^{1/2}{\cal H}^{5/2}=1\quad\hbox{at}\quad y=0.
\end{equation}
The location $y=1$ is a singular point of the differential equation \eqref{hori_sim}; we therefore start the numerical integration at $y=1-\epsilon$ $(\epsilon\ll 1)$, noting that
\begin{equation}
{\cal H}(1-\epsilon)=\left(\frac{30}{7}\right)^{1/2}\epsilon^{1/4}\left(1-\frac{1}{60}\epsilon +\frac{43}{28880}\epsilon^2+\ldots\right).
\end{equation}
It is then straightforward to integrate the differential equation \eqref{hori_sim} numerically and evaluate ${\cal H}(0)=2.038$, ${\cal H}'(0)=-0.479$, and so $K=0.753$.  The dimensional expression for distance covered by a horizontal, fluidised current with time is then
\begin{equation}
\label{eq:horzsima}
x_f=0.753\left(\frac{gF^2q_0^4}{d^2\phi^4}\right)^{1/7}t^{6/7}.
\end{equation}
We plot in Figure~\ref{fig_unsteady_hori} the similarity solution for the height profile along the current noting that, again, the model predicts a blunt-nosed current that advances along the channel.

\begin{figure}
  \centering
	\includegraphics[width=.6\linewidth]{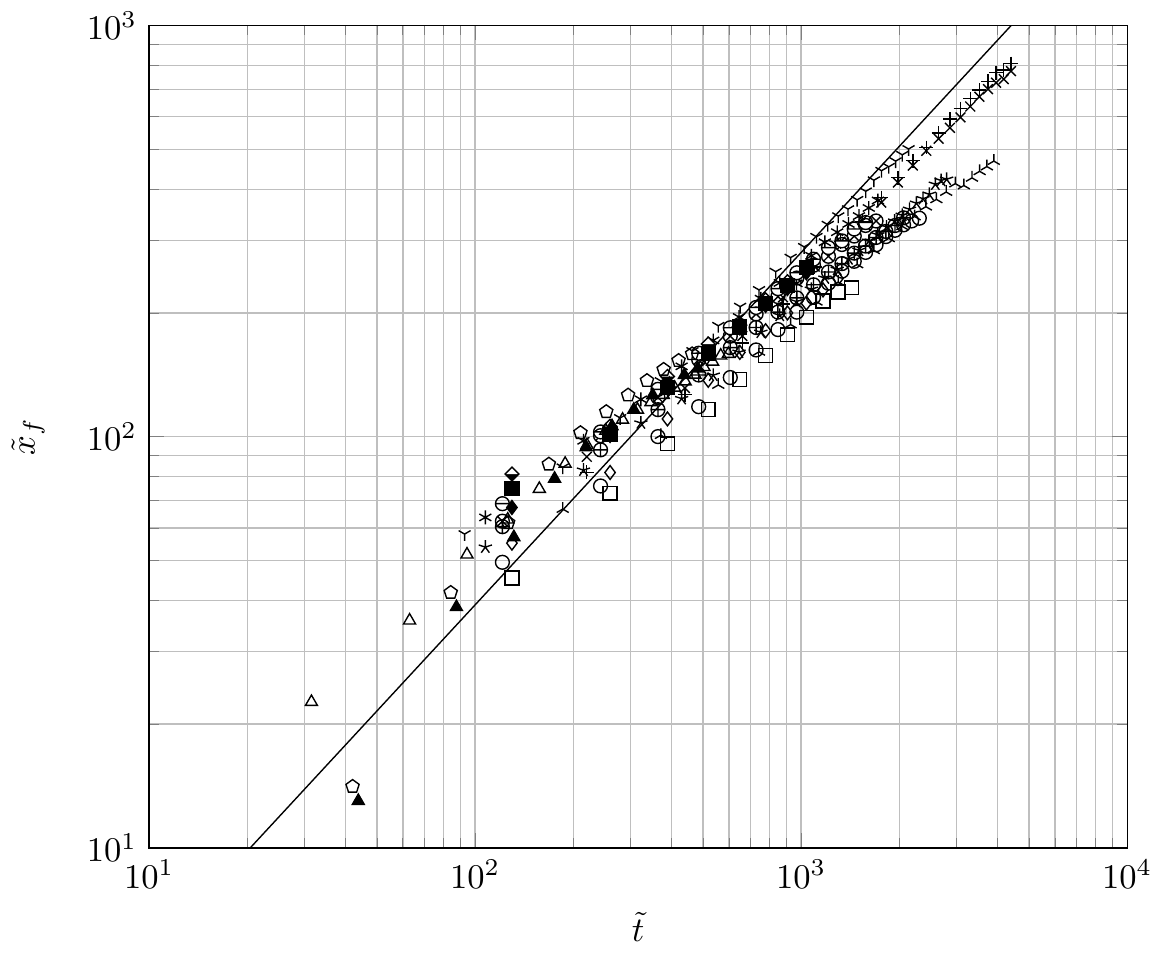}
	\caption{The scaled position of the front of the current, $\tilde{x}_f$ as a function of scaled time. The data is drawn from figure \ref{horzraw} with distance and time scales according to (\ref{slopescale}) and theoretical line given by (\ref{horzsim}) (solid line).  Key is as per figure \ref{horzraw}.  } %
	\label{horzscaledist}
  \end{figure}

\begin{figure}
	\centering
    \includegraphics[width=.6\linewidth]{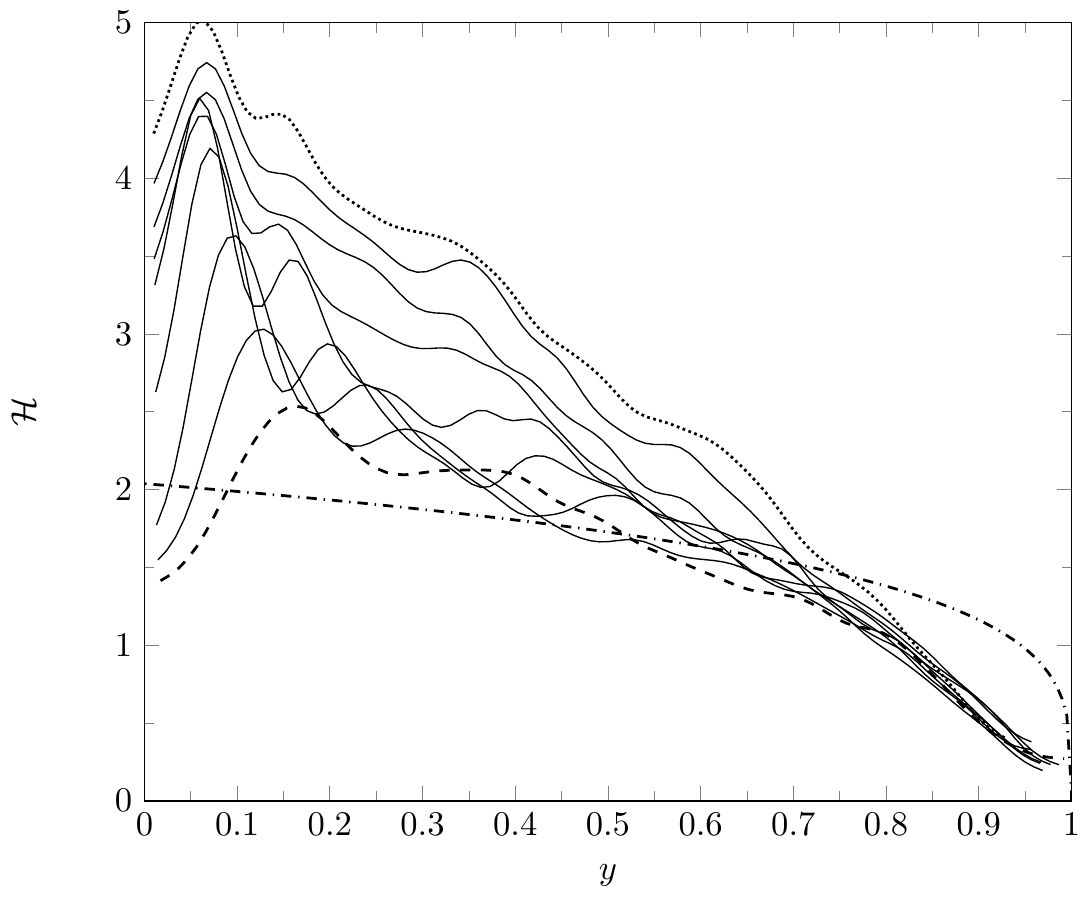}
	\caption{The scaled height of the current, ${\cal H}$, as a function of the scaled distance $y=\tilde{x}/\tilde{x}_f$ at various instants of time.  The data is drawn from figure \ref{horzshape} scaled using (\ref{horzsim}).  The model solution (\ref{horzsim}) is represented by the chain-dotted line.}
	\label{horzscaleshape}
  \end{figure}

The scaled distance against time is shown in figure~\ref{horzscaledist}, and the data is collapsed sufficiently for the power-law form of the curve to appear to be reasonable, though the value of the exponent is different from that predicted.  However, the shape of the current predicted by the scaled model is very different from the experimental measurements, taking the form of nearly flat current with a snub nose (figure~\ref{horzscaleshape}).

\subsection{Flow within a narrow channel}
\label{narrow}

One of the differences between the experimentally-realised flows over horizontal surfaces and those down slopes is that the former are significantly thicker than the latter, and so it is possible for the side walls to have a strong influence on their development.  Here, we analyse the motion of a fluidised current as it flows within a narrow channel of width $B$, between sidewalls for which the streamwise extent of the flow far exceeds the depth of the current $(L \gg H)$, which in turn far exceeds the width of the flow $(H \gg B)$.  For this regime, it is possible to simplify the governing equations for the unsteady motion down an incline on the basis that gradients across the flow are much greater than those in any other direction.  In this scenario, the dynamical balance is somewhat different from that analysed in \S\ref{wide} and the resulting governing equations for the unsteady evolution of the thickness of the flow are also different (\S\ref{unsteady}). 

Our derivation of the governing equation in a narrow channel is developed from the dimensional expressions presented in \S\ref{sec:model} and then depth-integrated to establish a shallow layer model; however, it will be shown that it leads to a similarity solution with a different gearing between the spatial and temporal variables.  Here we only present the governing equations for flows along horizontal channels, but the inclusion of a channel gradient is a straightforward generalisation and could lead to travelling wave solutions analogous to \S\ref{unsteady}.  

It is assumed that the solid particles are fully fluidised by an imposed gas flow and attain a state in which the volume fraction is uniform, $\phi=\phibar$.  Since the flow is relatively thin, vertical accelerations are negligible and the motion is governed by hydrostatic balance given by
\begin{equation}
  \pd{}{z}\left(p-\sigma_{zz}\right)=-\rho_s\phibar g.
  \label{narrow_hydro}
\end{equation}
 In this expression we have neglected the contribution due to the weight of the gas phase since $\rho_g/\rho_s\ll 1$. In the downslope direction, after neglecting terms proportional to the density and viscosity of the gas, the combined momentum equation of both phases to leading order is given by
\begin{equation}
  \rho_s\phibar\frac{Dv}{Dt}=-\pd{}{x}\left(p-\sigma_{zz}\right)+\pd{\sigma_{xy}}{y},
  \label{narrow_mom1}
\end{equation}
where $y$ is the distance across the channel. 
Here it has been assumed that the normal stresses of the solid phase are isotropic $(\sigma_{xx}=\sigma_{zz})$ and that gradients across the flow dominate all others.  To complete this model, we introduce the granular temperature, which provided the channel is much wider than the grain size, is in local equilibrium between its production and dissipation.  Then we may write
\begin{equation}
  f_1d^2\left(\pd{v}{y}\right)^2=f_3 T,
  \label{narrow_temp}
\end{equation}
and the constitutive law for the shear stress is given by
\begin{equation}
  \sigma_{xy}=f_1 d\rho_sT^{1/2}\pd{v}{y}.
  \label{narrow_constit}
\end{equation}
Finally, by eliminating the fluid pressure from the normal force balances of each phase (see \eqref{fluidised}), we find that
\begin{equation}
  \pd{\sigma_{zz}}{z}=\phibar\rho_s g-\frac{\beta w_g}{(1-\phibar)^2}.
\end{equation}
The volume fraction, temperature, and therefore the normal stress component, $\sigma_{zz}$, are independent of $z$ to leading order, and thus we deduce that
\begin{equation}
  0=\phibar\rho_s g-\frac{\beta w_g}{(1-\phibar)^2}.
\end{equation}
This expression determines the average volume fraction as a function of the fluidising gas flux and marks a important departure from the shallow layer model of \S\ref{unsteady}, because to leading order the solid stresses do not contribute to the support of the granular layer.  

We progress by assuming that the velocity field of the solid phase exhibits cross-stream dependence, which is identical to that found in fully developed flows with the shear in the vertical plane.  Thus we write
\begin{equation}
v=\overline{v}\frac53 \left(1-\left|1-\frac{2y}{B}\right|^{3/2}\right),
\end{equation}
and consequentially $\partial v/\partial y=5\overline{v}/B$ at $y=0$.  The dimensional governing equations then express conservation of mass and after sufficient time has passed so that inertia is negligible, a balance between the streamwise pressure gradient and the side wall stresses is given by
\begin{eqnarray}
\frac{\p h}{\p t}+\frac{\p}{\p x}\left(\overline{v} h\right)&=&0,\\
gh\frac{\p h}{\p x}&=&-\frac{50d^2\overline{v}^2 h}{F^2B^3}.
\end{eqnarray}
For these flows over a horizontal surface we adopt the dimensionless variables given by
\begin{equation}
\label{narrowscale}
  {\overline h}=h/{\overline H},\qquad
  {\overline x}= x/{\overline H}\qquad\hbox{and}\qquad
  {\overline t}=(q_0t)/(\phibar\, {\overline H}^2 B),
\end{equation}
where the height scale scale, ${\overline H}$, is determined by
\begin{equation}
  \overline{H}=\frac{q_0d}{F\phibar(gB^5)^{1/2}}.
\end{equation}
The dimensionless governing equation then becomes
\begin{equation}
  \pd{{{\overline h}}}{{\overline t}}+\frac{1}{5\sqrt{2}}\pd{}{{\overline x}}\left({\overline h}\left(-\pd{{\overline h}}{{\overline x}}\right)^{1/2}\right)=0,
  \label{narrow_pde_hori}
\end{equation}
subject to 
\begin{equation}
  \frac{1}{5\sqrt{2}}{\overline h}\left(-\pd{{\overline h}}{{\overline x}}\right)^{1/2}=1\qquad\hbox{at}\qquad x=0.
  \label{narrow_bc}
\end{equation}

\begin{figure}
	\centering
	\includegraphics[trim=5cm 10cm 6cm 10cm,clip,width=0.6\columnwidth]{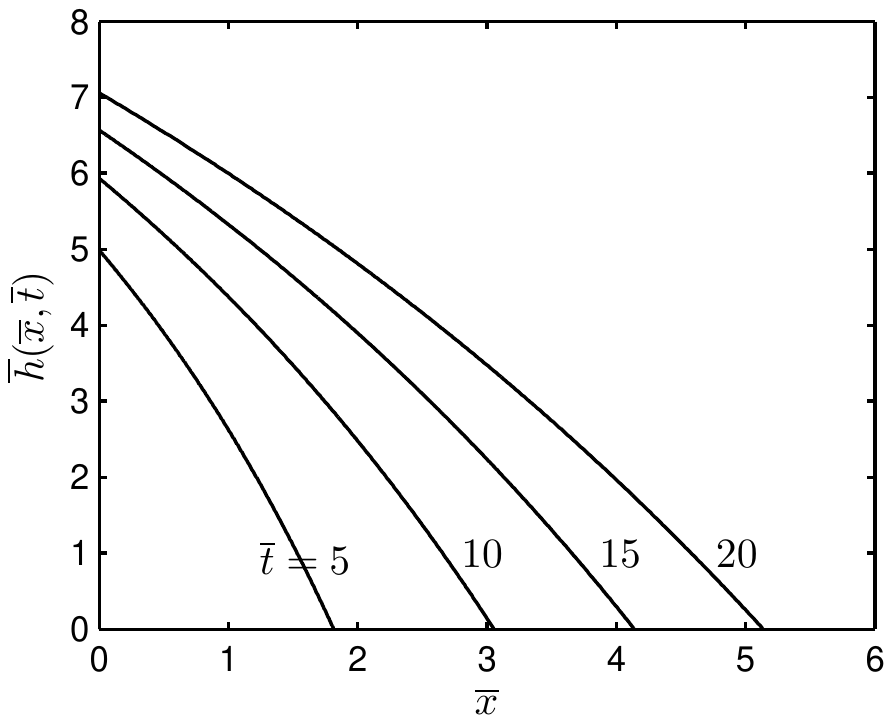}%
	
	\caption{The height of the current as a function of distance along a narrow horizontal channel at various instances of time}
	\label{unsteady_hori_narrow}
\end{figure}

We may construct a similarity solutions to this governing equation \eqref{narrow_pde_hori} and boundary condition \eqref{narrow_bc} by first deducing the gearing between the spatial and temporal scales.  The governing equation demands ${\overline h}/{\overline t}\sim {\overline h}^{3/2}/{\overline x}^{3/2}$, while the boundary condition leads to ${\overline h}^{3/2}\sim {\overline x}^{1/2}$.  Thus we deduce that ${\overline x}\sim {\overline t}^{3/4}$, ${\overline h}\sim {\overline t}^{1/4}$ and that the similarity solutions of the following form may be sought
\begin{equation}
  {\overline h}=C^3{\overline t}^{1/4}{\cal H}(y)\qquad\hbox{and}\qquad
  {\overline x}_f=C {\overline t}^{3/4},
  \label{thinheight}
\end{equation}
where $y={\overline x}/{\overline x}_f({\overline t})$ and $C$ is a constant to be determined.  In dimensional variables, 
	\begin{equation}
	\label{thindistance}
	x_f=C \left(\frac{q_0F^2 g B^2}{\phibar d^2}\right)^{1/4}t^{3/4}.
	\end{equation}
On substitution of $\bar{h}$ into the governing equation \eqref{narrow_pde_hori}, we deduce that
\begin{equation}
  \tsf 14 {\cal H}-\tsf 34 y{\cal H}'+\tsf {1}{5\sqrt{2}}\left\lbrack{\cal H} (-{\cal H}')^{1/2}\right\rbrack'=0,
  \label{narrow_sim}
\end{equation}
subject to $\mathcal{H}(1)=0$ and $C^4\mathcal{H}(0)[-\mathcal{H}'(0)]^{1/2} = 5\sqrt{2}$.  The similarity differential equation \eqref{narrow_sim} is singular at $y=1$ and the numerical solution may be initiated from a series solution valid close to that value, given by 
$$\mathcal{H}(1-s) = \frac{225}{9}\left(s-s^2/5+\ldots\right).$$
when $s\ll 1$.  It is then straightforward to integrate \eqref{narrow_sim} to compute the height profile, shown in figure~\ref{unsteady_hori_narrow}, and the dimensionless constant $C=0.5434$.

\begin{figure}
	\centering
	\includegraphics[width=0.6\linewidth]{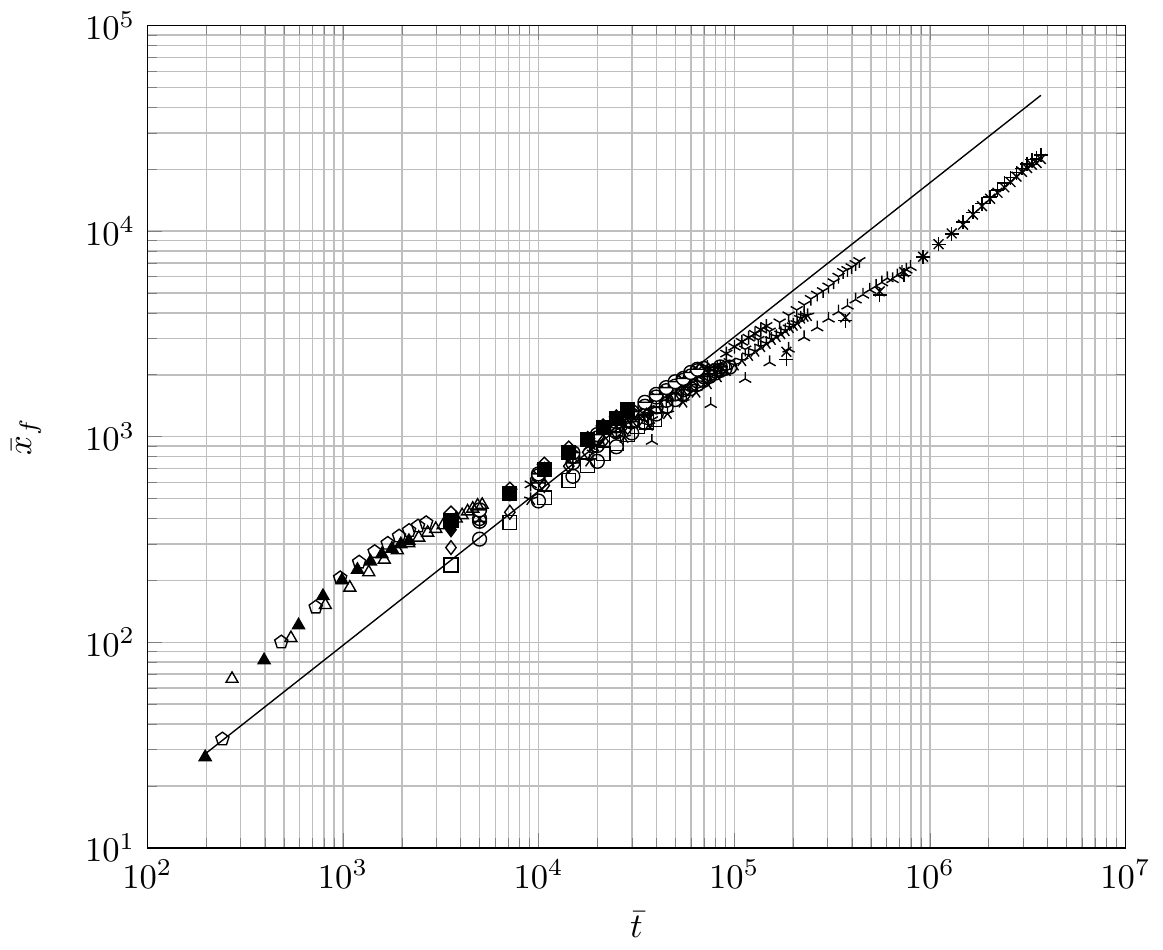}
	\caption{The scaled position of the front of the fluidised current in a horizontal channel as a function time.  The data is that plotted in figure~\ref{horzraw} and scaled using \eqref{narrowscale} with the model prediction from (\ref{thindistance}) (solid line).  The key is given in  figure~\ref{horzraw}.  }
	\label{horzthindist}
\end{figure}

\begin{figure}
    \centering
    \includegraphics[width=.6\linewidth]{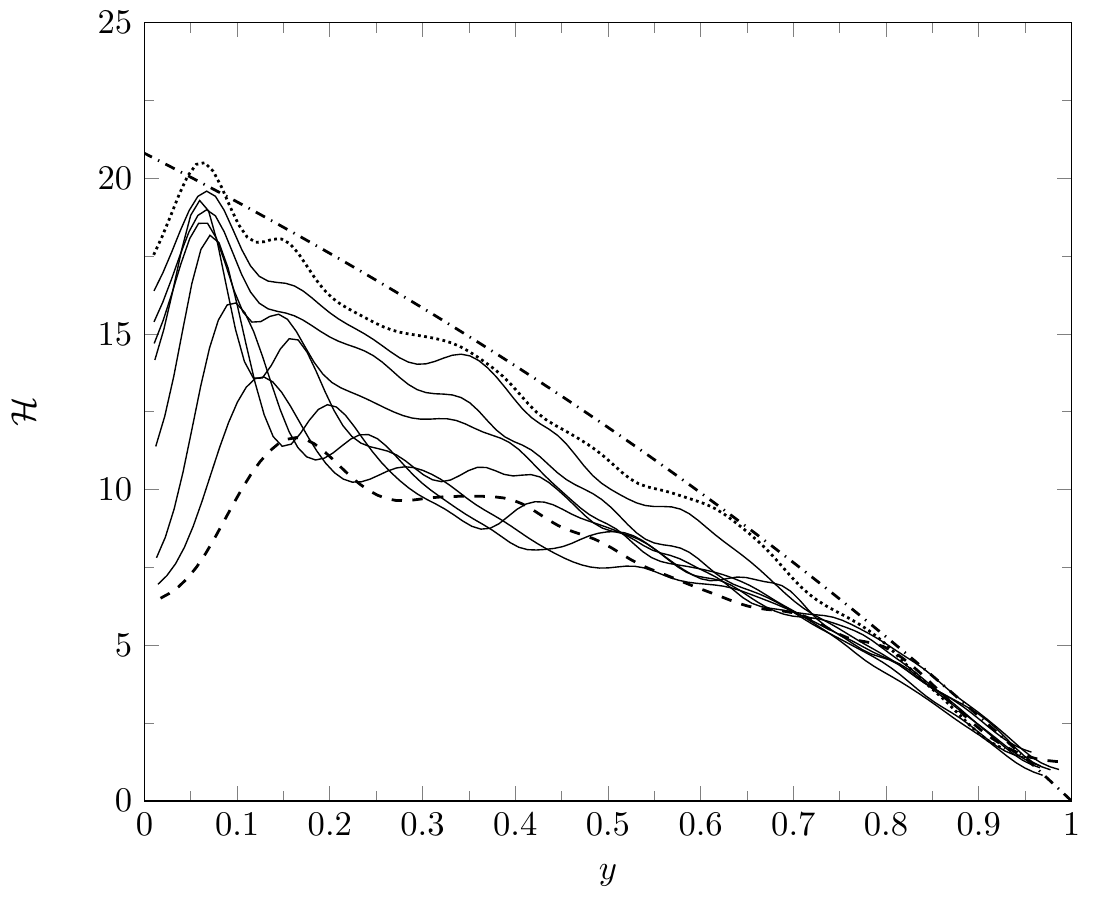}
    \caption{The scaled depth of the fluidised current along a horizontal channel, ${\cal H}$, as a function of the scaled position, $y=\overline{x}/\overline{x}_f$. The data is drawn from figure \ref{horzshape} scaled using (\ref{slopescale}).  The model solution is from (\ref{thinheight}) and is represented by the chain-dotted line. }
    \label{horzthinshape}
  \end{figure}

The scaled distance against time for the experiments is shown in figure~\ref{horzthindist}. There is reasonably good collapse of the data (better than in figure \ref{horzscaledist}).  Furthermore, it appears to follow a power law, though the exponent of the power law is slightly small than that predicted.  The measured, scaled shape of the currents is plotted in figure~\ref{horzthinshape}.  The predicted shape is now much closer to that of the experiments, although  the scaling does not collapse completely all the measured data.

\section{Discussion and conclusions}
\label{discussion}

This investigation of fluidised granular currents reveals important distinctions in their dynamical properties from both dry granular flows and static fluidised beds. Most significantly, there is substantial and sustained shear in the velocity profiles. Consequentially particles are driven into each other  and this provides a mechanism for the generation of stresses.  In the regime we investigated, the inertia of individual grains remains relatively high and thus the particles interact with each other through dissipative collisions, and it these interactions that lead to the shear stress that balances the downslope gravitational acceleration.  Granular flows in the absence of fluidisation must generate sufficient normal stresses to support the weight of the flowing layer and thus typically their shear stresses are also relatively large; however, fluidisation changes the balance of forces acting on the current.  The fluidising gas flow provides most of the normal support to the flowing layer and thus both normal and shear stresses from the particulate phase are reduced relative to their non-fluidised counterparts, leading to flows that are much more mobile.

In this study we have formed a framework of modelling fluidised currents based on the solid-phase stresses that is generated from collisions between the particles.  The degree of agitation in the system is measured through the granular temperature and constitutive laws are employed to determine the stress tensor in terms of the gradients of the velocity field, the granular temperature, and the volume fraction of solids, as well as several material parameters.  For the regime studied here in which the flows are many particles thick, a local balance emerges between the generation and dissipation of granular temperature.  This leads to an accurate asymptotic model for the complete dynamics, in which the flowing material is essentially modelled by a non-linear, local rheology.  Furthermore, this reduction leads to a Bagnold-like expression between the flow depth and the flux of particles carried by the current, with the volume fraction, determined by the fluidising gas flow, contributing to this relationship.  Although this approximation is a simplified description of the more complete dynamics, it embodies the key processes of these flows: the fluidised currents are granular flows in which the fluidisation affects the normal support of the layer.

The experimental measurements provide encouraging support for the model.  For example, without tuning through empirical factors, the predictions are quite close to the measured flow depths and flow speed in both the uniform steady state and the transient state as it becomes established.  Additionally, when applied to flows along horizontal channels, the model is able to predict the unsteady motion to reveal both the progressive deceleration and the growth in flow depth.  There are, however, some systematic features in the measurements that are not reproduced in the model.  Perhaps the most significant of these is the decrease in particle velocity towards the top of the layer.  This feature is absent from the model and presumably corresponds to particles in the `free-board' of the fluidised layer (i.e. the region above the dense current within which the volume fraction of the particles is reduced). Such a dilute layer is subject to slightly different dynamical interactions: the role of particle collisions becomes much reduced and the particles may saltate, and are potentially intermittently suspended above the denser layer below.  The model predictions are also dependent upon the material properties that characterise the collisions between the particles and the boundaries.  These can be difficult to measure directly, but the specularity coefficient, $\psi$, and the boundary coefficient of restitution, $e_w$, only play a significant role for a relatively thin boundary layer in the dynamical regime considered in this study.  It is arguable that the boundary conditions require further research to refine and sharpen their formulation.

One important feature that emerges from the modelling framework is the determination of the volume fraction of particles in the fluidised current.  Here we have assumed that the flows are relatively dense and that the Ergun equation provides an appropriate representation of the volume fraction dependence of the drag due to the fluidising gas flow.  Other expressions could easily be used in its place \citep[see][]{nott_frictional-collisional_1992, agrawal_role_2001,Oger201322}.  However, perhaps of greater significance is whether there are `bubbles', or inhomogeneities in the volume fraction within the fluidised current.  Patches of increased voidage locally provide paths through which the fluidising gas can more readily flow and thus it is possible for the layer to exhibit fluctuations or instabilities on relatively rapid timescales.  Since the local volume fraction affects the mobility of the flowing layer, one might expect fluctuations in volume fraction and velocity to be correlated and consequentially to influence the bulk dynamics.  Bubbles may also affect the particle volume fraction in the bed.  The classical model of fluidised beds \citep{Toomey1952} proposes that all the gas in excess of that necessary to fluidise the particles forms bubbles so that the particle volume fraction in the bulk of the flow is insensitive to $w_g$:  this is contrast with (\ref{phiapprox}).  The lack of dependence on $\theta$ of the experimental scaled velocity profiles in figure~\ref{velscale} also suggests that $\bar{\phi}$ may change less with conditions than might be expected.  \added{In contrast to static fluidised beds, the stability of fully developed fluidised flow down inclines has not been assessed} \citep{jackson_dynamics_2000} \added{and this appears to be an interesting topic for future research.  Indeed it is intriguing that linear shear flows of unfluidised granular materials appear to exhibit transient linearised growth, but asymptotic stability} \citep{sava92,schm94}\added{.  It would be interesting to investigate whether these properties carry over to sheared fluidised motions.}

Our modelling framework and experimental methods could be extended to a number of related flow problems.  First, one could investigate fluidised currents that are generated by instantaneous or non-sustained releases,  that are not fully fluidised and for which the fluidising gas flow is localised to the region close to the source.  These flows would be largely unsteady and, in situations where the fluidisation is not maintained, would introduce additional mechanisms for generating resistive shear stresses as the contact friction begins to become important.  Non-monodisperse granular materials would also be interesting to investigate because the onset of full fluidisation is dependent upon grain-size and the proportion of each particle component \citep{Formisani1991}. It is possible that mixtures of particles segregate according to size and  generate an inhomogeneous flowing current in terms of composition and therefore, the average volume fraction ($\bar{\phi}$).  Finally, we comment that liquid-fluidised systems may pose additional challenges since it is likely that viscous forces at the particle scale are non-negligible and that collisions are strongly affected by lubrication pressure in the the fluid between particles.

Although direct applications have not been the focus of our study, our model formulation could be readily applied to larger-scale flows, either in industrial contexts or in nature.  There are a number of practical implications of our results.  For example, the transport of granular materials when they are fluidised is likely to be more efficient on even shallowly inclined surfaces than on horizontal surfaces.  In addition, the transport is unlikely to be greatly improved by an increase in the gas flow rate, $w_g$, once the granular materials are fully fluidised.   This is because it only directly affects the average volume fraction, $\bar{\phi}$, and for practical materials $\bar{\phi}$ can strongly depend on their characteristics (e.g.\ the bubble-free expansion seen in small, light \citet{geldart_types_1973} group~A particles), as can the value of the coefficient of restitution $e$.   

The assumption at the heart of the modelling framework is that inelastic particulate collisions generate stresses that provide the resistance to motion and this is likely to be the case for larger-scale flows.  Of particular note is that for steady flow, no assumption is made for the relative importance of inertial and resisting forces and hence the resulting model is valid for flows of arbitrary scale.  The results show that, at least for some granular flows, full understanding of their nature can only be reached if full account is taken of both the interactions between particles and those between particles and the interstitial fluid.

\section*{Acknowledgements}
\added{The authors thank three anonymous reviewers who helped to improve our manuscript, as well as O.~Pouliquen for his handling of it.}  
This work was funded from a grant under the UK NERC Environmental Mathematics and Statistics programme (NER/S/E/2004/12600). This research was supported also in part by the National Science Foundation under Grant No. NSF PHY11-25915 and AJH also acknowledges support from Max Planck Institute for the Physics of Complex Systems (Two-Phase Continuum Models for Geophysical Particle-Fluid Flows).  MAG carried out part of this work while holding a University Research Fellowship provided by the Institute of Advanced Study at the University of Bristol.  \added{This paper is LabEx Clervolc contribution no.~259.}  
\appendix
\section{Extended kinetic theory}\label{extendke}
\begin{figure}%
  \centering
  \includegraphics[trim=4cm 10cm 4cm 10cm, width=0.7\columnwidth]
    {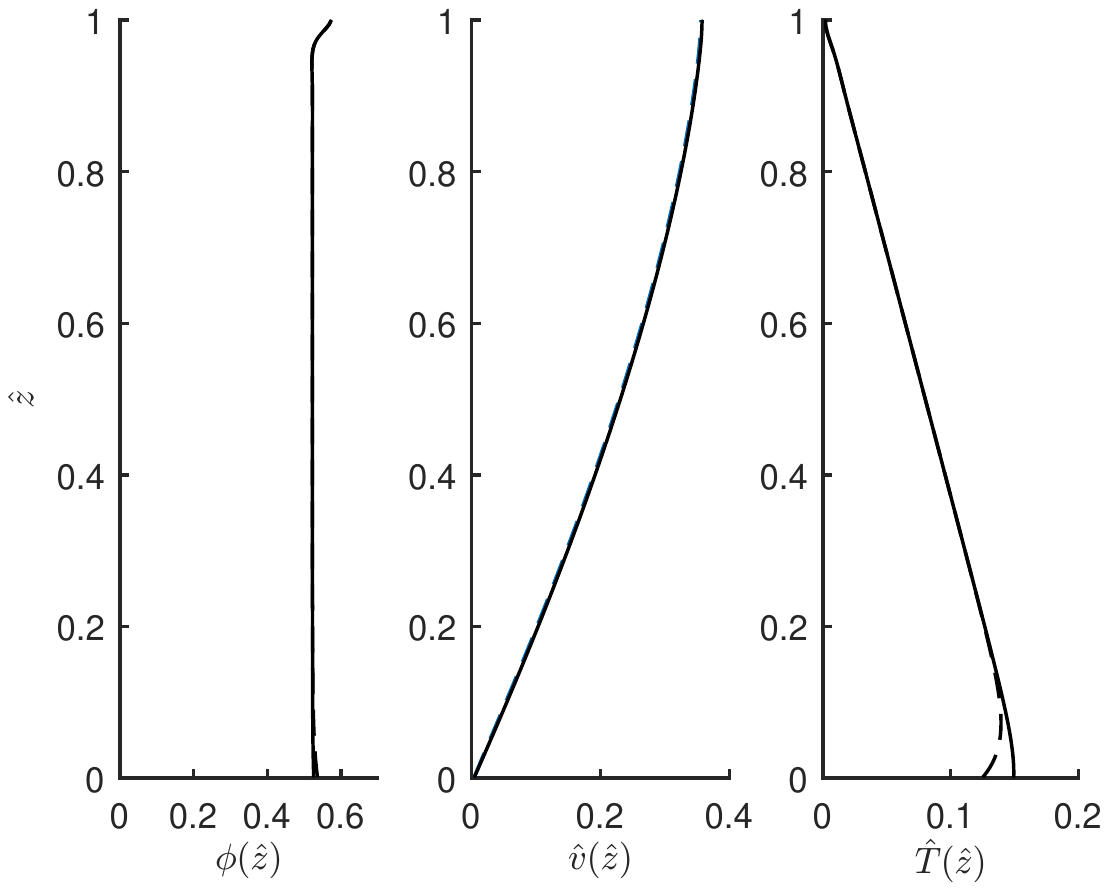}%
  \caption{The volume fraction, $\phi(\hat{z})$, velocity of the solid phase,  $\hat{v}({\hat z})$ and the granular temperature, $\hat{T}({\hat z})$, as functions of the dimensionless depth within the current for parameter values $R=10^{-3}$, $\psi=0.5$, $\phi_m=0.63$, $e=0.85$, $e_w=0.75$, $S=0.1$, $St=10^3\delta^2$, $\delta=0.01$ and $W_g=10^{-3}$ for extended kinetic theory (solid liness) and `standard' kientic theory (dashed lines).}
  \label{fig_ekt1}
\end{figure}
\added{In this appendix we analyse the consequences for the predicted flow field of employing the extended kinetic theory proposed by} \citet{jenk07} \added{and recently used to compute unfluidised flow down inclined planes by} \cite{jenk10,jenk12} and \cite{berz14}\added{.  In essence, the extension to kinetic theory is based upon the realisation that at higher concentrations, particles begin to form structures in the flow that have a correlation length in excess of their own diameter.  Thus the rate of dissipation is reduced - and in terms of the expression of the evolution of granular temperature \eqref{thermal}, the dissipation term is now given by $\rho_sf_3T^{3/2}/L_c$.}  \cite{jenk07} \added{suggested a phenomenological model for the length, $L_c$, in which its magnitude is proportional to the rate of compression that occurs along at least one axis in shear flows and inversely proportional to the agitation (the granular temperature) that can destroy these structures.  Thus in dimensional form for simple shear flows ${\bf v}=v(z){\bf\hat x}$,} \citet{jenk10} \added{propose}
\begin{equation}
\frac{L_c}{d}=\hbox{max}\left(1,\frac{\hat{c}(\phi g_0)^{1/3}d}{2T^{1/2}}\frac{\partial v}{\partial z}\right),\label{defL}
\end{equation}
\added{where $\hat{c}$ is a dimensionless constant of order unity (often $\hat{c}=1/2$).}  \citet{jenk10} \added{validate this formulation empirically for unfluidised granular flows.  We are not aware of any studies that have tested formulae for fluidised flows, but we can nevertheless employ this formulation \eqref{defL} to compute profiles of the volume fraction of particles, the velocity field and the granular temperature for typical parameter values used in this study (figure \ref{fig_ekt1}).  For a dimensionless fluidising gas flow rate, $W_g$ equal to $10^{-3}$ and a slope $S$ of $0.1$, we find negligible differences in the profiles apart from very close to the base of the flow.  Moreover the dimensionless volume flux per unit width for the `standard' kinetic theory $\hat{q}=0.1117$, while for the extended kinetic theory $\hat{q}=0.1127$.  }

\begin{figure}%
  \centering
  \includegraphics[trim=4cm 10cm 4cm 10cm,width=0.7\columnwidth]{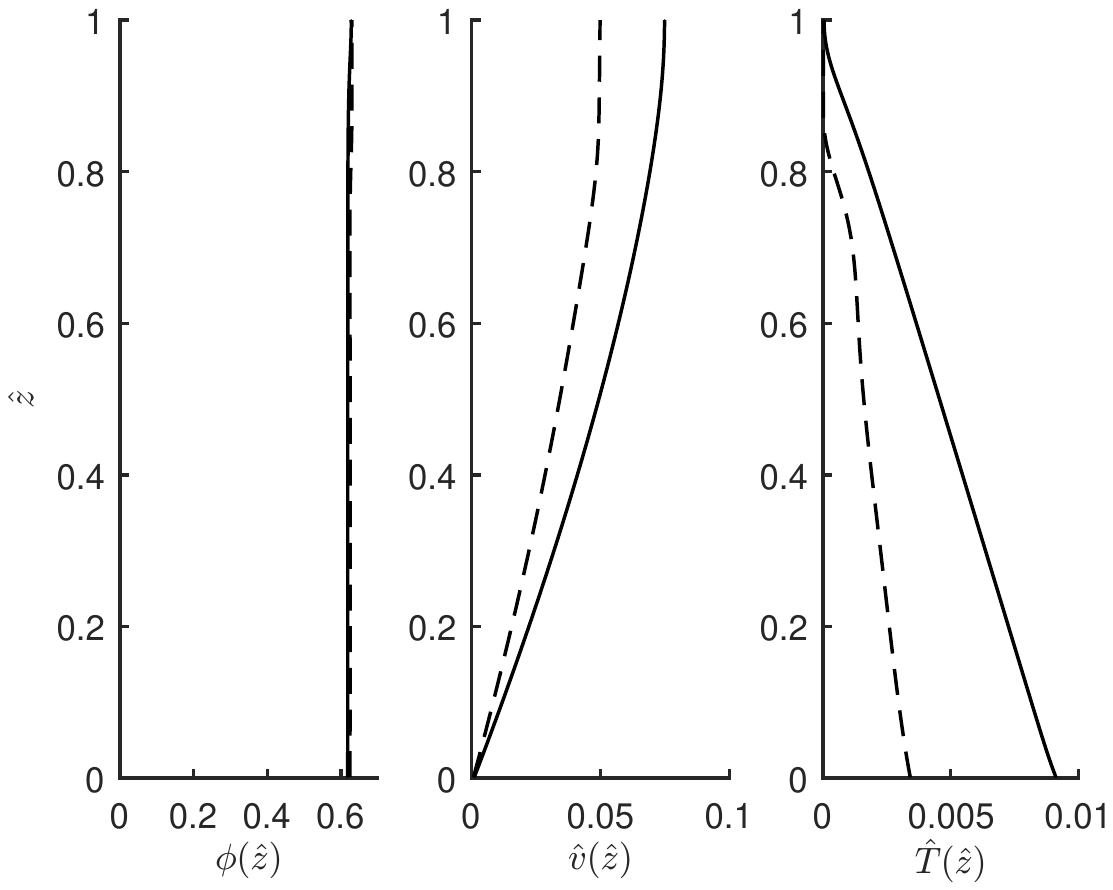}%
  \caption{The volume fraction, $\phi(\hat{z})$, velocity of the solid phase,  $\hat{v}({\hat z})$ and the granular temperature, $\hat{T}({\hat z})$, as functions of the dimensionless depth within the current for parameter values $R=10^{-3}$, $\psi=0.5$, $\phi_m=0.63$, $e=0.85$, $e_w=0.75$, $S=0.1$, $St=10^3\delta^2$, $\delta=0.01$ and $W_g=4.1\times 10^{-4}$ for extended kinetic theory (solid liness) and `standard' kinetic theory (dashed lines).}
  \label{fig_ekt2}
\end{figure}

\added{For more weakly fluidised flows, there can be a significant difference between the predictions of the two theories, because in these situations the concentration of particles is higher and thus $L_c/d$ exceeds unity in many parts of the flow.  For example when $W_g=4.1\times 10^{-4}$ and $S=0.1$, we find that extended kinetic theory predicts more energetic and faster moving flows (see figure \ref{fig_ekt2}). For these parameter values, the dimensionless volume flux, $\hat{q}=0.0198$ for the `standard' kinetic theory, whereas $\hat{q}=0.0284$ for the extended kinetic theory.  The flows that we consider in this study are more strongly fluidised than this example and thus we find it unnecessary to include this phenomenon in our analysis in the main body of this paper because it introduces negligible difference to the computed flow.}

\section{The effects of side-wall stresses}\label{sidewall}

In this appendix we analyse the effects of side-wall resistance on the motion of shallow fluidised flows down inclined channels (see \S\ref{unsteady}) and derive the first-order correction to the prediction of the front speed for flows that are unaffected by side walls.  We show that the reduction in front speed is proportional to $(H/B)^2$, where $H$ is the scale depth of the current given by \eqref{Hdef} and $B$ is the channel breadth.

The downslope flows studied experimentally are realised within a channel, the width of which is usually greater than the flow depth, but not far in excess of the depth.  Thus, it is feasible that side wall stresses may play a non-negligible role in the overall dynamics and may further retard the motion.  Indeed for flows along horizontal surface  for which the flow depth is much greater than the width of the channel, we postulate that the side wall stresses may even play a dominant role in the resisting the driving forces (see \S\ref{horizontal}).

We analyse the motion in a channel of width $B$ in a regime for which the volume fraction is spatially uniform and given by $\overline{\phi}$.  On depth- and width-averaging the streamwise balance of momentum \eqref{solidslopemom}, we find that the dimensional governing equation is given by
\begin{equation}
\begin{split}
\rho_s\overline{\phi}\left(\pd{}{t}\int_0^B\int_0h v\;\rd z \rd y+\pd{}{x}\int_0^B\int_0^h v^2\;\rd z\rd y
+g\cos\theta\int_0^B\int_0^h\pd{h}{x}\;\rd z\rd y\right)=\\
\rho_s \overline{\phi}g\sin\theta Bh -\int_0^B\sigma_{xz}(0,y)\;\rd y-\int_0^h\left(\sigma_{xy}(z,0)-\sigma_{xy}(z,B)\right)\rd z,
\label{eq:sidewall}
\end{split}
\end{equation}
where $\sigma_{xz}(y,0)=f_1\rho_s dT^{1/2}\partial v/\partial z$ denotes the basal shear stress and $\sigma_{xy}(0,z)=-\sigma_{xy}(B,z)=f_1\rho_s dT^{1/2}\partial v/\partial y$ denotes the side wall stresses.  

In general, to include these side wall effects, even if the flow had adjusted to a local balance that was independent of the streamwise coordinate, we would have to resolve the variations of the dependent fields in the $(y,z)$ plane (see, for example, \citealp{Oger201322}). In this subsection we take a different strategy and develop a model which is appropriate to the regime $H/B\ll 1$, where $H$ is the scale depth of the flow and given by \eqref{Hdef}.  In this regime, we treat the granular temperature and flow field as predominantly varying with the distance from the basal boundary  and assume that these flow fields adopt the form established in \S\ref{asymptotic} (see \citealp{jenk10}).  This approach was used above to derive the depth-averaged model above, but is now generalised to include lateral gradients in order to model the side wall stresses.

Adopting the dimensionless variables using the scales of \eqref{slopescale}, we estimate the velocity gradient at the side wall $\partial {\hat{v}}/\partial \hat{y}=\alpha H \hat{v}/B$, where $\alpha$ is a dimensionless constant of order unity.  We may then compute the depth and width averages to deduce a governing equation for a travelling wave solution $\hat{h}(x,t)=\hat{h}(x-ct)$ that features the additional stresses due to the side walls  ({\it cf.} \eqref{travel_ode}); it is given by
\begin{equation}
  \frac{{\cal R}c^2}{4}{\hat h}'+{\hat h}{\hat h}'=\hat{h}-\frac{25}{4}\frac{c^2}{{\hat h}^2}-\frac{25\alpha}{9} \left(\frac{H}{B}\right)^2 c^2.
  \label{sidewalltravel_ode}
\end{equation}
The final term of \eqref{sidewalltravel_ode} represents the extra stress due to the side walls.  Then using the uniform conditions far from the front $(\hat{x}-c\hat{t}\to-\infty)$ we deduce that
\begin{equation}
0=\frac{1}{c}-\frac{25c^4}{4}-\frac{25\alpha H^2c^2}{9B^2}.
\end{equation}
Thus the speed, $c$, is reduced by the action of the side-wall stresses and in the regime $\alpha \left(H/B\right)^2\ll 1$, we find that
\begin{equation}
c=\left(\frac{2}{5}\right)^{2/5}\left(1-\frac{4\alpha}{45}\left(\frac25\right)^{-4/5}\left(\frac{H}{B}\right)^2+\ldots\right)
\end{equation}

The value of the coefficient of the second term of the expansion is $\sim$0.2.



\bibliographystyle{jfm}

\end{document}